\documentclass[usenatbib]{mnras}

\usepackage{amssymb}
\usepackage{amsmath}
\usepackage{relsize}
\usepackage[pdftex]{graphicx}
\usepackage{hyperref}
\usepackage{mathptmx}
\usepackage{color}

\newcommand{\be}{\begin{equation}}
\newcommand{\ee}{\end{equation}}

\makeatletter
\newcommand*{\rom}[1]{\expandafter\@slowromancap\romannumeral #1@}
\makeatother

\newcommand{\cosmosis}{\textsc{cosmosis}}
\newcommand{\acamb}{\textsc{axionCAMB}}
\newcommand{\emcee}{\textsc{emcee}}

\newcommand{\phbar}{\ensuremath{\bar{\phi}}}

\newcommand{\jhep}{JHEP}		

\hypersetup{draft}

\begin{document}
\newcommand{\tkDM}[1]{\textcolor{red}{#1}}  
\newcommand{\tkRH}[1]{\textcolor{magenta}{#1}}  
\newcommand{\tkDG}[1]{\textcolor{green}{#1}}  

\title{Using the Full Power of the Cosmic Microwave Background to Probe Axion Dark Matter}
\author[R. Hlo{\v z}ek, D. J. E. Marsh, D. Grin] {Ren\'{e}e Hlo{\v z}ek$^1$, David~J.~E.~Marsh$^2$\thanks{Corresponding Author: david.marsh@kcl.ac.uk}, and Daniel~Grin$^3$\\
$^1$Dunlap Institute for Astronomy and Astrophysics \& Department of Astronomy and Astrophysics\\ University of Toronto, 50 St. George Street, Toronto, Ontario, Canada M5S 3H4,\\ $^2$King's College London, Strand, London, WC2R 2LS, United Kingdom, \\$^3$Department of Physics and Astronomy, Haverford College, 370 Lancaster Avenue, Haverford, PA 19041, United States.}

\date{\today}
\maketitle

\begin{abstract}

The cosmic microwave background (CMB) places stringent constraints on models of dark matter (DM) that deviate from standard cold DM (CDM), and on initial conditions beyond the scalar adiabatic mode. Here, the full \textit{Planck} data set (including anisotropies in temperature, $E$-mode polarisation, and lensing deflection) is used to test the possibility that some fraction of the DM is composed of ultralight axions (ULAs). This represents the first use of CMB lensing to test the ULA model. We find no evidence for a ULA component in the mass range $10^{-33}\leq m_a\leq 10^{-24}\text{ eV}$. We put percent-level constraints on the ULA contribution to the DM, which improve by up to a factor of two compared to the case with temperature anisotropies alone.  Axion DM also provides a low-energy window onto the high-energy physics of inflation through the interplay between the vacuum misalignment production of axions and isocurvature perturbations. 
We perform the first systematic investigation into the parameter space of ULA isocurvature, using an accurate isocurvature transfer function at all $m_{a}$ values.
We precisely identify a ``window of co-existence" for $10^{-25}\text{ eV}\leq m_a\leq10^{-24}\text{ eV}$ where the data allow, simultaneously, a $\sim10\%$ contribution of ULAs to the DM, and $\sim 1\%$ contributions of isocurvature and tensors to the CMB power. ULAs in this window (and \textit{all} lighter ULAs) are shown to be consistent with a high-energy-scale inflationary Hubble parameter, $H_I\sim 10^{14}\text{ GeV}$. The window of co-existence will be fully probed by proposed CMB Stage-IV observations with increased accuracy in the high-$\ell$ lensing power and low-$\ell$ $E$ and $B$-mode polarisation. If ULAs in the window exist, this could allow for two independent measurements of $H_I$ in the CMB using the axion DM content and isocurvature, and the tensor contribution to $B$-modes. 


Preprint: KCL-PH-TH/2017-39

\end{abstract}

\section{Introduction}
\label{sec:intro} 

Observations of the cosmic microwave background (CMB) over the last twenty five years have been pivotal in establishing the standard cosmological model~\citep[e.g.][]{1992ApJ...396L...1S,1994ApJ...420..439M,2001PhRvL..86.3475J,2013ApJS..208...20B,2013ApJ...779...86S,2016A&A...594A..13P,2016PhRvL.116c1302B,2016arXiv161002360L}, and will continue to play a key role in the coming decades~\citep[e.g.][]{2016arXiv161002743A}. A central role in the standard cosmological model is played by Cold Dark Matter (CDM). Using a combination of temperature ($T$), $E$-mode polarisation, and weak gravitational lensing power spectra of the CMB, the CDM density is measured to be $\Omega_ch^2=0.1199\pm 0.0022$~\citep{2016A&A...594A..13P}. 

CDM is the only form of Dark Matter (DM) required by the data, providing the backbone of the ``cosmic web'' of large scale structure on linear scales. One well-motivated theoretical possibility for the CDM is the axion, a new hypothetical particle motivated by the charge-parity conjugation problem of quantum chromodynamics (QCD)  \citep{pecceiquinn1977,weinberg1978,wilczek1978,1983PhLB..120..133A,1983PhLB..120..137D,1983PhLB..120..127P}. On scales of relevance to cosmology, QCD axions with mass $m_a \lesssim 10^{-4}~\mathrm{ eV}$ are produced non-thermally, have the correct cosmological relic density, and behave as CDM. There are very few direct constraints on the QCD axion CDM parameter space~\citep[e.g.][]{2010PhRvL.104d1301A}. The lightest supersymmetric particle also provides a canonical CDM candidate \citep{1996PhR...267..195J}, though limits on ``natural'' regions of its parameter space are strong~\citep[e.g.][]{2017PhRvL.118b1303A}.

The CMB places stringent limits on a variety of theoretical models for DM beyond CDM, including limits on the mass of standard model neutrinos~\citep{2016A&A...594A..13P}, thermal axions~\citep{2013JCAP...10..020A}, DM interactions~\citep{Wilkinson:2013kia,2014JCAP...05..011W}, and generalized models~\citep{2016PhRvD..93l3527C,2016ApJ...830..155T}. A particularly interesting bound comes from constraints on non-thermally produced ultralight axions (ULAs), as this establishes an absolute lower-bound on the DM particle mass from linear observables, $m_a\gtrsim 10^{-24}\text{ eV}$~\citep{2015PhRvD..91j3512H}. 

The constraints of \cite{2015PhRvD..91j3512H} (hereafter, H15) not only bound the particle mass of the dominant component of DM, but also place stringent limits on the axion DM density over many orders of magnitude in particle mass ($10^{-33}~\mathrm{ eV}\lesssim m_{a}\lesssim 10^{-24}~\mathrm{ eV}$) for models in which the DM is a mixture of ``standard" CDM and ULAs. Cosmology thus probes not only the abundance but also the composition of the dark sector. ULAs are expected to be ubiquitous in string theory \citep[e.g.][]{2006JHEP...06..051S,2006JHEP...05..078C,axiverse} and provide candidates for dark matter (if $m_{a}\gtrsim 10^{-27}~\mathrm{ eV}$) \textit{or} dark energy (if $m_{a}\lesssim 10^{-27}~\mathrm{ eV}$) components, depending on their mass.\footnote{The boundary between the two regimes is a matter of convention. Here it has been chosen as approximately the inverse Hubble parameter at matter-radiation equality.} For fairly natural values of the axion initial field value in string models, $\bar{\phi}_i\sim f_a\lesssim 10^{17}\text{ GeV}$, where $f_a$ is the axion ``decay constant'', ULAs contribute at the percent level to the cosmic critical density~\citep{2016PhR...643....1M,2017PhRvD..95d3541H}.

H15 used the \cite{2014A&A...571A..15P} CMB $TT$ auto-correlation power spectrum and the \cite{Parkinson:2012vd} galaxy-galaxy auto-correlation power. In the present work, we use the full \cite{2016A&A...594A..11P} CMB data release, including $T$, $E$, and lensing power spectra and cross correlations to give the up-to-date precision constraints on the ULA relic density from the CMB. This work thus includes the first application of CMB weak lensing data to test the ULA hypothesis, a first step towards the sensitivity to ULAs of future efforts like CMB Stage-IV~\citep{2017PhRvD..95l3511H}, a proposal for a nearly cosmic-variance limited ground-based CMB polarisation experiment.

Another key piece in the standard cosmological model is the theory of inflation~\citep{1981PhRvD..23..347G,1982PhRvL..48.1220A,1982PhLB..108..389L}. Inflation establishes the initial conditions of the hot big bang, including a spectrum of nearly scale-invariant, nearly Gaussian density fluctuations, and the CMB can be used to test a vast array of models of inflation~\citep{2014PDU.....5...75M,2014JCAP...03..039M}. CMB observations significantly constrain the model space, for example ruling out many ``large field'' models. In spite of the impressive progress in constraining inflation, the inflationary energy scale, as parameterised by the Hubble parameter during inflation, $H_I$~\citep{1984PhLB..147..403L}, remains unknown. The parameter $H_{I}$ determines the ratio $r$ of primordial tensor-to-scalar perturbations, which could be determined by a detection of large angle, primordial, CMB $B$-mode polarisation correlation~\citep{1997PhRvD..55.7368K,1997PhRvL..78.2054S,1997PhRvL..78.1861L}.

\begin{figure} 
\includegraphics[width= \columnwidth]{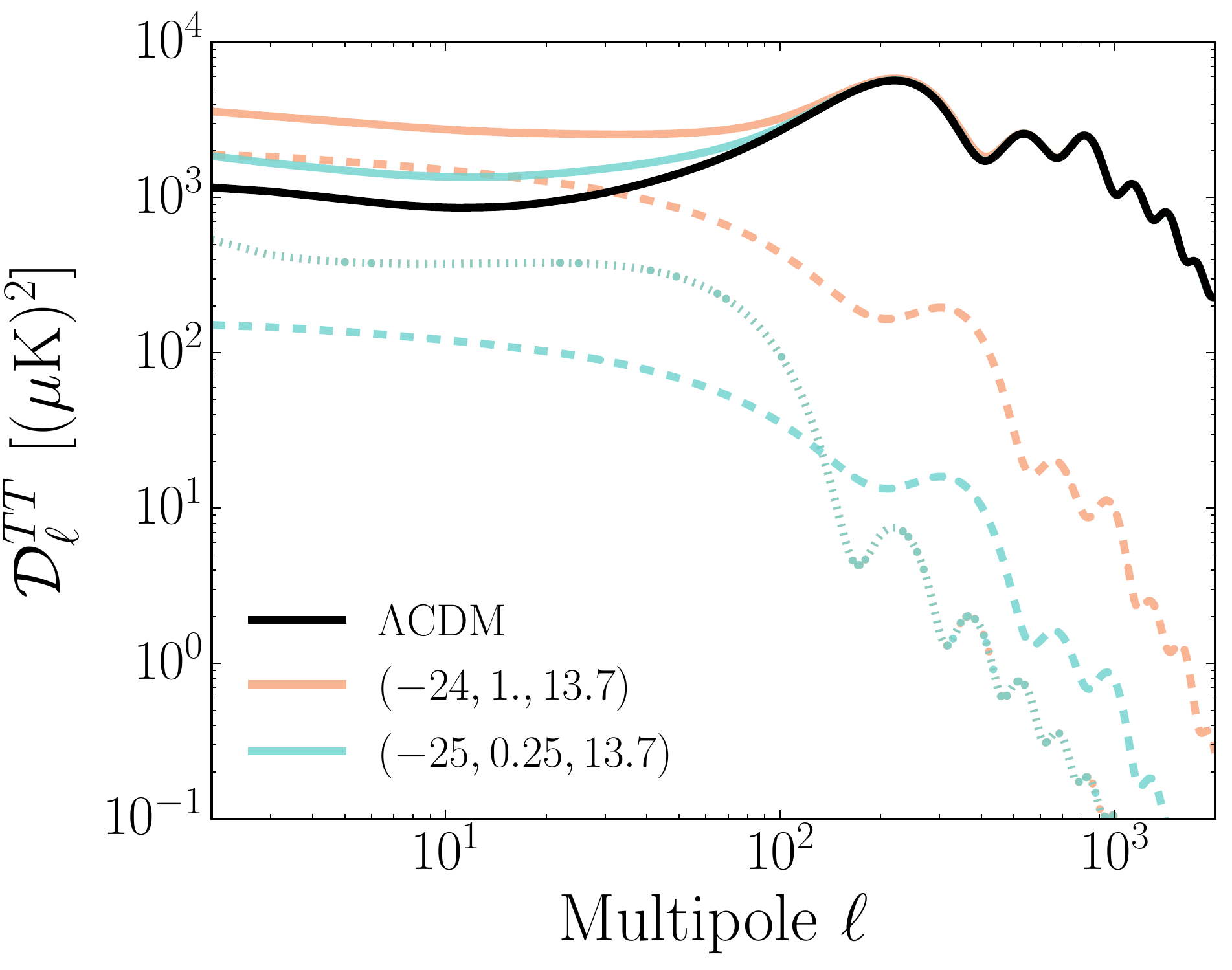} 
 \caption{{\bf Isocurvature and Tensor Modes:} The Legend labels for ULA models give $(\log_{10}[m_a/\text{eV}],\Omega_a/\Omega_d,\log_{10}[H_I/\text{GeV}])$. The curve marked $\Lambda$CDM has zero isocurvature power and zero tensor power, with parameters set to fiducial $\Lambda$CDM values. Temperature power spectra are the sum (solid) of those generated by the scalar adiabatic mode (dot-dashed, indistinguishable from $\Lambda$CDM for the models and scales shown, as evidenced by the perfect overlap of the cyan dot-dashed and black solid curves), the scalar isocurvature mode (dashed), and the tensor power (dotted). The isocurvature power is lowered by reducing \textit{either} $\Omega_a$ \textit{and/or} $H_I$, allowing a significant axion isocurvature component to co-exist with a significant tensor component for masses $m_a\sim \mathcal{O}(\mathrm{ few})\times 10^{-25}\text{ eV}$, and $\Omega_a/\Omega_d\sim \mathcal{O}(0.1)$. With a consistent $TT$ spectrum, the tensor component can be measured in low-$\ell$ $BB$ (Fig.~\ref{fig:iso_modes}), while the effects of axions can be detected in high-$\ell$ lensing (Fig.~\ref{fig:adi_DM}), and isocurvature measurements improved with low-$\ell$ $EE$ (Fig.~\ref{fig:iso_mass_plot}).
\label{fig:opener}}
\end{figure}

If axions contribute significantly to DM and the symmetry-breaking that sets their relic density took place during the inflationary epoch, \textit{isocurvature} perturbations would be produced, as for any nearly massless field during inflation ~\citep[e.g.][]{1983PhLB..126..178A,1985PhRvD..32.3178S,1991PhRvL..66....5T,1992PhRvD..45.3394L,2004hep.th....9059F,2008PhRvD..78h3507H,2009ApJS..180..330K}. In that case, the amplitude of the uncorrelated isocurvature contribution to the CMB power spectrum, $\beta_\mathrm{ iso}$, can also be used to determine $H_I$. Such a determination, however, depends on the axion DM density. For ULAs to have cosmologically relevant relic densities today, initial field values $\phi_{i}$ must be large, that is, $\bar{\phi}_i\sim f_a\gg 10^{14}\text{ GeV}$ (H15) (in contrast to QCD axions). This more or less guarantees that $f_{a}\gg H_{I}$, implying that isocurvature perturbations will be produced. The CMB can thus be used to probe both the ULA DM density and isocurvature amplitude, implying the existence of a window in which detectably large component of mixed axion DM can co-exist with high energy-scale inflation. This potentially  allows for a simultaneous detection of isocurvature, tensor modes, and ULAs using the CMB~\citep{2013PhRvD..87l1701M,2014PhRvL.113a1801M,2016arXiv161002743A}, as shown schematically in Fig.~\ref{fig:opener}. In this window a determination of $H_I$ is possible using two complementary physical mechanisms, namely: the isocurvature amplitude plus the distinctive effects of ULAs on high-$\ell$ anisotropies, and the low-$\ell$ $B$-mode power of tensors.

Here we extend the methods of H15 to include the axion isocurvature mode, and use this to present the first combined constraints on $H_I$ and $\Omega_a$. This requires computing the shape of the CMB isocurvature power spectrum induced by ULAs, which is different from a CDM power spectrum due to scale-dependent growth and a non-standard cosmological expansion history \citep{2009PhLB..680....1H,2013PhRvD..87l1701M,2014PhRvL.113a1801M,2016PhR...643....1M}, leading to modified ULA isocurvature power spectrum. This spectrum smoothly varies from the CDM to quintessence-type result as the ULA mass $m_{a}$ is lowered. 

This computation requires extending the standard cosmological initial conditions \citep{2000PhRvD..62h3508B} to ULAs. We determine these initial conditions using an eigenmode analysis \citep{2003PhRvD..68f3505D}, correcting past work on quintessence-type isocurvature \citep{1999PhRvD..59l3508P}, and obtaining the full early-time power-series solutions for ULA isocurvature mode evolution. In \cite{2013PhRvD..87l1701M} and \cite{2014PhRvL.113a1801M}, we used accurate spectra to estimate constraints, without a full parameter-space analysis. Here we use accurate ULA isocurvature spectra and a comprehensive sweep of parameter space to probe the inflationary energy scale in ULA cosmologies.

We summarise our data, methodology and introduce parameter definitions in Section \ref{sec:methods}. We introduce basic axion theory, as well some qualitatively new aspects relating to CMB polarisation, lensing, and isocurvature in Section~\ref{sec:theory}. For a more complete description, we expand on the details of axion physics and computation of initial conditions in Appendices \ref{appendix:axion_basics}-\ref{appendix:power_law}.
The main results of our ULA constraints using the full \textit{Planck} CMB data set for the adiabatic mode are presented in Section \ref{sec:dm_constraints}, and the full constraints including the axion isocurvature signature are shown exhaustively in Section~\ref{sec:iso_constraints}. As one might have guessed, we both discuss the results and speculate on the future in Section~\ref{sec:discussion}. Finally, those interested in the nitty gritty of the sampling and analysis should head over to Appendix \ref{appendix:sampling} for some interesting discussion on priors, particularly in the inflationary parameter space. The main computational tools used in this analysis are the Boltzmann code \acamb, the analysis suite \cosmosis, and the sampler \emcee. The data used are \cite{2016A&A...594A..11P} and \cite{Kazin:2014qga}.

\section{Data and Analysis Methods}
\label{sec:methods}
\begin{table*}
    \begin{tabular}{c|c|c|c}
    {\bf Model choice} & {\bf axion mass $m_a$} & {\bf Data combination} & {\bf abbreviation}\\
    \hline
    adiabatic &  $\log(m_a) = -28$ & \textit{Planck} temperature & $\mathrm{adi}\mathrm{T}$\\
    adiabatic & $\log(m_a) = -28$ & \textit{Planck} temperature + polarisation & $\mathrm{adi}~\mathrm{T+P}$\\
    adiabatic &  $\log(m_a) = -28$ & \textit{Planck} temp.+pol.+lensing+BAO & $\mathrm{adi}~\mathrm{T+P+lens+BAO}$ \\
     adiabatic &  $-33\leq \log(m_a/\mathrm{ eV}) \leq -24$ & \textit{Planck} temperature+ polarisation + lensing & $\mathrm{adi}~ \mathrm{T+P+lens}$ \\
    adiabatic + isocurvature & $-33\leq \log(m_a/\mathrm{ eV}) \leq -24$ & \textit{Planck} temperature + polarisation+lensing &$\mathrm{iso}~\mathrm{T+P+lens}$ \\ 
    \end{tabular}
    \caption{{\bf Data combinations used for different axion masses and initial conditions.} The constraints over the full range of masses are shown in Figure~\ref{fig:lens_u_plot}. We choose our baseline data combination without BAO, since isocurvature power could alter the BAO peak shape, and the standard likelihood does not account for this. In order to study the improvement of constraints with the inclusion of polarisation, lensing and distance measures to the temperature constraints, we test a well constrained model with $m_a=10^{-28}~\mathrm{eV}$ for a variety of combinations of data. This is shown in Figure~\ref{fig:m28_constraints}. 
    \label{table:runs}}
\end{table*}

\begin{table}
    \begin{tabular}{c|c|c|c}
    \multicolumn{1}{c}{\bfseries Parameter} & 
    \multicolumn{3}{|c|}{\bfseries Mass range}\\ \cline{2-4}
    &Dark energy & Belly & Dark Matter \\ \hline 
    $\Omega_ah^2$ & $<0.5$ & $< 0.05$ & $ < 0.15$ \\ 
     $\Omega_ch^2$ & $0.09-0.15$ & $0.09-0.15$ & $ < 0.15$ \\ 
    \end{tabular}
    \caption{{\bf Prior ranges for the axion energy density:} to efficiently explore the full parameter space, we set our prior ranges based conservatively on the upper bounds from H15. In the `belly' region, we expanded the upper bound to $0.05$ for completeness.
    \label{table:priors}}
\end{table}

Before detailing our investigation, we lay out some conventions. The cosmic scale factor is $a$, and the Hubble rate is $H=(\mathrm{ d}a/\mathrm{ dt})/a$. The standard synchronous-gauge metric fluctuations $h_{m}$ and $\eta_{m}$ of \cite{1995ApJ...455....7M} are used (the subscript $m$ is used to avoid confusion with either the dimensionless Hubble constant $h$ or the conformal time $\eta$). Wavenumbers are denoted by $k$, and we use the same symbol for a field and its Fourier transform. We define $\mathcal{D}_\ell=\ell (\ell+1)C_\ell/2\pi$, where $C_\ell$ is the angular power spectrum. For some particle physics quantities we use natural units where $\hbar=c=1$, for example the reduced Planck mass is $M_\mathrm{ pl}=1/\sqrt{8\pi G}=2.435\times 10^{18}\text{ GeV}$. Throughout we work to first order in cosmological perturbation theory.

The computations presented in this paper are performed using \acamb~\citep[H15,][]{2017PhRvD..95l3511H}, a modified version of the \textsc{camb} Boltzmann code~\citep{2000ApJ...538..473L}. Data analysis is performed using the \cosmosis~suite~\citep{cosmosis}, which we have modified to sample the axion DM isocurvature parameter space. This suite of codes is ideal for computing constraints over a wide range of data sets and is streamlined to input spectra and nuisance parameters self-consistently throughout. We use Bayesian techniques of Markov Chain Monte Carlo (MCMC) sampling with \emcee~\citep{emcee}.\footnote{\acamb~is available for download at \url{https://github.com/dgrin1/axionCAMB}. The isocurvature modifications and \cosmosis~module will be available shortly. The MCMC chains for the present paper will also be made available at \url{http://www.dunlap.utoronto.ca/~hlozek/AxiChains}.}

This research made use of \href{http://www.astropy.org}{Astropy}, a community-developed core Python package for Astronomy \citep{astropy}.

The results of H15 used the \textit{Planck} (2013) $TT$ CMB auto-power and the WiggleZ galaxy-galaxy auto-power. Here we use the 2015 \textit{Planck} data release. In particular,
we include the \textit{Planck} \texttt{plik\_lite\_v18\_TTTEEE} foreground-marginalised likelihood for multipoles $\ell=30-2508$, and use the \texttt{lowl\_SMW\_70\_dx11d\_2014\_10\_03\_v5c\_Ap} low-$\ell$ TEB likelihood for the lowest multipoles \citep{2016A&A...594A..11P}. We add in information from CMB lensing with \textit{Planck} through the lensing likelihood \texttt{smica\_g30\_ftl\_full\_pp}. For some pemutations of our results, we also include measurements of Baryon Acoustic Oscillations (BAO) from the WiggleZ survey \citep{Kazin:2014qga}.\footnote{We note, however, that BAO are only included through geometric constraints to the acoustic scale, without modelling how isocurvature could change the shape of the power spectrum and thus the BAO bump in the galaxy correlation function \citep{2010JCAP...10..009M,2012JCAP...07..021M}; we will explore this issue further in future work. Due to this, we choose the conservative option not to use BAO in our isocurvature constraints.}  The data combinations for different model assumptions are summarised in Table~\ref{table:runs}. We discuss details of our sampling methodology in Appendix~\ref{appendix:sampling}, including the starting point procedures in the different `regimes' of axion mass parameter space. We also describe our convergence tests, which make use of \cite{2005MNRAS.356..925D}.

In H15 we presented a discussion on the ULA parameter space and the mass-dependent covariance between ULAs, dark matter, and dark energy. The U-shape of the posterior makes this parameter space difficult to sample, and in the present work we restrict ourselves to constraining the ULA density at fixed mass. This means that we cannot discuss whether or not the data show a preference for a given axion mass, but we can bound the axion energy density \textit{given} a specific mass $m_a$. H15 showed that the binned analysis and the full analysis yield consistent results bin-by-bin. 

The cosmological parameters are:
\be
\Xi_\mathrm{ standard} = (A_s,n_s,\tau,\Omega_b h^2, h,\Omega_c h^2)\,
\label{eqn:param_vec_standard},
\ee
where $A_s$ is the amplitude of scalar fluctuations and $n_s$ is the scalar spectral index (both defined at a pivot scale $k_0=0.05$ Mpc$^{-1}$), $\tau$ is the optical depth to reionisation (which we assume to be instantaneous), $\Omega_bh^2$ is the baryon density, $\Omega_c h^2$ is the CDM density and $H_0=100\,h\,\mathrm{ km\, s}^{-1}\mathrm{ Mpc}^{-1}$ is the Hubble constant. In addition to these standard parameters, we include the axion parameters:
\be
\Xi_\mathrm{ axion} = (\Omega_a h^2, m_a, H_I)\,
\label{eqn:param_vec_axion},
\ee
namely the axion energy density,  $\Omega_a h^2$, the axion mass, $m_a$, and the inflationary Hubble rate, $H_I$, defined when the pivot scale exits the horizon. 

Derived parameters are the cosmological constant density, $\Omega_\Lambda$, the total dark sector density, $\Omega_d=\Omega_a+\Omega_c$, the axion fractional density, $\Omega_a/\Omega_d$, the isocurvature amplitude and spectral index, $A_\mathrm{ iso}$, $n_\mathrm{ iso}$, the isocurvature fraction, $\beta_\mathrm{ iso}=(\Omega_a/\Omega_d)^2A_\mathrm{ iso}/(A_s+A_\mathrm{ iso})$, the axion initial field value, $\phbar_i$, the tensor-to-scalar ratio, $r$. To avoid confusion, when $r$ is strictly a derived parameter (i.e. in the posterior PDF) we append a superscript, $r^\mathrm{ (d)}$. We hold the parameters of the neutrino sector fixed, with $N_\mathrm{ eff}=2.046$ massless species, and $N_\mathrm{ eff}=1$ massive species with $m_\nu=0.06\text{ eV}$, fiducial choices consistent with \cite{2016A&A...594A..20P}.

We also include the \textit{Planck} calibration parameter $a_p,$ which we vary around unity. We use the foreground-marginalised \textit{Planck} likelihood, thus allowing us to ignore CMB foreground parameters \citep{2016A&A...594A..11P}. We consider fixed axion mass bins and vary axion density (and all other parameters) independently in each bin. Because of the distinct physics of ULAs, we choose a different prior range for $\Omega_a h^2$ depending on the axion mass range: the `dark-energy-like' axions ($10^{-33}\text{ eV}\leq m_a \leq10^{-31}\text{ eV}$), the constrained `belly region' ($10^{-29}\text{ eV}\leq m_a \leq10^{-26}\text{ eV}$) and the `dark-matter-like' axions ($10^{-25}\text{ eV}\leq m_a \leq10^{-24}\text{ eV}$). The prior ranges for the parameters of interest are shown in Table~\ref{table:priors}. These choices are informed by the results of H15 and speed up the numerical convergence of our results without otherwise affecting them.

\begin{figure*}
\includegraphics[width=\textwidth]{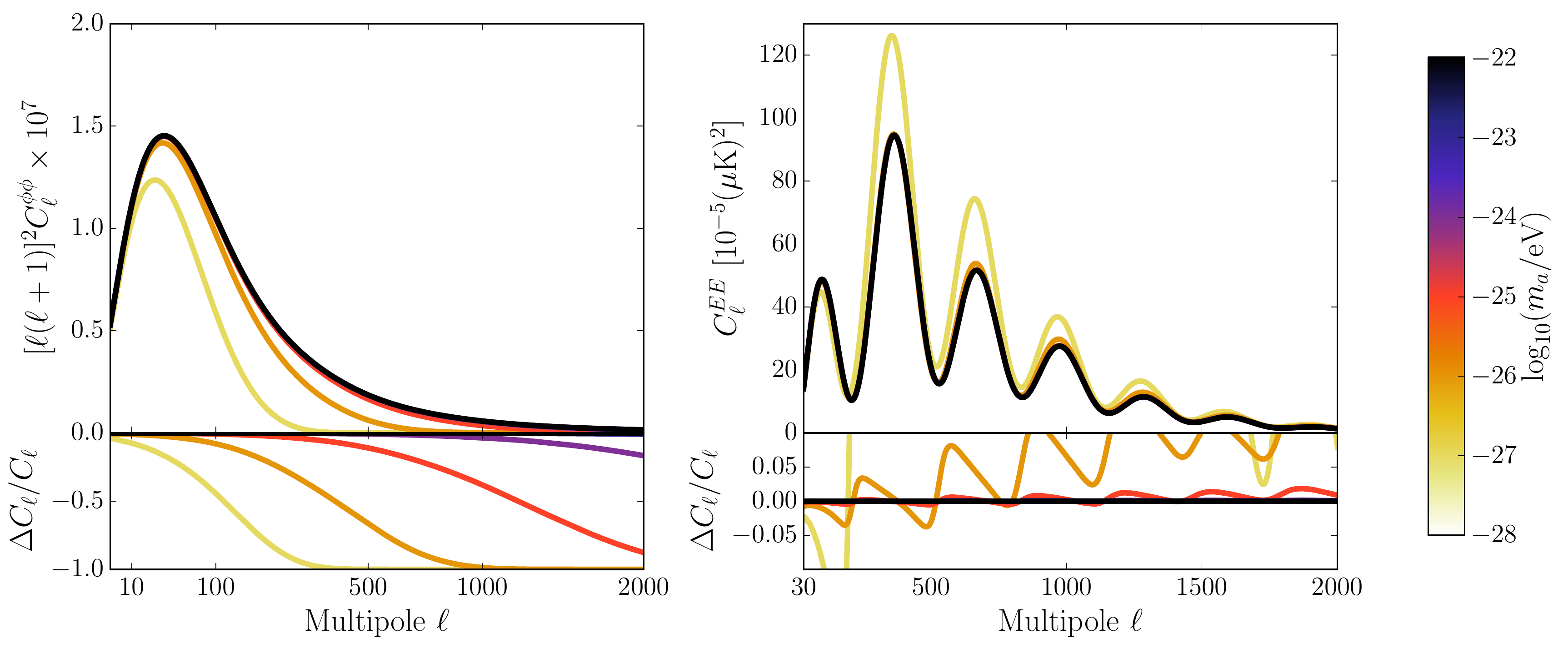} 
 \caption{{\bf Adiabatic Power Spectra for DM-like Axions:} We fix axions to be all of the DM, with $\Omega_a h^2=0.12$ \textit{Left Panel:} CMB lensing deflection auto power. The dominant effect visible is the axion Jeans scale, which suppresses power for $\ell>\ell_J$, with $\ell_J$ increasing with increasing axion mass. Effects for $\ell<2000$ are as large as 10\% up to $m_a=10^{-24}\text{ eV}$. However, the reference \textit{Planck} lensing likelihood only uses multipoles $\ell<500$. \textit{Right Panel:} $E$-mode polarisation auto power. The dominant effect comes from the $w_a$-transition, which affects the diffusion damping scale, as well as the amplitude of the early integrated Sachs-Wolfe effect. Effects are larger than 1\% up to $m_a=10^{-25}\text{ eV}$.}
\label{fig:adi_DM}
\end{figure*}

\section{Ultralight Axions and CMB Observables}
\label{sec:theory}

\subsection{Introduction to Axions}

We begin with a qualitative overview of axion cosmology, and refer the reader to H15, \cite{2016PhR...643....1M}, and Appendix~\ref{appendix:axion_basics} for a more detailed discussion. 

Axions (or more generally ``axion-like particles'', ALPs) are pseudoscalar, pseudo Nambu-Goldstone bosons of a spontaneously broken global $U(1)_A$ chiral symmetry. The symmetry is broken when the temperature (or Gibbons-Hawking temperature in the case of quasi de-Sitter space) of the Universe drops below the critical temperature $T_c\approx f_a$, with $f_a$ being the vacuum expectation value (vev) of a complex scalar field. The vev is also known as the ``axion decay constant'', and it is expected to be a high energy quantity significantly larger than the weak scale, but less than the Planck scale. The angular degree of freedom, $\theta=\phi/f_a$, is the massless axion field and it is invariant under the continuous shift symmetry, $\theta\rightarrow \theta+\xi$ for any real number $\xi$. Therefore, after spontaneous symmetry breaking there is no preferred value of the axion field, and $\theta\in \mathcal{U}[-\pi,\pi]$ with $\mathcal{U}$ the uniform distribution.

The existence of the shift symmetry enforces that in the classical Lagrangian the axion couples only derivatively to other fields, i.e. $\mathcal{L}\supset (\partial_\mu\phi)\hat{O}^{\mu}$ for any operator $\hat{O}^\mu$. Because of this symmetry structure, axions mediate no long range forces between matter, except in the presence of a spin-polarizing magnetic field~\citep[][]{Moody:1984ba,Srednicki:1985xd,2013PhRvD..88c5023G,2014PhRvL.113p1801A}. These couplings allow for a variety of possible laboratory searches for axions, and constraints from astrophysical processes. For reviews see \cite{2008LNP...741...51R,2015ARNPS..65..485G}. We will not concern ourselves with axion interactions, since they are highly model dependent.

The shift symmetry also forbids axions from possessing any \textit{perturbative} mass term. Axions develop masses via \textit{non-perturbative} means. For example, if the axion is coupled to fermions charged under a non-Abelian gauge group like QCD, then quantum anomalies and instantons generate a potential and thus a mass for the axion \citep[see e.g.][]{1988assy.book.....C,2003qftn.book.....Z}. Other non-perturbative effects apart from gauge theory instantons can also generate axion masses~\citep[e.g.][and references therein]{2006JHEP...06..051S,2017arXiv170607415A}. Quantum non-perturbative effects, by definition, depend exponentially on constants and fields in the classical theory. This means that non-perturbatively generated masses can span many orders of magnitude in a technically natural manner. The perturbative shift symmetry protects these masses from any further corrections.

Once the potential has ``switched on'' \citep[some non-perturbative effects such as those in QCD have non-trivial temperature dependence, e.g.][]{1981RvMP...53...43G}, the axion begins to minimize its energy by ``rolling'' down its potential to the minimum. Cosmologically, this motion is damped by ``Hubble friction''. When friction dominates, the field is underdamped, and the axion energy density remains approximately constant. The axion only begins to roll on relevant time scales once $H(t_{\rm osc})<m_a(t_{\rm osc})$~\citep{1983PhLB..120..133A,1983PhLB..120..137D,1983PhLB..120..127P}. After this time, the axion field undergoes coherent damped oscillations in an effectively quadratic potential, and the energy density in the coherent state scales like that of pressureless matter, $\rho\sim a^{-3}$~\citep{1983PhRvD..28.1243T}. We consider only the minimal assumption for the axion potential of a temperature independent mass term.

Thus, the cosmological life of the axion is determined by two times:
\begin{itemize}
\item Symmmetry Breaking: $T(t_{\rm SSB})\approx f_a$. This defines the epoch when initial conditions are generated. For reasons that will become clear, we focus exclusively on the case when symmetry breaking occurs before or during inflation (and it is not restored later). This is always the case for $f_a>10^{14}\text{ GeV}$.
\item Oscillations: $H(t_{\rm osc})\approx m_a$. This divides axion behaviour in two. For $t<t_{\rm osc}$ axions are ``DE-like''; for $t>t_{\rm osc}$ axions are ``DM-like'' (assuming monotonicity of $H$). Since in the standard cosmological model the epoch of matter-radiation equality is well determined, we fix the split between DM and DE-like axions to be given by $t_{\rm eq}$. Axions are DM-like if $m_a>10^{-27}\text{ eV}$, and DE-like otherwise.
\end{itemize}

Being bosons, axions can macroscopically populate their energy states. For the energy densities of interest in cosmology, the occupation numbers are huge, and the field can be treated classically (which accounts for Bose enhancement). This means that the axion de Broglie wavelength becomes a macroscopic property, manifest in the effective soundspeed of the axion (super)fluid. Above the de Broglie wavelength axions can be treated as pressureless dust, while below it the effects of the ``quantum pressure'' must be accounted for \citep[see e.g. ][ and references therein]{2016PhR...643....1M}.

The above properties lead to the following important points for axions in cosmology:
\begin{itemize}
\item Axion initial conditions contain vacuum fluctuations imprinted during inflation~\citep[see e.g. ][]{2010LNP...800....1L}, which manifest as isocurvature perturbations in observables. There is a smooth background field value across the observable Universe. In the adiabatic mode, the initial density perturbations are negligible but they later grow to match CDM on large scales (H15).
\item The axion equation of state changes from $w_a\approx -1$ to $w_a\approx 0$ at $t_{\rm osc}$. This causes the expansion rate to differ from that in the standard cosmological model, affecting the Silk damping scale and Sachs-Wolfe effects (early or late depending on axion mass) in the CMB, and in the evolution of the Hubble rate and other distance measures (e.g. H15).
\item The axion fluid has non-negligible pressure on scales at and below the de Broglie wavelength. This suppresses the axion density power spectrum compared to CDM, which manifests in all observables sensitive to structure formation \citep[e.g.][]{2010PhRvD..82j3528M}.
\end{itemize}

Most of the preceding discussion applies to any scalar or pseudoscalar field with a (small) mass term and (comparatively) large cosmic energy density: the moniker ``axion'' simply gives context and rigour to the theoretical interpretation of our results. Having thus introduced the theory behind our model, we now discuss the practicalities of the CMB observables in ULA models with adiabatic and isocurvature initial conditions.

\subsection{Adiabatic Power Spectra}

In the adiabatic mode, initial conditions in a species with equation of state $i$ are related to those in the (dominant) photon density perturbation, $\delta_\gamma$, as:
\be
\delta_i=\frac{3}{4}(1+w_i)\delta_\gamma \, .
\ee
Since at early times axions have $w_a\approx -1$, the adiabatic initial condition for $H\gg m_a$ is $\delta_a\approx 0$. The full super-horizon adiabatic initial conditions are given in the Appendix of H15. When $H\sim m_a$ the equation of state transitions to $w_a\approx 0$, the axion begins to cluster, and the density perturbations above the Jeans scale grow and ``lock-on'' to their CDM counterparts (H15). 

The effects described in Appendix~\ref{appendix:axion_basics} lead to differences in the adiabatic CMB power spectra in the presence of ULAs compared to a pure $\Lambda$CDM universe. These effects are caused by the integrated effect on the expansion rate (when $w_a\approx -1$, axions behave as DE), and the axion Jeans scale (axions do not cluster for $k>k_J$) \citep{2000PhRvL..85.1158H,axiverse,2015PhRvD..91j3512H,2016PhR...643....1M}.

Figure~\ref{fig:adi_DM} shows the lensing deflection and $E$-mode polarisation auto power spectra for DM-like ULAs with $\Omega_a h^2=0.12$ (these have not been presented elsewhere before). The dominant effect in the lensing power is caused by the axion Jeans scale, which suppresses structure formation, and thus reduces the total amount of gravitational lensing of the CMB. The structure suppression scale is given by $\ell_J$, which increases for increasing axion mass. For $\ell<500$ there is a greater than 10\% suppression of lensing power relative to CDM for $m_a\leq 10^{-25}\text{ eV}$. For larger $\ell<2000$, differences to CDM become sub percent for $m_a>10^{-24}\text{ eV}$. 

We use the lensing power spectrum computed from the linear theory matter power spectrum. In the range of $\ell$ covered by \textit{Planck}, non-linear corrections can be safely neglected~\citep[e.g.][]{2006PhR...429....1L}. For a discussion of non-linear effects for axions at high-$\ell$ see~\cite{2017PhRvD..95l3511H}.

For the $m_a\geq 10^{-27}\text{ eV}$ axion DM models shown, the dominant effect in the $E$-mode power is caused by the $w_a$ transition. The different early time equation of state alters the diffusion damping scale compared to CDM and changes the decay rate of gravitational potential wells, which drives the redshifting photons through the early integrated Sachs-Wolfe (ISW) effect. Both effects change the ratio of the acoustic peaks. The same effect is observed in the temperature anisotropies (H15). Deep in the radiation dominated era, both the axion and CDM energy densities becomes increasingly irrelevant. Differences between the axion and CDM equation of state at $z\gg z_{\rm dec}$ (where $z_{\rm dec}$ is the decoupling redshift) thus have little effect on the diffusion damping, and the CDM and ULA models become virtually indistinguishable in the $EE$ power at the percent level for $\ell<2000$ for $m_a> 10^{-25}\text{ eV}$. This is consistent with the size of the effects in the $TT$ power in H15.

The adiabatic $TT$ constraints of H15 on the axion energy density relative to CDM are rather permissive for $m_a\geq 10^{-25.5}$. Fig.~\ref{fig:adi_DM} demonstrates that the addition of lensing and $EE$ power (not to mention cross-correlations) will tighten constraints relative to $TT$ alone for all $m_a< 10^{-24}\text{ eV}$, but will not improve constraints at higher mass.

\begin{figure*}
\center
\includegraphics[width=\columnwidth]{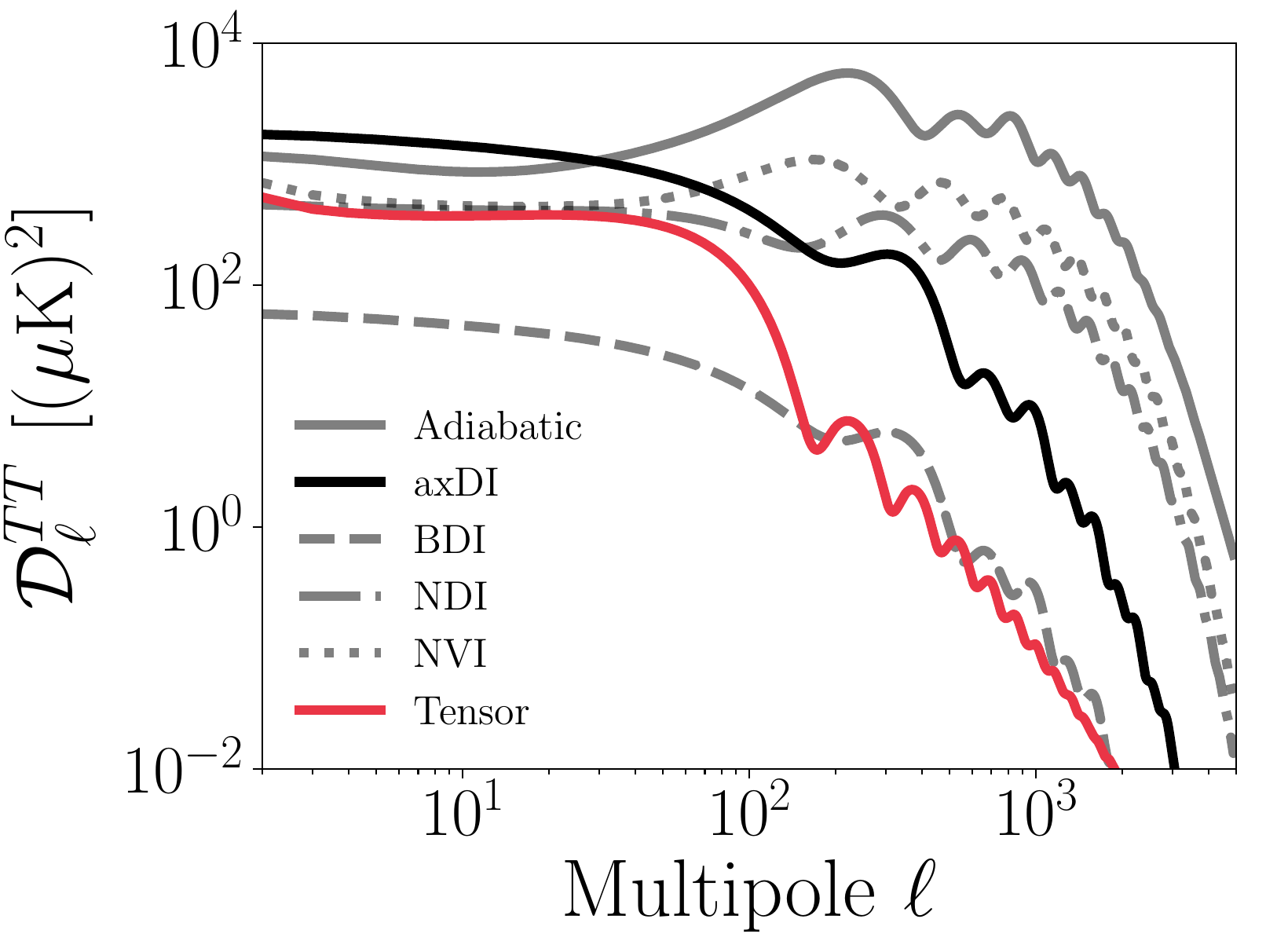}
\includegraphics[width=\columnwidth]{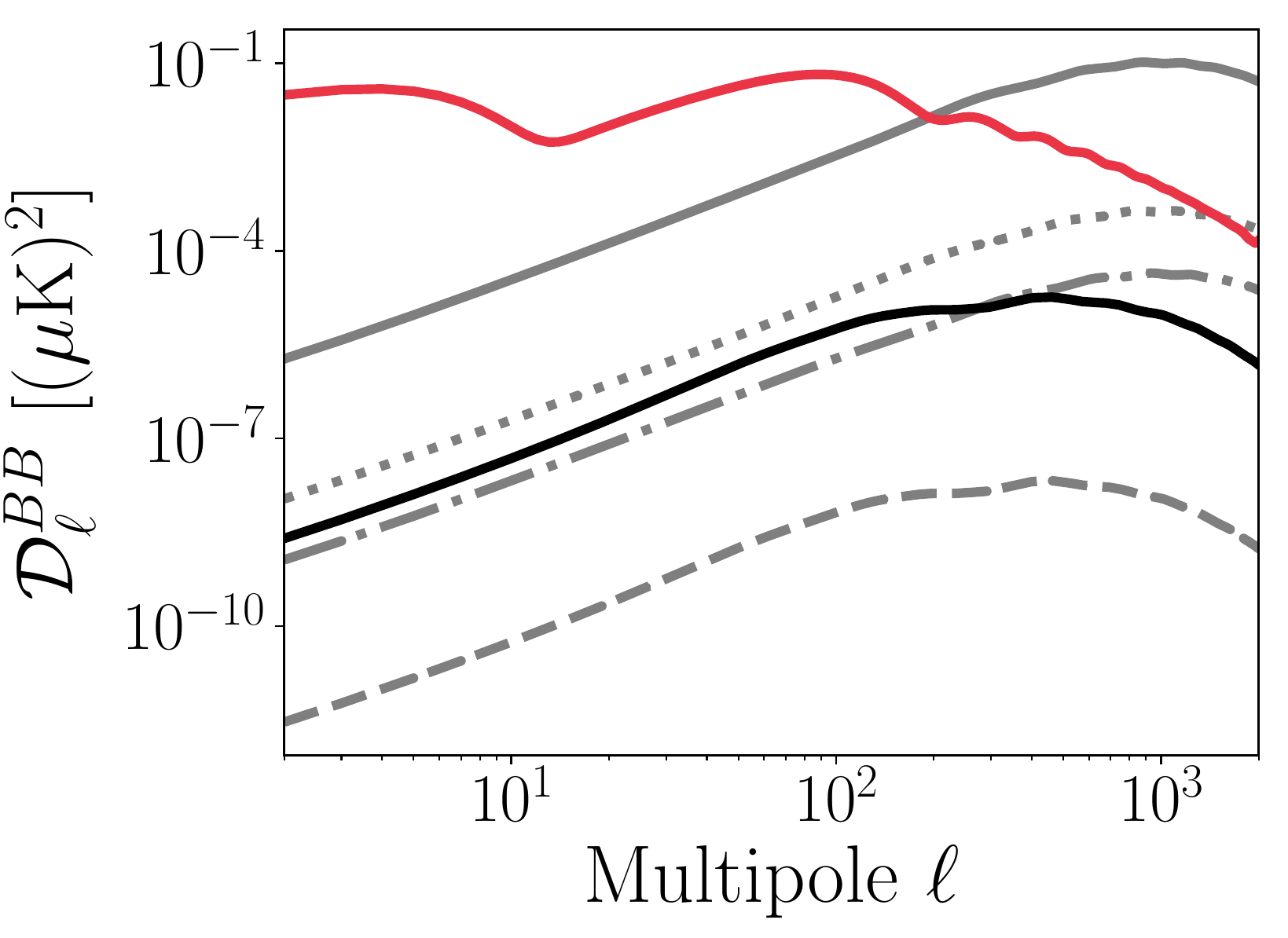}
\caption{{\bf Isocurvature Modes:} We show the different possible isocurvature modes (with $n_\mathrm{ iso}=1$, $A_\mathrm{ iso}=A_s$), along with the usual adiabatic mode (with $n_s=0.96$), and the tensor mode (with $r=1$, $n_\mathrm{ T}=0$). The axDI (axion) mode has $m_a=10^{-25}\text{ eV}$ and is degenerate with the CDI mode on the scales shown, $\ell<5000$. Up to an amplitude factor, it is degenerate with BDI. The axDI mode is distinct from the NDI and NVI (neutrino) modes. The right panel shows the $B$-mode polarisation power. Only the tensor mode generates low-$\ell$ $BB$ power, breaking any degeneracy with the isocurvature modes. Isocurvature and adiabatic modes generate large-$\ell$ $BB$ power via lensing.}
  \label{fig:iso_modes}
  \end{figure*}
\subsection{Isocurvature in axion models}
\subsubsection{Inflationary physics of isocurvature}
\label{sec:iso_theory}
In the canonical cosmological model, perturbations in all species are seeded by curvature fluctuations laid down during inflation, and these perturbations are adiabatic, which is to say entropy fluctuations between the $i^\mathrm{ th}$ species and photons vanish:
\begin{equation}
S_{i\gamma}\equiv \frac{\delta n_{i}}{n_{i}}-\frac{\delta n_{\gamma}}{n_{\gamma}}=0.
\end{equation}
In the adiabatic mode the axion fluctuation $\delta_a\approx 0$ at leading order when $H\gg m_a$ and the axions only later (when $H\ll m_a$) develop fluctuations in $\delta_a$ from the curvature perturbations in the other species (H15).

If the Peccei-Quinn symmetry is broken during inflation, that is if $f_a<H_I/2\pi$,  and $m_{a}\ll H_{I}$, then the axion is a spectator field during inflation. The axion field then carries deSitter-space vacuum fluctuations with dimensionless power spectrum \citep[e.g.][]{1983PhLB..126..178A,1985PhRvD..32.3178S,1991PhRvL..66....5T,1992PhRvD..45.3394L,2004hep.th....9059F,2008PhRvD..78h3507H,2009ApJS..180..330K}
\be
\Delta_{\delta \phi} = A_{\phi}\left(\frac{k}{k_0}\right)^{n_\mathrm{ iso}-1}\, . \label{eqn:delta_phi_power}
\ee
The amplitude and spectral index are given by:
\be
A_{\phi}=\left(\frac{H_I}{2\pi}\right)^2\,;\,\, n_\mathrm{ iso}=1-2\epsilon\, ,\label{eq:adpsp}
\ee
and the inflationary slow-roll parameter $\epsilon=A_{\phi}/(2M_\mathrm{ pl}^2A_s)$. As long as the Peccei-Quinn symmetry is not restored after inflation (e.g. by thermal fluctuations or parameteric resonance) the axion field fluctuations source isocurvature axion density perturbations with spectrum:
\be
\Delta_{\delta_{a}}= A_\mathrm{ iso}\left(\frac{k}{k_0}\right)^{n_\mathrm{ iso}-1}\, ,\label{eq:isopsp}
\ee
where $A_\mathrm{ iso}=4A_\phi/\phbar_i^2$, with $\phbar_i$ the background field value. For standard slow-roll inflation we have $\epsilon \ll 1$ and so axions carry a nearly scale-invariant power spectrum. Since they are energetically subdominant (and thus don't contribute to to the total curvature fluctuation $\zeta$), axions source primordial isocurvature fluctuations, with primordial power-spectrum $\Delta_{S_{a}}=\Delta_{\delta_{a}}$. This isocurvature mode is called the axion density isocurvature (axDI) mode.

Because the Peccei-Quinn symmetry in this scenario is broken during inflation, the axion field has a nearly uniform initial field value across horizon-size patches:
\be
\bar{\phi}_i^2=f_a^2\theta_i^2+(H_I/2\pi)^2\approx f_a^2\theta_i^2\sim f_a^2\, ,
\label{eqn:initial_phi_exact}
\ee
where the ``initial misalignment angle'' is $\theta_i\in [-\pi,\pi]$. The initial field value fixes the ULA relic density as described in Appendix~\ref{appendix:axion_basics}. 

The second term in the first equality in Eq.~\eqref{eqn:initial_phi_exact} emerges from the variance of the vacuum fluctuations [Eqs.~(\ref{eqn:delta_phi_power}) and (\ref{eq:adpsp})] and can be safely neglected for the models of interest since one requires $\bar{\phi}_i\gg 10^{14}\text{ GeV}\gg H_I$ for ULAs with non-negligible relic density, implying $f_a\gg H_I$. This guarantees that all the ULA models we study in the $(m_a,\Omega_a h^2)$ plane must, by necessity, be accompanied by the axDI mode. We have checked that this approximation is self-consistent, as shown in Fig. \ref{fig:phi_hinf} and discussed in Section \ref{sec:iso_constraints}. 

Furthermore, in this scenario residual populations of thermal axions and axion topological defects are diluted away by inflation. Thus the ULA models we study have relic density sourced entirely by the classical evolution of the axion field.\footnote{We ignore the possible production of axions from decays of heavy particles after inflation. Such axions can be modelled in \acamb~using the effective neutrino parameter $N_\mathrm{ eff}$. For interesting consequences of relativistic thermal axions in the CMB and direct detection, see \cite{2016PhRvL.117q1301B}.}

Just like axions, gravitons are massless spectator fields during inflation, and therefore the same parameters that set the axion isocurvature spectrum also determine the dimensionless tensor power spectrum \citep[e.g.][]{2009arXiv0907.5424B}:
\be
\Delta_{t}^{2}= rA_s\left(\frac{k}{k_0}\right)^{n_\mathrm{ T}},
\ee 
The tensor to scalar ratio and tensor spectral index are given by:
\begin{align}
r&=16\epsilon\approx 0.17\left(\frac{2.1\times10^{-9}}{A_s}\right)\left(\frac{H_I}{10^{14}\text{ GeV}}\right)^2\, ,\\ n_\mathrm{ T}&=2\epsilon \, .
\label{eqn:r_H_relation}
\end{align}
In all of our constraints we enforce the consistency relations for $n_\mathrm{ iso}$, $n_\mathrm{ T}$, $r$, and $A_\mathrm{ iso}$ by using $H_I$ as our primary cosmological parameter. By bounding the possible contribution of the combined tensor and isocurvature modes to the CMB power, we bound $H_I$.

Due to the axion being a stable spectator field, the axDI isocurvature mode is \emph{uncorrelated} with the adiabatic mode, $\langle\zeta\delta\phi\rangle\approx 0$, and so the total CMB power is given by the sum, $C_\ell^\mathrm{ adi}+C_\ell^\mathrm{ iso}+C_\ell^\mathrm{ tens}$, where $C_\ell^\mathrm{ tens}$ is the power (in our case $T$, $E$, or lensing convergence and cross spectra) sourced by the tensor mode. These considerations lead to the total spectra shown in Fig.~\ref{fig:opener}. It is the close agreement of observed CMB anisotropy power spectra with the adiabatic model that allows severe constraints to isocurvature and tensor modes to be imposed from \textit{Planck} satellite measurements and other CMB data \citep{2016A&A...594A..13P}. 

For the well-known case of QCD axion isocurvature \citep[see e.g.][]{1983PhLB..129...51S,1986PhRvD..33..889T,1991PhRvL..66....5T,2009ApJS..180..330K,2010PhRvD..82l3508W,2010PhRvD..81f3508V,2013ApJS..208...20B}, taking the DM to be entirely composed of the QCD axion, and assuming standard formulae for the axion relic density, one can translate bounds on the fractional isocurvature amplitude, $\beta_{\rm iso}$, into a bound on the two free parameters of the model, namely $f_a$ and $H_I$. The most recent bound of $\beta_{\rm iso}<0.038$~\citep[95\% C.L.][]{2016A&A...594A..20P} translates into the bound $H_I<0.87\times 10^7\text{ GeV}(f_a/10^{11}\text{ GeV})^{0.408}$. This implies a tensor-to-scalar ratio of $r\lesssim 10^{-14}$. Thus, in the case when the PQ symmetry is broken during inflation, the QCD axion is incompatible with a measurably large value of $r$ \citep[although various exotica can be invoked to avoid this conclusion, e.g. ][and references therein]{2014PhLB..734...21H,2016PhRvL.116n1803N}.

For ULAs, it is safe to ignore the temperature dependence of the mass in many cases, yielding simpler expressions, namely \citep{2013PhRvD..87l1701M,2014PhRvL.113a1801M}\footnote{See e.g. \cite{2017arXiv170202116D,2017arXiv170308798V} for the formulae including temperature dependence.}
\begin{eqnarray}
\frac{\Omega_{a}}{\Omega_{d}}&\lesssim\left\{ \begin{array}{l}3\times
10^{-8}\left(\frac{M_\mathrm{ pl}}{H_{I}}\right)^{2}\mbox{if $m_{a} \lesssim 10^{-27}~\mathrm{ eV}$.}\\3\times 10^{-8} \left(\frac{M_\mathrm{ pl}}{H_{I}}\right)^{2}\left(\frac{10^{-27}~\mathrm{ eV}}{m_{a}}\right)^{1/2}\mbox{if $m_{a} \gtrsim 10^{-27}~\mathrm{ eV}$,}
\end{array}\right.
\end{eqnarray}under the assumption that \textit{Planck} limits \citep{2016A&A...594A..13P} to the observed isocurvature power are unchanged when the species carrying the isocurvature is a ULA. Direct bounds on $r$ imply that $H_I\lesssim 10^{14}\text{ GeV}$. Thus we see that isocurvature improves sensitivity to $\Omega_a/\Omega_d$ only if $m_a\gtrsim 10^{-25}\text{ eV}$.

Since low-mass ($m_{a}\lesssim 10^{-27}~\mathrm{ eV}$) ULAs behave more like DE than DM, we cannot in fact treat the axDI mode simply as CDI and naively apply the \textit{Planck} bounds on $\beta_\mathrm{ iso}$~\citep{1995PhRvL..75.2077F,2015PhRvD..91j3512H}. Indeed, in what follows, when we use the appropriate ULA isocurvature initial conditions we will find that the axDI mode has suppressed small-scale power compared to the CDM isocurvature case. In prior work, we estimated the resulting degradation to ULA constraints from a simple mode-counting argument \citep{2014PhRvL.113a1801M}. In this work, we actually compute ULA isocurvature power spectra using a new version of \acamb~that implements the initial conditions of Sec. \ref{isinit:theory}, and systematically compare with CMB data to obtain robust constraints. 

\subsubsection{Modified isocurvature initial conditions for ULAs}\label{isinit:theory}
\begin{figure*}
\begin{center}
\includegraphics[width=\textwidth]{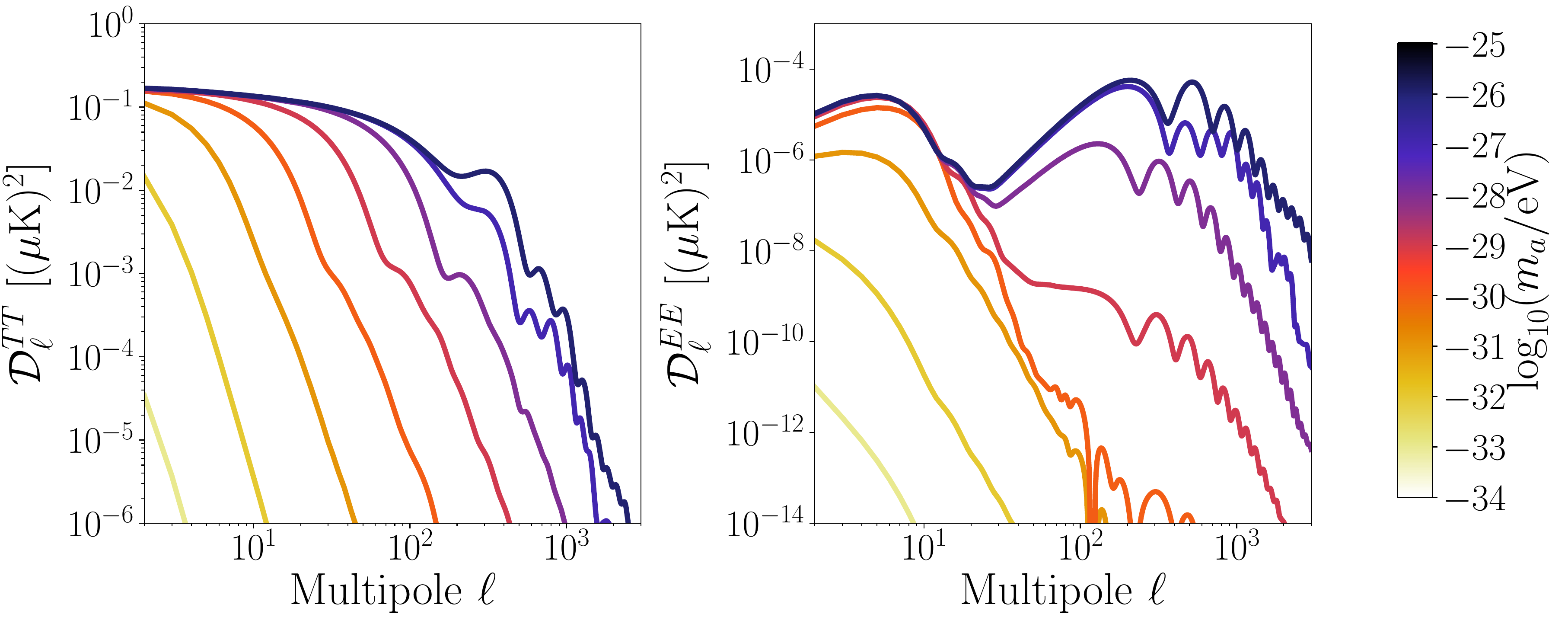} 
 \end{center}
\caption{{\bf Ultralight Axion Isocurvature:} ULAs are introduced as a small fraction, $\Omega_a/\Omega_d=0.01$, of the total density. The axDI shows evidence of the Jeans scale, suppressing the power above some value of $\ell_J$, which decreases with decreasing axion mass. On the scales shown, the $m_a=10^{-26}\text{ eV}$ and $m_a=10^{-25}\text{ eV}$ spectra are degenerate. \textit{Left Panel:}~Temperature auto-power. \textit{Right Panel:} E-mode polarisation auto-power. \textit{Note the different scale on the axion mass colour bar compared to Fig.~\ref{fig:adi_DM}}.}
\label{fig:iso_mass_plot}
\end{figure*}

In order to initialise perturbations in all species in a Boltzmann code like \textsc{camb}, super-horizon perturbations must be evolved for some relatively short but finite time increment analytically, so that numerical quantities are all finite when the code is initialized \citep{2000PhRvD..62h3508B}. ``Initial conditions" are given as power-series solutions for all density fluctuations $\delta_{i}$ as a function of conformal time $\eta$ and comoving wavenumber $k\eta$ relative to the horizon. To obtain accurate isocurvature transfer functions for low-mass ULAs, we must consider the early-time behaviour of the axion field. 

For the QCD axions with $m_{a}\gg 10^{-18}~\mathrm{ eV}$, this transition occurs very early in the radiation-dominated epoch at $T\approx 1\text{ GeV}$. The well known isocurvature limits on the QCD axion (summarised shortly in Section~\ref{sec:iso_theory}) are consistently obtained by treating axions as standard CDM in both mode evolution and derivation of isocurvature initial conditions. For ULAs, the $w_{a}$ transition occurs later and the analytic initial conditions are modified, a new result that we detail below. 

For the standard cosmological species (photons $\gamma$, neutrinos $\nu$, baryons \textit{b}, and CDM \textit{c}), we use the usual synchronous-gauge equations of motion (EOMs), with the accompanying Einstein equations as given in \cite{1995ApJ...455....7M}. The EOMs for the axion field $\phi=\overline{\phi}(\tau)+\delta \phi(\tau,\vec{k})$ are given in Appendix~\ref{appendix:axion_basics}.

The usual technique for deriving analytic initial conditions \citep{2000PhRvD..62h3508B} is easiest to implement if the EOMs of all species are recast as first-order fluid equations, and this can be easily done using the generalized DM (GDM) formalism of \cite{1998ApJ...506..485H}. The GDM fluid EOMs for the ULAs in synchronous gauge are
\begin{align}
\dot{\delta}_{a}=&-ku_{a}-\left(1+w_\mathrm{ a}\right)\dot{h}_m/2-3\mathcal{H}\left(1-w_{a}\right)\delta_{a}\nonumber \\-&9\mathcal{H}^{2}\left(1-c_\mathrm{ ad}^{2}\right)u_\mathrm{ a}/k,\label{eqn:eoma}\\
\dot{u}_{a}=&~2\mathcal{H}u_{a}+k \delta_{a}+3\mathcal{H}\left(w_{a}-c_\mathrm{ ad}^{2}\right)u_{a}
\label{eqn:eomb},\end{align}
where the ULA density contrast is $\delta_{a}=\delta \rho_{a}/\rho_{a}$ and $u_{a}=(1+w_{a})v_{a}$ is the dimensionless ULA momentum flux. The ULA equation of state is given in Eq.~\ref{eqn:wdef} while the adiabatic sound speed $c_\mathrm{ ad}$ is given by
\begin{align}
c_\mathrm{{ad}}^2\equiv \frac{\dot{P}_{a}}{\dot{\rho}_{a}}= &w_{a}-\frac{\dot{w}_{a}}{3\mathcal{H}\left(1+w_{a}\right)},\label{eqn:adiabat_cs} 
\end{align} where $w_{a}$ is the time-dependent equation-of-state parameter for ULAs. Both $w_{a}$ and $c_\mathrm{ ad}$ are set by the solution to the homogeneous ULA field equation [Eq.~(\ref{eqn:homo_eom})]. The ULA source terms to the Einstein equations [Eqs.~(\ref{eqn:deltarho})-(\ref{eq:comove_find})] can be recast as 
\begin{align}
\delta P_{a}=&\rho_{a}\left[\delta_{a}+3\mathcal{H}(1-c_\mathrm{ ad}^{2})v_{a}/k\right]\label{dpdef},\\
\delta \rho_{a}=&\rho_{a}\delta_{a},\\
\left(\rho_{a}+P_{a}\right)v_{a}=&\rho_{a}u_{a}.\label{dqdef}
\end{align}
Equivalent expressions for scalar-field variables are given in \cite{2003MNRAS.346..987W,2004PhRvD..69h3503B}. The power-series solutions for these background quantities are given in Appendix~\ref{appendix:early_axion}. 

The leading-order early-time evolution of the isocurvature mode is given by $\delta_{a}=\mathrm{ constant}$. Higher orders are obtained as described in Appendix~\ref{appendix:power_law}. The initial conditions for all perturbations including next-to-leading order terms are then
\begin{eqnarray}
\delta_{\gamma}&=&\delta_{\nu}=-\frac{\mathcal{A}^{\left(0\right)}\eta_\mathrm{ b}^{4}}{3}=\frac{4}{3}\delta_\mathrm{ c}=\frac{4}{3}\delta_\mathrm{ b},\label{aia}\\
\frac{\theta_{\gamma}}{\kappa \mathcal{C}}&=&\frac{\theta_{\nu}}{\kappa \mathcal{C}}=-\frac{\mathcal{A}^{\left(0\right)}\kappa\eta_\mathrm{ b}^{5}}{60},\\
\theta_\mathrm{ c}&=&0,\\
\sigma_{\nu}&=&\frac{6R_\mathrm{ b}\mathcal{A}^{\left(0\right)}\left(K-1\right)\eta_\mathrm{ b}^{5}}{5\left(75+4R_{\nu}\right)},\\
F_{\nu}^{\left(3\right)}&=&\frac{6R_\mathrm{ b}\mathcal{A}^{\left(0\right)}\left(K-1\right)\kappa \eta_\mathrm{ b}^{6}}{35\left(75+4R_{\nu}\right)},\\
\delta_{a}&=&1-\frac{\left(\kappa\eta_\mathrm{ b}\right)^{2}}{10},\\
u_{a}&=&-\frac{\mathcal{A}^{\left(0\right)}\mathcal{C}\kappa^{2} \eta_\mathrm{ b}}{5},\\
h_{m}&=&\frac{\mathcal{A}^{\left(0\right)}\eta_\mathrm{ b}^{4}}{2},\\
\eta_{m}&=&-\frac{\mathcal{A}^{\left(0\right)}\eta_\mathrm{ b}^{4}}{12},\\
\mathcal{A}^{\left(0\right)}&\equiv&\frac{\rho_\mathrm{ a}^{\left(0\right)}}{a_{0}^{4}\Omega_{r}\rho_\mathrm{ crit}}.\label{aig}
\end{eqnarray}
Fractional density perturbations in the $i^\mathrm{ th}$ species are denoted by $\delta_{i}$, while the corresponding velocity perturbation $\theta_{i}$ and higher moments $F_{\nu}^{\left(i\right)}$ in the Boltzmann hierarchy (like the neutrino/photon anisotropic stress) are defined as in \cite{1995ApJ...455....7M}. As a matter of convention, $\delta_{a}=1$ is chosen here; all initial mode amplitudes are later rescaled in \textsc{camb} for consistency with the desired primordial power spectrum. Here 
\begin{eqnarray}R_\mathrm{ b}=\Omega_{b}/\left(\Omega_{b}+\Omega_{c}\right),\\
R_{\nu}=\Omega_{\nu}/(\Omega_{\nu}+\Omega_{b}),
\end{eqnarray} 
$\rho_\mathrm{ a}^{\left(0\right)}$ is the initial value of the axion energy density when $a \ll a_\mathrm{ osc}$, $\Omega_{r}=\Omega_{\gamma}+\Omega_{\nu}$ is the sum of relic densities in photons/neutrinos, and $\eta_\mathrm{ b}=\Omega_{m}H_{0}\eta/(4\sqrt{\Omega_{r}})$ is a dimensionless conformal time. The wave number is similarly made dimensionless by using $\kappa=k/C$, where $C=\sqrt{4\pi G \rho_\mathrm{ eq}a_\mathrm{ eq}^{4}/4}$, $\rho_\mathrm{ eq}$ is the cosmic energy density at matter-radiation equality, and $a_\mathrm{ eq}$ is the scale factor at equality. The normalization constant $K$ is given by Eq.~\ref{eq:kval}, parameterises the reduced matter density after equality when CDM is swapped out for ULAs when they have $w_{a}\simeq -1$, and asymptotically approaches $1$ as $m_{a}\to\infty$. Further explanation is given in Appendix~\ref{appendix:power_law}.

\subsubsection{Comparison of initial conditions with other results in the literature}

The evolution of $\delta_{i}$ in the early time ULA isocurvature mode agrees with the behaviour of the quintessence isocurvature described in \cite{cambnotes}, as should be the case for $\eta\ll \eta_\mathrm{ osc}$. This is long before the axion field has started to coherently oscillate, and so it behaves like a quintessence component rolling gently down a quadratic potential. We go beyond those results to obtain the leading-order behavior of perturbations in other species and metric perturbations.

Our solution for the ULA isocurvature mode for $\eta\ll\eta_\mathrm{ osc}$ does not agree with the expressions for quintessence isocurvature given in \cite{1999PhRvD..59l3508P}. The solution for the metric perturbation $h_{m}$ in \cite{1999PhRvD..59l3508P} is incorrect, obtained by taking the $00$ Einstein equation,
\begin{equation}
k^{2}\eta_{m}-\frac{\dot{a}}{2a}\dot{h}_{m}=-4\pi G a^{2}\delta \rho
\end{equation} and dropping the term proportional to $\eta$. Schematically, the equation that remains is $\dot{h}_{m}\propto 3 \delta_\mathrm{ R}/a^{2}+3a^{2}\mathcal{A}^{\left(0\right)}\delta_\mathrm{ a}$, where $\delta_\mathrm{ R}=R_{\gamma}\delta_{\gamma}+R_{\nu}\delta_{\nu}$. In  \cite{1999PhRvD..59l3508P}, the term proportional to the radiation energy density fluctuation is then dropped. As we can see, this term is of order $\eta^{2}$ when $a\simeq \eta$ during radiation domination, just like the axion-sourced term; it is thus inconsistent to drop this term. No such truncation takes place in our solution. Additionally, if the same truncation is performed in a different Einstein equation, a different solution altogether for $h_{m}$ results. 

Furthermore, the solution of \cite{1999PhRvD..59l3508P} does not satisfy the shear-sourced Einstein equation:
\begin{eqnarray}
\ddot{h}_{m}+6\ddot{\eta}_{m}+2\frac{\dot{a}}{a}\left(\dot{h}_{m}+6\dot{\eta}_{m}\right)-2k^{2}\eta_{m}\nonumber\\=-24 \pi G a^2 \left(\overline{\rho}_{\nu}+\overline{P}_{\nu}\right)\sigma_{\nu}\label{shearsource}.\end{eqnarray} 
Here $\overline{\rho}_{\nu}$ and $\overline{P}_{\nu}$ are the homogeneous neutrino energy density and pressure, while $\sigma_{\nu}$ is the neutrino anisotropic stress. We have checked explicitly that the power series in Eqs.~(\ref{aia})-(\ref{aig}) \textit{does} (once the next-to-leading order corrections to $h_\mathrm{ m}$ and $\eta_\mathrm{ m}$ are included) satisfy Eq.~(\ref{shearsource}) order by order, and is thus self-consistent. 

While this work was in preparation, \cite{2016PhRvD..94d3512K} used similar methods to derive initial conditions for the generalized dark matter (GDM) scenario, obtaining power-series solutions for the adiabatic, CDI, BDI, NDI, NVI, and GDM isocurvature modes. That work is valid in the limit $w\simeq 0$ (to be compared with the $w\simeq -1$ scenario valid for ULAs with the initial conformal time choices of \acamb), which yields more significant GDM-driven changes to the expansion history from the usual radiation-matter mixture, and thus, a different power series in $x$ for super-horizon initial conditions. This is, however, a rather different physical scenario than the one we consider, and so we do not discuss it further.

\subsection{CMB observables of ULA isocurvature}

The fractional contribution of isocurvature auto-power to the temperature variance is given by:
\be
\alpha^{TT}_\mathrm{ iso}=\frac{(\Delta T)^2_\mathrm{ iso}}{(\Delta T)^2_\mathrm{ tot}}\, ,
\ee
where $(\Delta T)^2_X=\sum (2\ell+1)C^{TT}_{X,\ell}$. We define the parameter
\be
\beta_\mathrm{ iso}=\left(\frac{\Omega_a}{\Omega_d}\right)^2\frac{A_\mathrm{ iso}}{A_s+A_\mathrm{ iso}}\, ,
\ee 
as an estimator for $\alpha_\mathrm{ iso}$ in the axDI mode. It is simpler to compute than $\alpha^{TT}_\mathrm{ iso}$, and still accounts for the amplitude suppression due to the energy density, but not the multipole structure of the power spectra. It can be used because the axion isocurvature mode is uncorrelated with the adiabatic mode. In addition, $\beta_\mathrm{ iso}$ depends only on initial conditions and can also be used as a measure of $\alpha^{XY}_\mathrm{ iso}$ for the other spectra that we use beyond temperature. 

\begin{figure}
\includegraphics[width= 1.0\columnwidth]{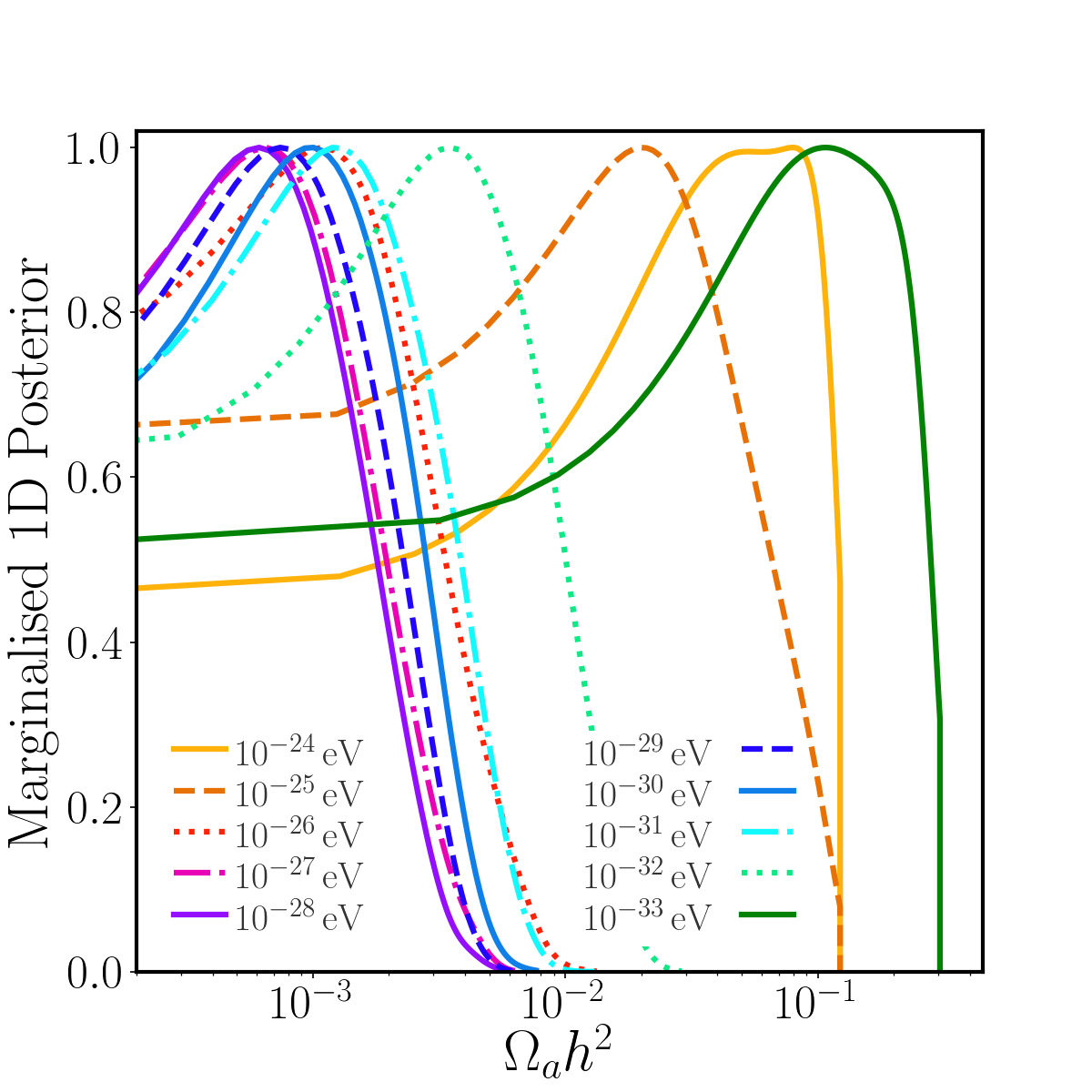}
\vspace{-0.5cm}
 \caption{{\bf Constraints on the axion energy density for different masses.} The marginalised 1D distributions on the axion energy density, $\Omega_ah^2.$ The highest and lowest mass bins ($m_a=10^{-24}, m_a=10^{-33}\,\mathrm{eV}$) have a flat distribution as the degeneracy between dark matter and dark energy respectively becomes complete. However, for the intermediate masses the constraints become well behaved simple upper limits. 
\label{fig:adi_constraints} }
\end{figure}

In \cite{2016A&A...594A..20P}, a different, model-independent parameterisation is used, with $\beta$ and $\alpha$ treated as derived parameters; this treatment is more robust to the possibility of an extremely blue isocurvature mode. The analysis in that work cannot be naively used to impose limits to ULAs, however, due to the differences between CMB power spectra for the CDI and axDI modes at low $m_{a}$. Because we work in the ULA context, we justifiably can ignore extremely blue isocurvature spectra and restrict our attention to the ULA-relevant case of nearly scale-invariant isocurvature, as appropriate for a light spectator during the inflationary era.
\begin{figure*}
\includegraphics[width= \columnwidth]{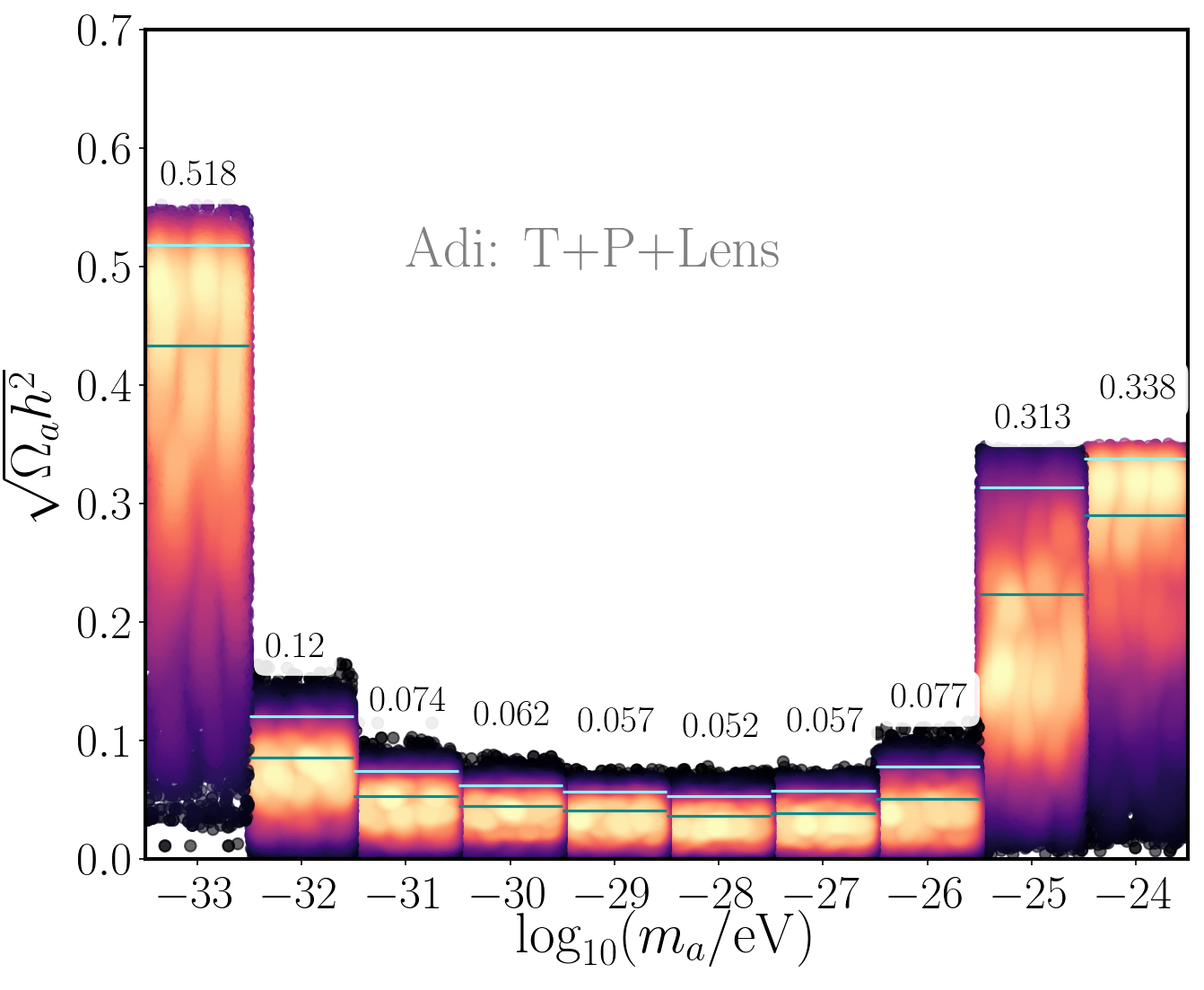}
\includegraphics[width=0.9\columnwidth]{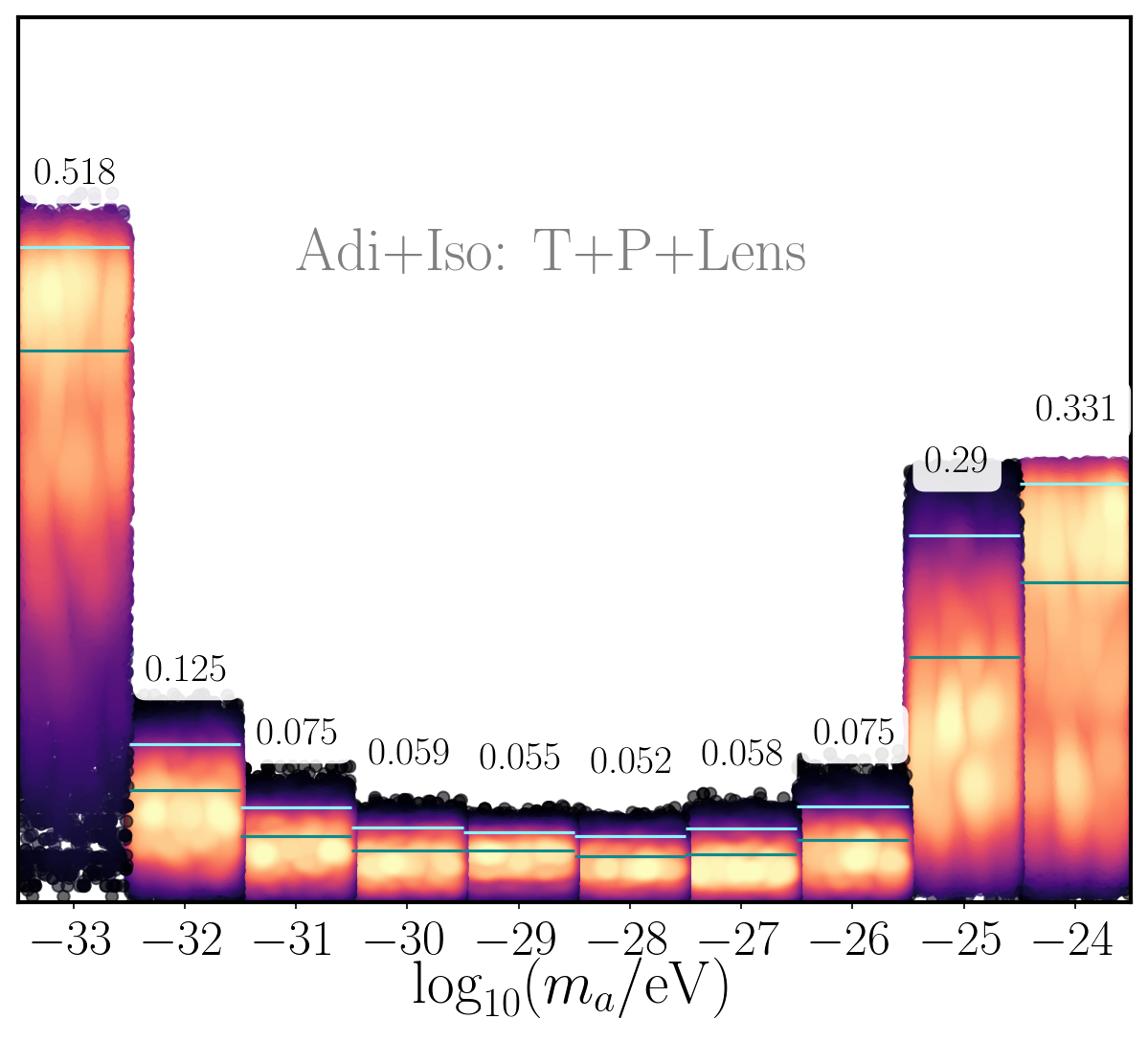} 
\vspace{-0.5cm}
 \caption{{\bf Constraints on mixed dark matter}: the constraints on $\sqrt{\Omega_a h^2}$ in each mass bin for both the adiabatic only case, and the combined adiabatic and isocurvature constraints, indicated in the left and right panels respectively. The 68\% and 95\% upper limits are shown as dark and light blue lines, and the 95\% upper limit is also indicated numerically. Adding isocurvature tightens the constraints by 10-20\% for the heaviest axions (indicating that isocurvature is relevant and disfavoured for these masses at high density fraction), and does not affect the limits for lighter axions. To aid visualisation, the MCMC samples for each $m_a$ bin are plotted with a random horizontal scatter within the bin. Points are coloured by the point density defined by a Gaussian kernel density estimate, with brighter colours indicating higher density. The data combination used is the \textit{Planck} 2015 temperature and polarisation high-$\ell$ and low-$\ell$ likelihoods, as well as the \textit{Planck} minimum variance lensing likelihood for $40\leq \ell \leq 400$. 
\label{fig:lens_u_plot}}
\end{figure*}

Figure~\ref{fig:iso_modes} shows all CMB $TT$ power spectra given by adiabatic, axDI, BDI, NDI, and NVI modes. We compute the CMB power spectra for the axDI mode using  \acamb, applying the initial conditions given in Eqs.~(\ref{aia})-(\ref{aig}). The axDI mode is shown here with $m_a=10^{-25}\text{ eV}$; it is degenerate on the scales shown with the CDM Density Isocurvature (CDI) mode. The high-mass axDI mode is also degenerate up to an amplitude factor with the BDI mode. The axDI mode is distinct from the NDI and NVI modes. For comparison, we also show the tensor contribution to the power. Although not degenerate with the isocurvature modes, the tensor power has a similar low-$\ell$ contribution to the temperature power, and is similarly suppressed at high-$\ell$. 

Figure~\ref{fig:iso_modes} demonstrates that tensor initial conditions, while to isocurvature similar in their low-$\ell$ temperature power, are distinct in the $B$-mode polarisation power, where tensors generate low-$\ell$ power while isocurvature only generates high-$\ell$ power due to lensing. This highlight the importance of more precise measurements of CMB polarisation power spectra (and in particular, the primordial B-mode spectrum) to disentangle the relative contributions of isocurvature and tensor modes to CMB observations. In our analysis we do not use the $B$-mode power to constrain $r$, since there is no low-$\ell$ detection and the upper limits on $r$ are not significantly tighter that the constraints using only $T$ and $E$.

Figure~\ref{fig:iso_mass_plot} shows the axDI mode for $A_\mathrm{ iso}=2.3\times 10^{-9}$ for a series of axion masses at fixed low axion fraction. When the axion density is low the isocurvature mode is lower in amplitude than the pure CDI case, by a factor of $(\Omega_a/\Omega_d)^2$. The axDI mode shows evidence of the axion Jeans scale, with suppressed power above some value of $\ell$ for lower values of $m_a$~\citep{2013PhRvD..87l1701M}. This shape is a distinct feature of the axDI mode, and contributes to a large allowed isocurvature contribution for the lightest axions.

\section{Constraints on ultralight axions from CMB temperature, polarisation, and lensing}
\label{sec:dm_constraints}

\begin{table*}
    \begin{tabular}{l|l|l|l|l|l|l|l|l|l}
Mass ${m_a}$ & $\Omega_ah^2$&$\Omega_ch^2$&$\Omega_\Lambda$&$H_0$& $r$ &$\beta_\mathrm{iso}$&
$\bar{\phi}_i/M_\mathrm{ pl}$&$H_I$ &$\log\mathcal{P}$\\
$\mathrm{[eV]}$ & &&&${[\rm km\, s}^{-1}\mathrm{ Mpc}^{-1}\mathrm{]}$&&&&$\mathrm{[}10^{14}\mathrm{GeV]}$ &\\
\hline
   & & &  &{\bf Adiabatic} & &&&\\
$10^{-24}$ &$0.062 \pm 0.035$ &$0.057 \pm 0.035$& $0.695 \pm 0.008$ & $68.07 \pm 0.62$ & - & - & $0.09\pm 0.03$& -& -5530.19 \\
$10^{-25}$ &$0.039 \pm 0.029$ &$0.080 \pm 0.029$& $0.693 \pm 0.009$ & $67.96 \pm 0.67$ & - & - & $0.12\pm 0.05$& -&-5530.4 \\
$10^{-26}$ &$< 0.006$ &$0.117 \pm 0.002$ & $0.694 \pm 0.009$ & $68.01 \pm 0.66$ & - & - & $0.05\pm 0.02$& - &-5530.24\\
$10^{-27}$ &$< 0.003$ &$0.118 \pm 0.001$ & $0.691 \pm 0.009$ & $67.81 \pm 0.68$ & - & - & $0.06\pm 0.03$&-&-5530.27\\
$10^{-28}$ &$<0.003$ &$0.119 \pm 0.002$ & $0.682 \pm 0.014$ & $67.12 \pm 1.05$ & - & - & $0.08\pm 0.04$& - &-5530.23\\
$10^{-29}$ &$<0.003$ &$0.119 \pm 0.002$ & $0.676 \pm 0.017$ & $66.61 \pm 1.31$ & - & - & $0.11\pm 0.04$ & -&-5530.22\\
$10^{-30}$ &$<0.004$ &$0.120 \pm 0.001$ & $0.672 \pm 0.018$ & $66.26 \pm 1.33$ & - & - & $0.12\pm 0.05$ & -&-5530.47\\
$10^{-31}$ &$<0.005$ &$0.120 \pm 0.001$ & $0.669 \pm 0.021$ & $66.11 \pm 1.49$ & - & - & $0.15\pm 0.06$&- &-5530.37\\
$10^{-32}$ &$<0.014$ &$0.119 \pm 0.001$ & $0.661 \pm 0.036$ & $65.01 \pm 2.16$ & - & - & $0.24\pm 0.1$&- &-5530.33\\
$10^{-33}$ &$0.203 \pm 0.119$ &$0.119 \pm 0.001$ & $0.195 \pm 0.308$ & $65.88 \pm 1.46$ & - & - & $1.62\pm 0.57$ & -& -5530.36\\
\hline
    & & &  &{\bf Adiabatic} &{\bf + Isocurvature} & &&\\
$10^{-24}$ &$0.045 \pm 0.036$ &$0.073 \pm 0.036$& $0.696 \pm 0.009$ & $68.17 \pm 0.65$ & $< 0.01$ & $< 0.03$ & $0.07\pm 0.04$ &$< 0.30$&-5530.12\\
$10^{-25}$ &$0.08$ &$0.089 \pm 0.027$& $0.694 \pm 0.009$ & $68.02 \pm 0.67$ & $< 0.03$ & $<0.02$ & $0.10\pm 0.05$ & $<0.43$ &-5530.53\\
$10^{-26}$ &$< 0.006$ &$0.117 \pm 0.002$ & $0.694 \pm 0.009$ & $68.06 \pm 0.63$ & $<0.09$ &$<0.0009$ & $0.05\pm 0.02$& $<0.74$&-5530.49\\
$10^{-27}$ &$< 0.003$ &$0.119 \pm 0.001$ & $0.691 \pm 0.010$ & $67.82 \pm 0.69$ & $< 0.09$ & $< 0.0003$ & $0.06\pm 0.03$ & $<0.74$&-5530.37\\
$10^{-28}$ &$<0.003$ &$0.120 \pm 0.002$ & $0.683 \pm 0.014$ & $67.21 \pm 1.00$ & $<0.09$& $<0.0002$ & $0.08\pm 0.03$ &$<0.76$&-5530.14\\
$10^{-29}$ &$<0.003$ &$0.120 \pm 0.002$ & $0.676 \pm 0.017$ & $66.61 \pm 1.24$ & $<0.09$ & $< 0.0002$ & $0.11\pm 0.04$& $<0.75$&-5530.38 \\
$10^{-30}$ &$<0.003$ &$0.120 \pm 0.001$ & $0.676 \pm 0.016$ & $66.56 \pm 1.25$ & $<0.10$ & $<0.0002$& $0.11\pm 0.05$ &$<0.78$&-5530.63\\
$10^{-31}$ &$<0.006$ &$0.119 \pm 0.001$ & $0.671 \pm 0.022$ & $66.18 \pm 1.59$ & $< 0.10$ & $< 0.0004$ & $0.15\pm 0.06$ &$<0.81$&-5530.51\\
$10^{-32}$ &$<0.016$ &$0.119 \pm 0.001$ & $0.648 \pm 0.040$ & $65.07 \pm 2.36$ & $<0.09$ & $<0.001$ & $0.24\pm 0.10$ &$<0.76$&-5530.75\\
$10^{-33}$ &$< 0.38$ &$0.119 \pm 0.001$ & $< 0.64$ & $65.86 \pm 1.46$ & $< 0.10$ & $< 0.006$ & $1.38\pm 0.46$ & $<0.78$ &-5530.49 \\
    \end{tabular}
    \caption{{\bf Constraints on the axion parameter space:} the $1\sigma$ errors (for bounded parameters) and 95\% C.L. upper bounds (for constraints) for the data set of \textit{Planck} T+P and the lensing deflection power spectrum. 
    \label{table:results}}
\end{table*}

We show results in Fig.~\ref{fig:adi_constraints} for the full T+P+lens data combination. In each bin the value of $\Omega_a h^2$ is consistent with zero and we find no evidence for departures from one-component CDM across the entire range of scales probed by the CMB. The tightest constraint on the ULA fraction, $f_\mathrm{ ax}\equiv\Omega_a/\Omega_d$, is in the $m_a=10^{-28}\text{ eV}$ bin, where we find $f_\mathrm{ ax}<0.019$ at 95\% C.L. The bounds on all the relevant parameters in each mass bin are summarised in Table~\ref{table:results}.

Having established the consistency of the null hypothesis (the standard cosmological model with CDM only), we show the binned limits on axion energy density in the $(m_a,\sqrt{\Omega_a h^2})$ plane for the full data combination T+P+lens in Figure~\ref{fig:lens_u_plot}.~\footnote{In these plots the square root is chosen to aid visualistion by reducing the dynamic range of the density scale.} The upper $95\%$ limits are indicated numerically. Within a fixed mass bin, the data are randomly (horizontally) scattered, and the colour scale is proportional to the point density, smoothed with a Gaussian kernel density estimate. We see the familiar U-shaped degeneracy, with the lightest axions behaving as DE, and the heaviest ones as DM (see also Appendix~\ref{appendix:axion_basics}), with a mild but statistically insignificant preference for non-zero axion energy density at the lowest and highest $m_{a}$ values.

\begin{figure} 
\includegraphics[width= 1\columnwidth]{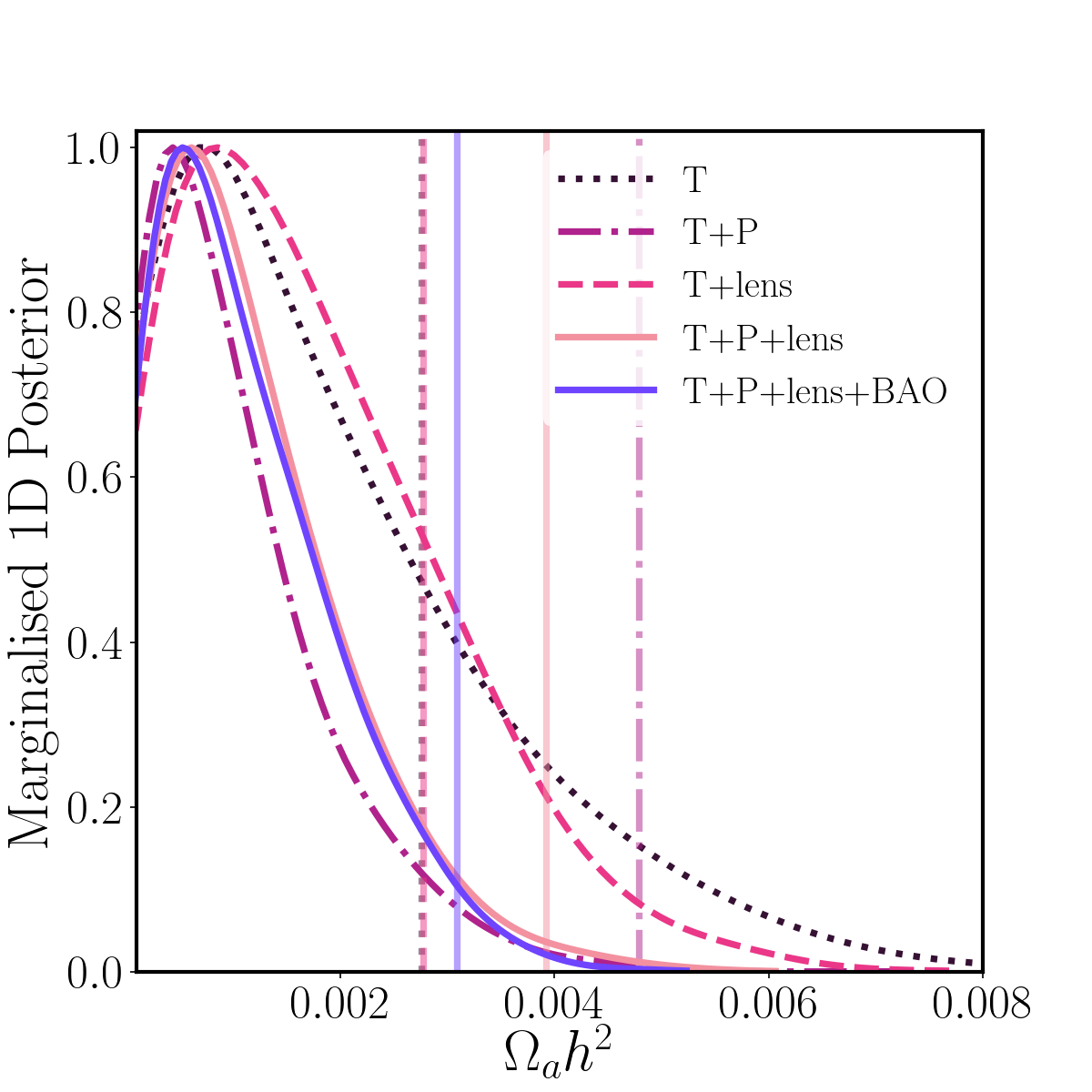}
\vspace{-0.5cm}
 \caption{{\bf Stringent upper bounds on intermediate mass axions}: for intermediate mass ULAs we obtain the tightest upper bounds on the axion energy density. We show the upper limits on the allowed energy density in axions, $\Omega_ah^2$, for the adiabatic mode for $m_a=10^{-28}\text{ eV}$ for different data combinations. The data combinations used are listed in Table~\ref{table:runs}. Adding in polarisation to the temperature data improves the constraint by a factor of two. Adding in CMB lensing from \textit{Planck}, however, pushes the peak of the distribution to slightly larger values of $\Omega_ah^2$ (see text for discussion). The upper $2\sigma$ bounds on the axion density are given in Table~\ref{table:m28}.  
\label{fig:m28_constraints}}
\end{figure}

Most of this work is devoted to the constraints on the axion parameters from the full set of CMB data, including the lensing deflection power spectrum. In Fig.~\ref{fig:m28_constraints} we show how our different data combinations affect the posterior distribution on $\Omega_a h^2$ for $m_a=10^{-28}\text{ eV}$.  The total data combination leads to an improvement in the 95\% upper limit by a factor of roughly two compared to the case with temperature anisotropies alone (H15). CMB lensing data as used does not alone lead to a significant improvement in the constraint, and in fact shifts the distribution to favour slightly larger values compared to data combinations not including lensing. This is similar to a trend seen in constraints on the neutrino mass when combining CMB lensing measurements with temperature-only measurements of the CMB \citep{2016A&A...594A..11P, Sherwin:2016tyf, 2017arXiv170801530D}. Such a shift could be due to observational systematic effects, or might also indicate that mixed DM could play a role in resolving (possibly physical) low-$z$ power spectrum amplitude tensions. The shifts seen here, however, are not statistically significant, and so additional data are needed to distinguish between new physics or systematics as the explanation for this shift. 
 
The log-posterior for the runs is given in the last column of Table~\ref{table:results}. Firstly, we note that the log-posterior is stable across mass ranges, suggesting that there isn't a preferred mass within the specific range tested. In addition, the value of $\log{\mathcal{P}} = -5530$ is comparable to the value obtained on the same combination of for a `vanilla' model of adiabatic CDM only in the \textit{Planck} analysis of  $\log{\mathcal{P}} = -5532.$

%
\begin{table*}
    \begin{tabular}{c|c|c|c|c}
 Data set& $\Omega_ah^2$&$\Omega_ch^2$&$\Omega_\Lambda$&$H_0~{[\rm km\, s}^{-1}\mathrm{ Mpc}^{-1}\mathrm{]}$\\
\hline
\textit{Planck} T & $< 0.0049$ & $0.1190\pm   0.0024$ &
$0.68 \pm  0.02$ &
$67.04 \pm 1.68$ \\
\textit{Planck} T+P & $< 0.0027$ & $0.1199 \pm 0.0016$ &
$0.68 \pm 0.01$ &
$67.12 \pm 0.99$ \\
\textit{Planck} T+lens & $< 0.0040$ & $0.1189\pm 0.0025$ &
$0.68 \pm 0.02$ &
$67.10 \pm 1.51$ \\
\textit{Planck} T+P+lens & $< 0.0028$ & $0.1197 \pm 0.0016$ &
$0.68 \pm 0.02$ &
$67.10 \pm 1.51$ \\
\textit{Planck} T+P+lens+BAO & $< 0.0027$ & $0.1196\pm   0.0016$ &
$0.68\pm 0.01$ &
$67.13\pm 1.05$ \\
  \end{tabular}
    \caption{{\bf Adiabatic constraints on $10^{-28}\, \mathrm{eV}$ axions:} 95\% C.L. upper bounds on the axion density, and $1\sigma$ errors on other parameters for one `belly-like' ULA, showing the effect of combining different data sets. The $\Omega_ah^2$ posterior distributions are shown in Figure~\ref{fig:m28_constraints}.
    \label{table:m28}}
\end{table*}
\section{Isocurvature and Constraints on Inflation}
\label{sec:iso_constraints}

The right-hand panel of Fig.~\ref{fig:lens_u_plot} and Table \ref{table:results} show how the binned constraints on the axion energy density are affected by the inclusion of the axDI mode (as compared to the adiabatic-only case in the left panel). We see that the constraints in the range $10^{-33}\text{ eV}\leq m_a\leq 10^{-26}\text{ eV}$ are largely unaffected. This indicates that isocurvature constraints do not play a significant role in the constraints to the energy density of the lightest ULAs. For $m_a=10^{-25}\text{ eV}$ and $m_a=10^{-24}\text{ eV}$, the bounds tighten by around $10-20\%$. This indicates the role of isocurvature in constraining heavier axions, and it further indicates that there is no preference for non-zero isocurvature. 

\begin{figure} 
\includegraphics[width=1.0\columnwidth]{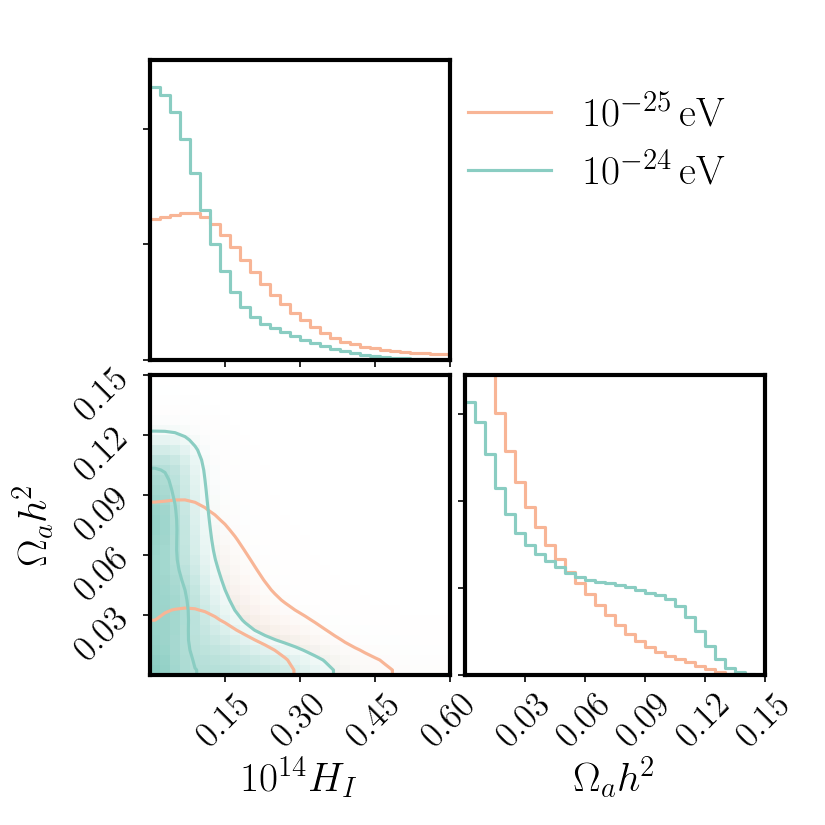} 
\vspace{-0.5cm}
 \caption{{\bf Distributions in $H_I-\Omega_ah^2$ space.} The value of $H_I$ induces an axDI mode and a tensor mode in the initial conditions, both of which contribute to low-$\ell$ $T$ and $E$ anisotropies. At large $\Omega_a h^2$, $H_I$ is tightly bounded by the isocurvature contribution, while at low $\Omega_a h^2$ the upper bound is weaker, and driven by the tensor contribution.
\label{fig:omaxh2_hinf}}
\end{figure}

In order to assess the importance of the ULA isocurvature mode in driving constraints on inflation, it is instructive to look at the 2d posterior correlations between $\Omega_a$ and inflationary parameters, and how these correlations depend on $m_a$. In our analysis we treated $H_I$ as a primary parameter, and enforced consistency between the isocurvature and tensor amplitudes (see Section~\ref{sec:iso_theory} and Appendix~\ref{appendix:hi_prior}). Figure~\ref{fig:omaxh2_hinf} shows constraints in the $(H_I,\Omega_a h^2)$ plane for the two highest axion masses we considered. In both cases there are regions of the allowed parameter space permitting large values of the axion density and in addition $10^{13}\text{ GeV}\lesssim H_{I}\lesssim 10^{14}\text{ GeV}$. The distribution has a slope, with large axion density requiring smaller values of $H_I$. As illustrated in Fig.~\ref{fig:opener}, at this value of $H_I$, the isocurvature and tensor spectra for these different masses are similar in amplitude on large scales, and so $H_{I}$ must take lower values if $\Omega_{a}h^{2}$ contributes significantly to the DM density.

\begin{figure*}
\includegraphics[width= 1.0\columnwidth]{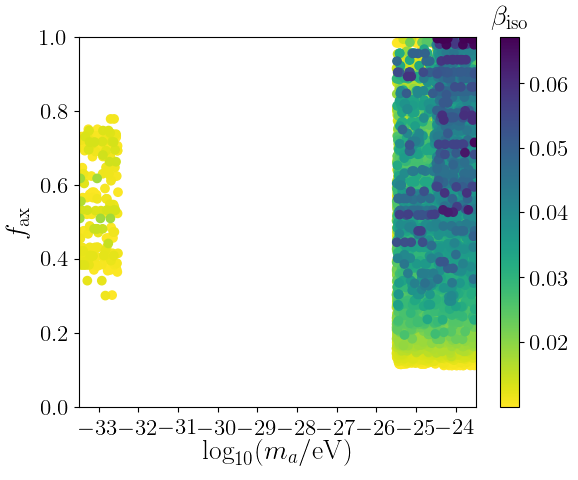}
\includegraphics[width=1.0\columnwidth]{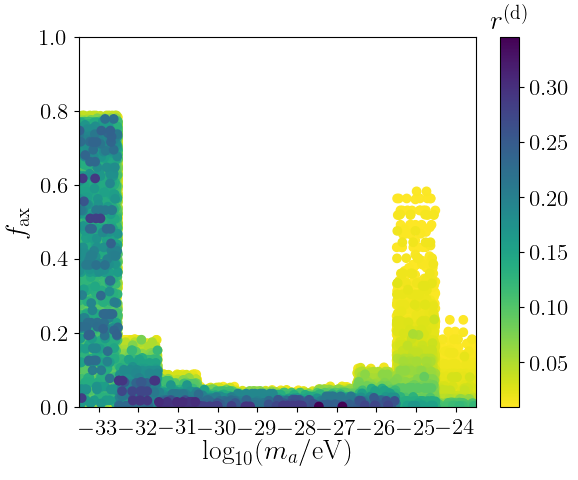} 
 \caption{
 {\bf Derived inflationary parameters:} the left and right panels show scatter plots from MCMC chains for all samples that satisfy either $\beta_\mathrm{iso}> 0.01$ or $r^{\rm (d)} > 0.01$ respectively. The isocurvature is only non-trivial for the DM-like axions with $m_a = (10^{-24}, 10^{-25})~\mathrm{eV},$ and the lightest DE-like axions $m_a=10^{-33}~\mathrm{eV}$. In the intermediate regime isocurvature is negligible and constraints are driven by limits to the tensor amplitude. This leads to constraints on the allowed value of the tensor-to-scalar ratio $r^{\rm (d)}$ as shown in Figure~\ref{fig:dm_iso_constraints}, and is also shown through a 2D contour plot in Figure~\ref{fig:beta_r}.   Note that the points are not explicitly weighted by the posterior (as in e.g. Figure~\ref{fig:two-u}), but the point density of the chains is representative of the goodness of fit.
\label{fig:window_coexist}
}
\end{figure*}

The bounds on the derived parameters $\beta_\mathrm{ iso}$ and $r^{(\rm d)}$ in each mass bin are summarised in Table~\ref{table:results}. Figure~\ref{fig:window_coexist} demonstrates how the distribution of each derived parameter depends on the axion mass and energy density. In this figure, we colour the MCMC sample points by the value of the derived parameter, and show only those points where the derived parameter is larger than $0.01$. The isocurvature mode dominates the constraints at high axion mass and fraction, forcing $r^\mathrm{(d)}<0.01$.

In the ``belly'' of the U, isocurvature constraints are negligible, and the constraints from tensor modes force $\beta_\mathrm{ iso}<0.01$~\citep{2013PhRvD..87l1701M}. In some regions there is a balance allowing both $r^\mathrm{(d)}$ and $\beta_\mathrm{ iso}$ of order a few percent. This is our so-called ``window-of-coexistence'', and allows for an interplay between constraints on DM and inflation in the CMB via the axDI mode. One region is for $m_a=10^{-25}\text{ eV}$ and $m_a=10^{-24}\text{ eV}$ with $f_\mathrm{ ax}\approx 0.1$. In this regime the axDI mode is virtually indistinguishable from the CDI mode at the accuracy the data. The second region is for $m_a=10^{-33}\text{ eV}$ and $0.4<f_\mathrm{ ax}<0.8$, which allows a large contribution of quintessence-type isocurvature.

\begin{figure} 
\includegraphics[width= 1.0\columnwidth]{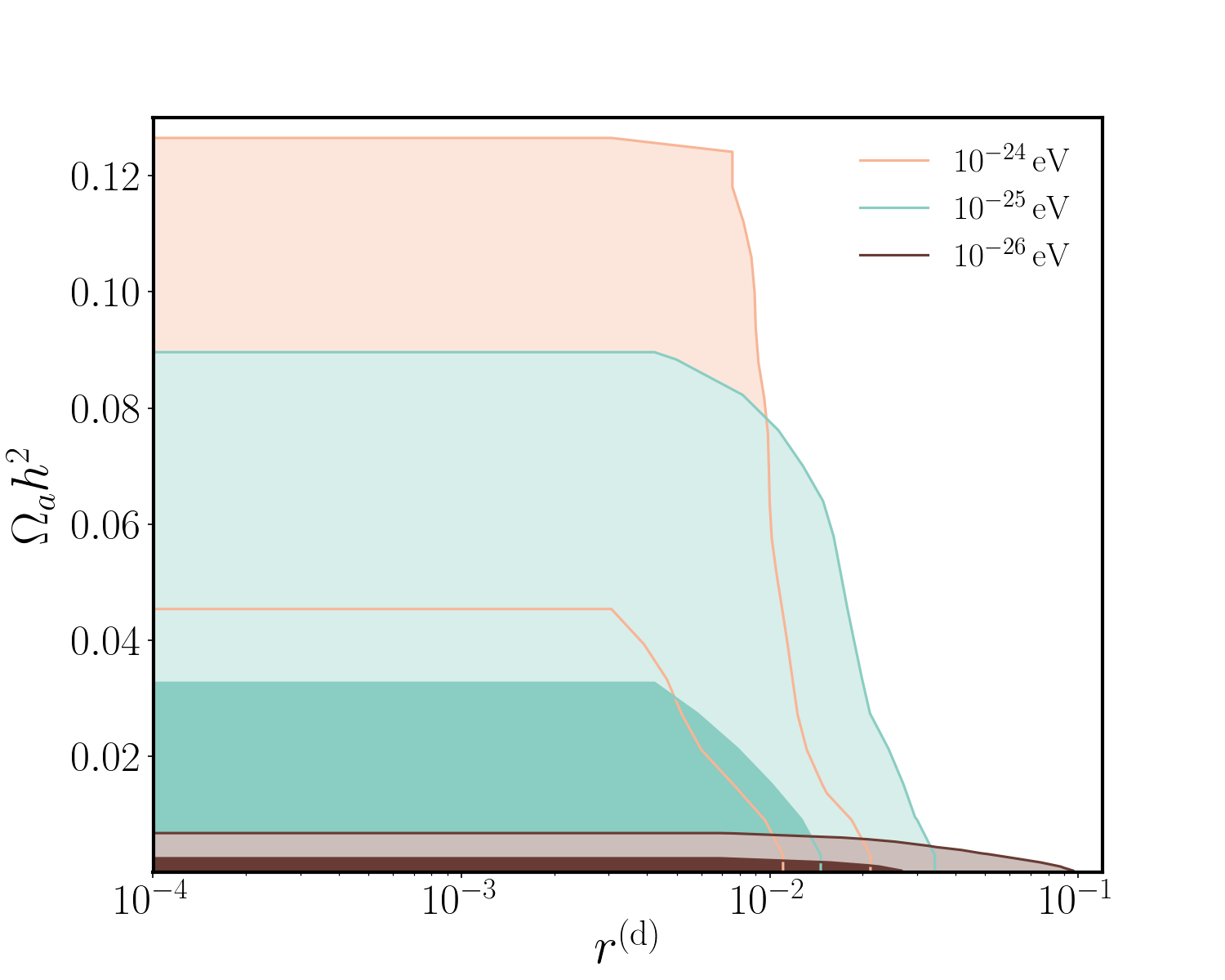}
\vspace{-0.5cm}
 \caption{{\bf Constraints on the derived tensor amplitude:} At low axion mass the isocurvature mode is negligible due to the bound on $\Omega_a h^2$ in the mixed DM model. The derived bound $r^\mathrm{ (d)}<0.09$ at 95\% C.L. for $m_a=10^{-26}\text{ eV}$ is driven by the tensor contribution to the $T$ and $E$ modes. As the axion mass, and thus energy density, increase, the axion isocurvature contribution becomes more important, tightening the derived constraint to $r^\mathrm{ (d)}\leq 0.04$ and $r^\mathrm{ (d)}\leq 0.01$ for $m_a=10^{-25}\text{ eV}$ and $m_a=10^{-24}\text{ eV}$ respectively. 
\label{fig:dm_iso_constraints}}
\end{figure}

Fig.~\ref{fig:dm_iso_constraints} shows the correlation between $r^\mathrm{ (d)}$ and $\Omega_a h^2$. For low axion mass, $m_a=10^{-26}\text{ eV}$, the bound on $r^\mathrm{ (d)} \lesssim 0.1$ is independent of $\Omega_a h^2$. The bound is comparable to the \textit{Planck} only constraints on an adiabatic+tensor (no isocurvature) model in the absence of additional constraints on $r$ from low-$\ell$ B-modes (e.g. from BICEP/KECK). Our posterior on $r^\mathrm{ (d)}$ leads to a slightly tighter constraint than the \cite{2016A&A...594A..13P} analysis due to the change of variables when using $H_I$ as the primary parameter (see Appendix~\ref{appendix:hi_prior}). For each of the higher axion masses, our constraint on $r^{\mathrm{(d)}}$ is tighter still, due to the importance of the isocurvature constraint and marginalisation over $\Omega_a h^2$. 

After marginalising over $\Omega_a h^2$ we find the 95\% C.L. bound
\be
r^\mathrm{(d)} < 0.01 ; \quad (m_a=10^{-24}~\mathrm{eV}) \, ,
\ee
significantly more stringent than direct constraints to $r$ from the tensor contribution to the $C_{\ell}$'s alone and highlights the constraining power of isocurvature on inflationary physics in the ULA scenario. The tighter limit is driven by the marginalisation and reflects that, on average, $m_a=10^{-24}\text{ eV}$ is consistent with a value of $r=0.01$. 

After marginalising over $\Omega_a h^2$ we find the 95\% C.L. bound
\be
\beta_\mathrm{iso} < 0.03 ; \quad (m_a=10^{-24}~\mathrm{eV}) \, ,
\ee
This is slightly tighter than the bound for axion type isocurvature in \cite{2016A&A...594A..20P},  $\beta_\mathrm{iso} < 0.038$, again this is due to our marginalisation, this time allowing lower values of $\Omega_a h^2$ (\emph{Planck} fix $\Omega_a h^2=\Omega_d h^2$). We also note that in our parameterisation $\beta_{\rm iso}$ is not strictly the same as that defined by \emph{Planck}.

\begin{figure} 
\includegraphics[width=0.49\textwidth]{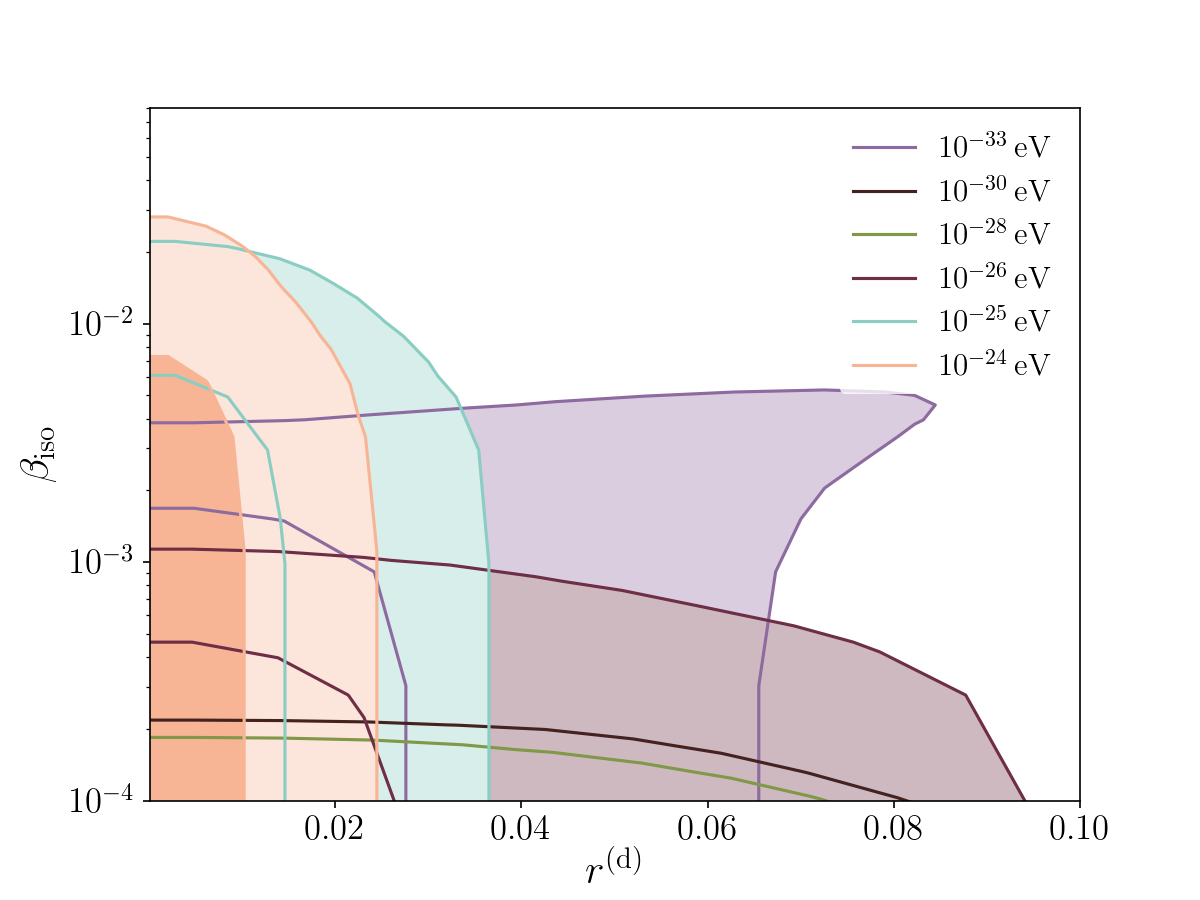} 
\caption{{\bf Tensor-isocurvature coexistence:} marginalised 2D contours in the $\beta_\mathrm{iso}-r_\mathrm{tens}$ plane. Only for the DM-like axions is there both a non-negligble tensor-mode contribution while also an appreciable amplitude of isocurvature, as also illustrated in Figure~\ref{fig:window_coexist}.}
\label{fig:beta_r}
\end{figure}

Fig.~\ref{fig:beta_r} shows the two-dimensional distribution for both the derived parameters, $r^\mathrm{ (d)}$ and $\beta_\mathrm{ iso}$. For $m_a=10^{-33}\text{ eV}$ we notice that the window of co-existence is disfavoured in the posterior, lying in the tail of the distribution at large $r^\mathrm{ (d)}$ and outside the 95\% C.L. contour. After marginalising over $\Omega_a h^2$ (which disfavours large $r^\mathrm{ (d)}$) there is a region in the allowed parameter space at 95\% C.L. allowing for $\beta_\mathrm{ iso}>0.01$ and $r^\mathrm{ (d)}>0.01$ for both $m_a=10^{-25}\text{ eV}$ and $m_a=10^{-24}\text{ eV}$, i.e. the window of co-existence. 

\begin{figure*}
\begin{center}
$\begin{array}{ccc}\hspace*{-2.5em}
\includegraphics[width=0.85\columnwidth]{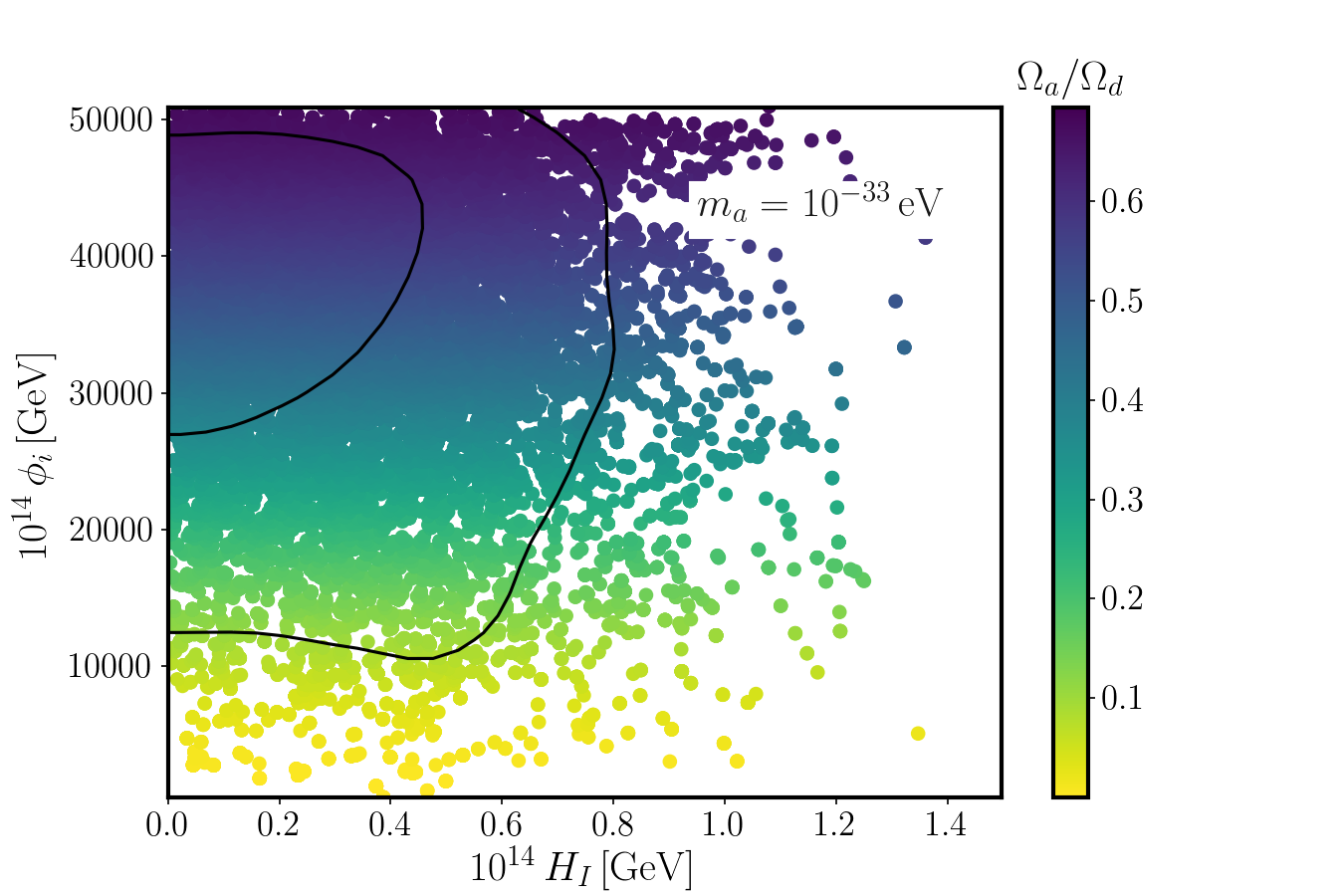}& \hspace*{-4.0em}
\includegraphics[width=0.85\columnwidth]{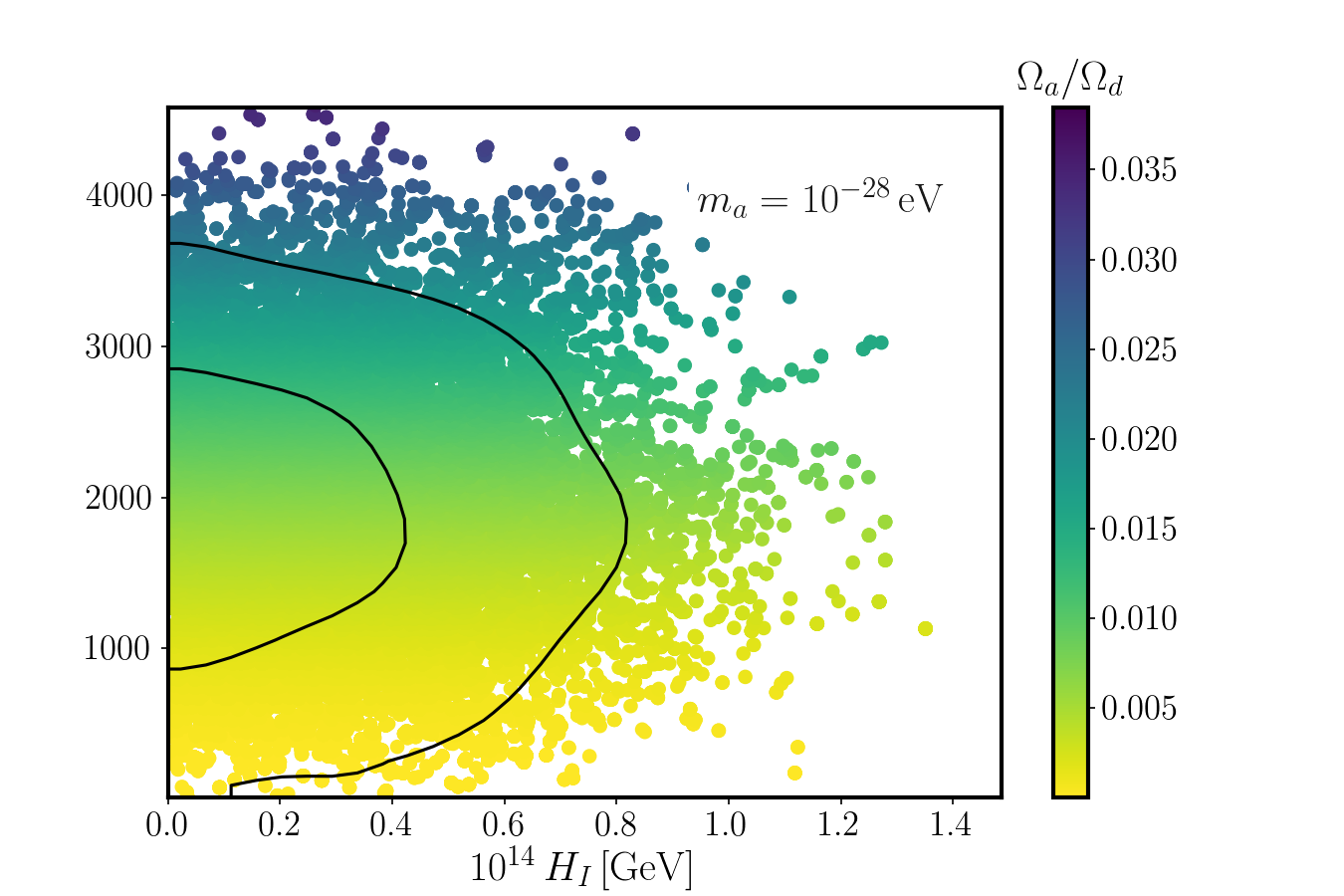}&\hspace*{-4.0em}
 \includegraphics[width=0.85\columnwidth]{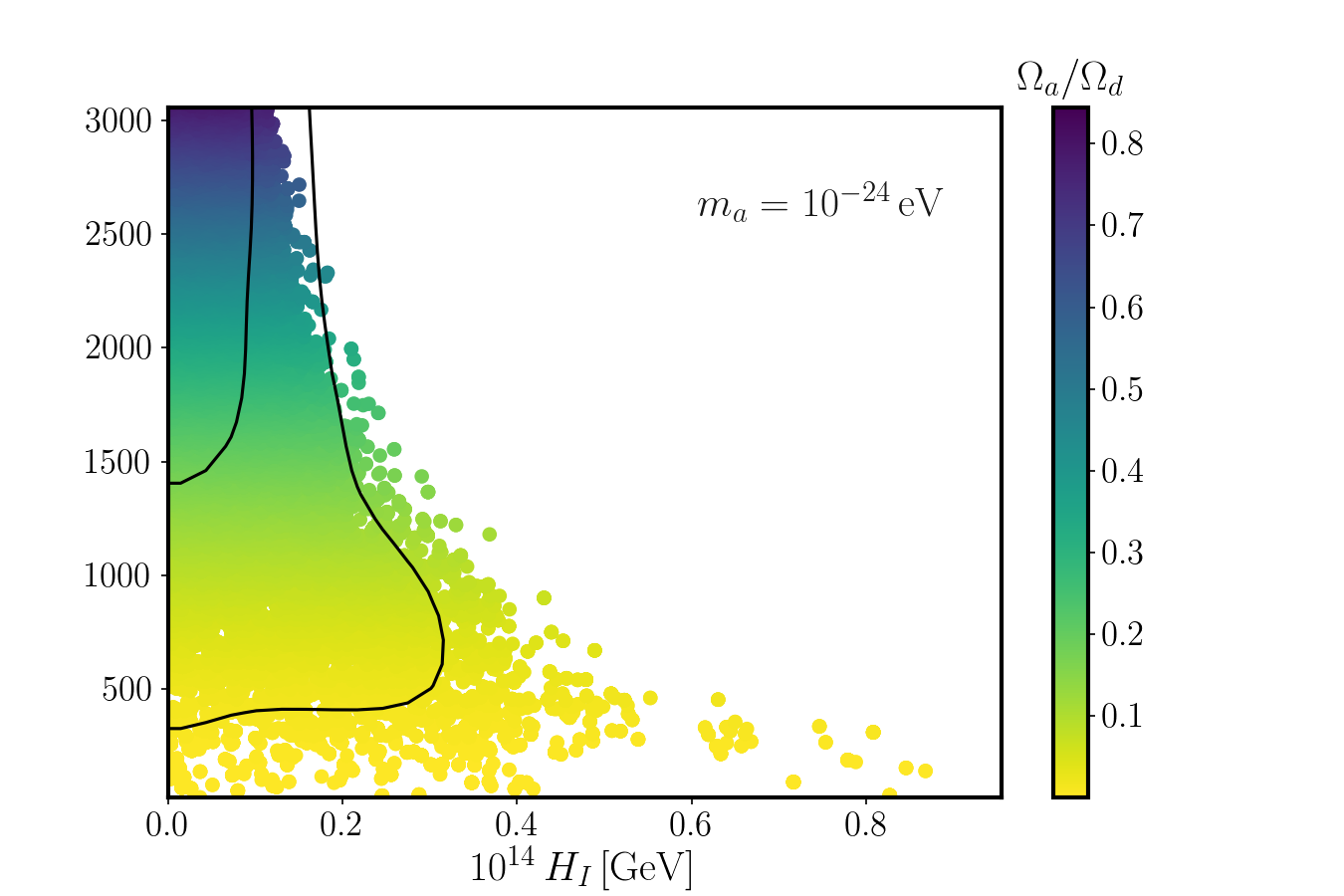}
 \end{array}$
\caption{{\bf Initial misalignment angle and the energy scale of inflation:} the marginalised 2D contours in the $\phi_i-H_I$ plane for DE-like, `belly' and DM-like axions in the left, middle and right panels. The y-axis is common to all three panels, while the x-axis shows the range in allowed values of $H_I.$ In all cases the field value is correlated with the axion density due to the misalignment production, with larger field values necessary for smaller masses. The allowed values of $H_I$ are larger for smaller masses. In all cases $\phi_i\sim f_a\gg H_I$ justifying the assumption that ULAs with a cosmologically relevant energy densities are accompanied by isocurvature perturbations.
\label{fig:phi_hinf}}
\end{center}
\end{figure*}

Finally in this section, we test our assumption that ULAs are indeed in the correct regime of spontaneous symmetry breaking to produce isocurvature perturbations. Fig.~\ref{fig:phi_hinf} shows the two dimensional posterior distribution for $H_I$ compared to the derived parameter $\phi_i$, the initial axion field value. The contours show the 68 and 95\% confidence intervals, and the scatter points show MCMC samples coloured by the value of the ULA DM fraction.

For the two extreme cases of $m_a=10^{-33}\text{ eV}$ and $m_a=10^{-24}\text{ eV}$ there is a slight preference for $\phi_i\geq 0$ driven by the strong degeneracies at these masses between ULAs and the cosmological constant and CDM respectively.

In all three mass bins we note that the scale on the $\phi_i$ axis is much larger that the scale on the $H_I$ axis. For all of our mass bins the axion energy density has to be far below the 95\% upper limit in order to achieve $\phi_i<H_I$, and within our sampling this condition is never satisfied for any masses we consider. That is, for all ULAs of cosmic relevance, the opposite inequality, $\phi_i\gg H_I$ is satisfied at a high level of confidence. The degree of confidence to which this is respected is dependent on our linear prior on $\Omega_a h^2$. A more physical prior would require modelling the distribution of decay constants in a particular model, e.g. M-theory as in \cite{2017arXiv170603236S}.

We recall that the initial field value is set by the decay constant as $\phi_i\approx\theta_i f_a$ (Eq.~\ref{eqn:initial_phi_exact}). Since $|\theta_i|\leq \pi$ we have that $|\phi_i|\leq \pi f_a$. The condition for production of isocurvature perturbations is approximately $f_a\gtrsim H_I/2\pi$, up to thermal corrections to the potential. If $|\phi_i|\gg H_I$ then $f_a\gg H_I$ and spontaneous symmetry breaking for a ULA field with a cosmologically relevant density must occur before or during inflation. Therefore our assumption that such a ULA is accompanied by isocurvature perturbations is valid.\footnote{This conclusion can be modified in the presence of ``monodromy'' \citep[e.g.][]{2008PhRvD..78j6003S,2017JCAP...01..036J} or multi-field mechanisms \citep[e.g.][]{1998PhRvD..58f1301L,2005JCAP...01..005K,2008JCAP...08..003D,2017arXiv170300453B,2017arXiv170603236S}. A particularly interesting possibility is the ``clockwork'' mechanism, which can lower the symmetry breaking scale as far as 1 TeV while maintaining very large field excursions, \citep[see e.g.][]{2016PhRvD..93h5007K,2016JHEP...08..044H}.}

\section{Discussion and Conclusions}
\label{sec:discussion}

Using the latest \emph{Planck} CMB data we have presented constraints on a mixed DM model where a component of the DM resides in a ULA. We constrained the DM fraction in ULAs to be less than a few percent across a wide range of masses. Thanks to the spontaneous symmetry breaking leading to its production, and the generation of initial conditions through vacuum fluctuations during inflation, the ULA carries uncorrelated isocurvature perturbations. We used constraints on the isocurvature amplitude, along with the concurrently produced tensor perturbations, to bound the energy scale of inflation in this model. Our results are the first to use CMB polarisation and lensing deflection to test the ULA hypothesis, and are the first systematic study of axion-type isocurvature where both the amplitude and DM fractions can be independently constrained. 

In the isocurvature analysis, the ``window of co-existence'' at high mass is particularly interesting, as the sensitivity of CMB data to the axion energy density in this regime will improve significantly with any high-$\ell$ lensing measurement. Furthermore, the corresponding tensor amplitude of $r\approx 0.01$ can be detected at high significance in the low-$\ell$ B-modes~\citep{2016arXiv161002743A}. The window of co-existence will be closed if CMB-S4 finds no evidence for mixed DM at the percent level in the mass range $10^{-25}\lesssim m_a \lesssim 10^{-24}\text{ eV}$ and/or excludes $r>0.01$.

The smaller window at $m_a=10^{-33}\text{ eV}$ depends on the correct computation of the axDI spectral shape, with the Jeans cut-off at low $\ell$. This smaller window is outside the 95\% confidence region and requires larger values of $H_I$ (and thus derived $r^{(\rm d)}$). This window is also much harder to probe using the isocurvature mode, as the CMB spectra are dominated by cosmic variance at low $\ell$. In this parameter space, future measurements of $T$ and $E$ (where isocurvature contributes) are unable to significantly improve over existing measurements due to cosmic variance. Furthermore, the large value of $r^\mathrm{ (d)}$ would be strongly disfavoured if we were to include direct constraints on the low-$\ell$ $BB$ amplitude. The quintessence type axDI in the CMB is thus likely of only theoretical interest, although there could be as yet unforseen consequences of such superhorizon DE perturbations. Axion quintessence can be further probed by observables sensitive to its effect on the Hubble expansion~\citep[e.g.][]{Amendola:2012ys,2014PhRvD..90j5023M,2016PhRvD..93l3005E,2017JCAP...01..023S}.

In addition to the contribution of axions to the isocurvature power, and their effect on lensing power, other interesting observational effects are worth exploring with upcoming CMB data. One such possibility is polarisation rotation from $E$-modes to $B$-modes caused by the birefringce induced by the axion coupling to the electromagnetic Chern-Simons term~\citep{1990PhRvD..41.1231C,1992PhLB..289...67H}. Isocurvature perturbations in an axion with $10^{-33}\text{ eV}<m_a<10^{-28}\text{ eV}$ cause such a term to generate low-$\ell$ $B$-modes~\citep{2009PhRvL.103e1302P}, which were most recently used in \cite{2017arXiv170502523A} \citep[building on the work of][]{2012PhRvD..86j3529G} to limit a combination of the ULA coupling to photons and $H_I$. Another affect of the axion-photon coupling pointed out by \cite{Liu:2016dcg} is the polarisation rotation sourced by gravitational anisotropies in adiabatic axion perturbations, an effect independent of $H_I$ and present for much heavier axions. In addition, axion-induced birefringence induces a non-negligible circular polarisation \citep{Payez:2009vi}, which is also entering the testable regime through experiments like SPIDER \citep{Nagy:2017csq}. Incorporating the ULA-photon coupling and the birefringence it induces into \acamb~and \cosmosis~will allow such studies to be combined with those we have presented here in global constraints to axion parameters, in particular the decay constant.

\cite{2006PhLB..642..192A} performed the first modern study of the mixed CDM and ultralight DM model, combining multiple probes of the CMB, galaxy, and Lyman-alpha forest flux power spectra \citep[for earlier work, see e.g.][]{1995PhRvL..75.2077F}. The CMB provides the most stringent bounds on $\Omega_a$ at low axion masses, while the Lyman-alpha forest probes smaller scales and larger axion masses. The CMB is a clean probe, depending very little on astrophysical and non-linear modeling. The Lyman-alpha forest has larger uncertainties from the modeling of the temperature evolution of the intergalactic medium~\citep{2017PhRvD..95d3541H,2015arXiv151007006G}, but can be used to place the strongest constraints on deviations from CDM on small scales~\citep[e.g.][]{2005PhRvD..71f3534V,2016JCAP...08..012B,2017arXiv170304683I,2017arXiv170309126A}. Very recently \cite{2017arXiv170800015K} studied the constraints on a multi-component axion DM model imposed by the measurements of the Lyman-alpha forest flux power spectrum by the XQ-100 survey at redshifts $3.5<z<4.5$~\citep{2016A&A...594A..91L,2017MNRAS.466.4332I}. They found $\Omega_a/\Omega_d\lesssim 0.2$ for $10^{-23}\text{ eV}\lesssim m_a\lesssim 10^{-21}\text{ eV}$. These constraints are highly complementary to those coming from the CMB that we have presented here. It would useful to perform a combined analysis of the CMB and Lyman-alpha forest, completing the work of \cite{2006PhLB..642..192A} with modern precision cosmological data. 

\cite{2017arXiv170800015K} also studied the effect of combining the Lyman-alpha forest bounds with isocurvature constraints, and found $r^\mathrm{( d)}<10^{-3}$ over the axion mass range covered. This is consistent with our results and previous work~\citep{2013PhRvD..87l1701M,2014PhRvL.113a1801M,2017arXiv170202116D,2017arXiv170308798V}. This, and our analysis presented here, implies that the currently popular ``Fuzzy DM'' model, with all DM composed of axions/ultralight scalars with $m_a\approx 10^{-22}\text{ eV}$~\citep{2000PhRvL..85.1158H,2016PhR...643....1M,2017PhRvD..95d3541H}, requires (in the simplest interpretation) low $H_I$, is not in the window of co-existence, and is inconsistent with a large tensor-to-scalar ratio. 

The lensing data we have used are from the \textit{Planck} minimum variance estimator, with the conservative lensing multipole range of $40\leq L \leq 400$ adopted in \cite{2016A&A...594A..13P}. Our sensitivity could improve significantly if we extended our analysis to a broader range of lensing $L$, an area for future exploration. Use of higher multipoles would lead to significant improvements to the constraints, particularly at higher values of $m_a$ by breaking the degeneracy between ULAs and CDM. 

The future offers exciting opportunities for tests of ULA DM using the CMB. CMB-S4 could improve limits to isocurvature by a factor of $\sim 4$ and improve sensitivity to the ULA mass fraction by a factor of $\sim 3$ in the currently constrained part of parameter space \citep{2016arXiv161002743A,2017PhRvD..95l3511H}. CMB-S4 could also move the potential detection window for a $\sim 1 \%$ mass fraction of DM in ULAs to axion masses as high as $m_{a}\sim 10^{-23}~\mathrm{ eV}$, thanks in large part to exquisite sensitivity to weak gravitational lensing of the CMB (in turn facilitated by the possible measurement of foreground subtracted CMB anisotropies at scales as small as $\ell=3000$ in polarisation).  If evidence for ULAs is seen through their signature on CMB weak lensing or cosmological birefringence, then observable imprints of isocurvature could complement $B$-mode signatures of tensor modes to offer new leverage on the inflationary energy scale and our knowledge of the origins of the Universe.

%
%
\section*{Acknowledgments}
The Dunlap Institute is funded through an endowment established by the David Dunlap family and the
University of Toronto. RH would like to acknowledge that the land on which the University of Toronto is built is the traditional territory of the Haudenosaunee, and most recently, the territory
of the Mississaugas of the New Credit First Nation. The territory was the subject of the Dish With
One Spoon Wampum Belt Covenant, an agreement between the Iroquois Confederacy and the Ojibwe
and allied nations to peaceably share and care for the resources around the Great Lakes.
This territory is also covered by the Upper Canada Treaties. RH is grateful to have the opportunity to work in the community, on this territory.
DJEM is supported by a Royal Astronomical Society Postdoctoral Fellowship at King's College London. Research at Perimeter Institute was supported by the Government of Canada through Industry Canada and by the Province of Ontario through the Ministry of Research and Innovation. DG was supported at the Institute for Advanced Study by the National Science Foundation (AST-0807044) and NASA (NNX11AF29G), at the University of Chicago by National Science Foundation Astronomy \& Astrophysics Postdoctoral Fellowship (Award No. AST-1302856), at the Kavli Institute for Cosmological Physics at the University of Chicago through grant NSF PHY-1125897 as well as an endowment from the Kavli Foundation and its founder Fred Kavli, and in part by the National Science Foundation under Grant No. NSF PHY-1125915 at the Kavli Institute for Theoretical Physics (KITP) at UC Santa Barbara. DG thanks KITP for its hospitality during the completion of this work. We thank E.~Calabrese and T.~L.~Smith for useful discussions, and J.~Zuntz for invaluable help with \textsc{cosmosis}, and the (longsuffering) demystification of compiler flags. We also thank the online community of Stack Overflow, without whom our use of \textsc{matplotlib} (and indeed many other things) would be much more limited.


\appendix

\section{Axion Basics}
\label{appendix:axion_basics}

We model axions as a classical scalar field, $\phi$, with a time independent mass, $m_a$, and no self-interactions, which evolves according to the Klein-Gordon equation: 
\begin{eqnarray}
\Box\phi-m_a^2\phi=0\label{eqn:kg_hom},
\end{eqnarray}
where $\Box$ is the D'Alembertian operator for the perturbed cosmological spacetime \citep{1995ApJ...455....7M}. The axion sources the Einstein equations via its classical energy momentum tensor, with fluid components, $\rho_a$, $P_a$, and $v_a$.

We treat the ULA mass as constant and ignore self-interactions as a minimal assumption that reduces the number of parameters that we must sample. Axions which obtain their mass from a gauge theory Chern-Simons interaction have a temperature (and thus time) dependent mass, and a calculable self-interaction potential \citep[e.g.][]{2016JHEP...01..034D,2017PhRvL.118n1801D}. The ULA mass, $m_a$, is effectively time-independent for geometric (closed string) axions in string theory~\citep{2006JHEP...06..051S}, and the self-interaction potential for string axions can take a wide variety of functional forms \citep[e.g.][]{2008PhRvD..78j6003S,2017JCAP...01..036J}. The effects of time dependent mass and self-interactions at the level of the relic density are well known~\citep{1992PhRvD..45.3394L}. The effects of self-interactions in \acamb~are under development, and have already been implemented in a version of \textsc{class}~\citep{2011arXiv1104.2932L} by other authors~\citep{2016JCAP...07..048U,2017arXiv170310180L}. See also \cite{2017PhRvD..96b3507Z}. 

The axion field is decomposed into a homogeneous part, $\phbar(t)$, and a (linear) perturbation, $\delta\phi(t,\vec{x})$. In a standard FRW spacetime Eq.~(\ref{eqn:kg_hom}) for the homogeneous field takes the form
\begin{equation}
\ddot{\phi}+2\mathcal{H}\dot{\phi}+m_a^2 a^2 \phi=0.\label{eqn:homo_eom}\end{equation}
The perturbed EOM in synchronous gauge (which is used by \textsc{camb}) is then
\begin{align}
{\delta \ddot{\phi}}+2\mathcal{H}{\delta \ddot{\phi}}+(m_a^2 a^2 +k^2)\delta \phi&=-\frac{1}{2}\dot{\overline{\phi}}\dot{h}_m,
\label{eqn:field_eoms}
\end{align}
where dots denote derivatives with respect to conformal time $\eta$ and the conformal Hubble parameter is $\mathcal{H}=\dot{a}/a=aH$. The initial homogeneous axion field value is set during inflation as discussed in Section \ref{sec:iso_theory}. 

The background equation of state is given by:
\be
w_a\equiv\frac{\bar{P}_a}{\bar{\rho}_a}=\frac{\dot{\phbar}^2-m_a^2\phbar^2}{\dot{\phbar}^2+m_a^2\phbar^2} \, .\label{eqn:wdef}
\ee
The perturbed axion density, pressure, and velocity fluctuations in synchronous gauge are \citep[e.g.][]{2004astro.ph..2060H}:
\begin{align}
\delta\rho_a =&~ a^{-2}\dot{\overline{\phi}}\delta \dot{\phi}+ m_{a}^2 \overline{\phi}\delta \phi \label{eqn:deltarho} \\
\delta P_a=&~a^{-2}\dot{\overline{\phi}} \delta \dot \phi - m_{a}^2\overline{\phi}\delta \phi \label{eqn:deltap}\\(\rho + P)v_a =&~a^{-2}k \dot{\overline{\phi}}\delta \phi\label{eq:comove_find}.
\end{align}

ULAs have two main characteristics that distinguish them from CDM in cosmology: the evolution of the background equation of state, $w_a$, and the presence of the axion Jeans scale, $k_J$, in linear perturbations. At early times, when $H\gg m_a$, the axion field is ``frozen'' by Hubble drag: $\phbar\approx\mathrm{ const.}$, and $w_a\approx-1$. At late times, when $H\ll m_a$, the field oscillates as $\phbar\propto a^{-3/2}\cos m_at$, and $w_a\propto \cos 2m_at$ such that on time average $\langle w_a\rangle_t\approx 0$. This defines a transitional scale-factor $a_\mathrm{ osc}$ and conformal time $\eta_\mathrm{ osc}$, defined by the condition $m_a\equiv 3H(a_\mathrm{ osc})$, which mark the time at which the axion begins to coherently oscillate.

This transition in the equation of state parameter $w_{a}$ fixes the axion relic density in terms of the initial field displacement, $\phbar_i$, which can be approximated as \citep{2015PhRvD..91j3512H}
\begin{equation}
\Omega_{a}\approx\left(\frac{\overline{\phi}_{i}}{M_\mathrm{ pl}}\right)^{2}\frac{m_{a}^{2}a_\mathrm{ osc}^{3}}{6H_{0}^{2}}.
\end{equation}
For all actual constraints and computations we use a numerical solution for the background evolution in \acamb. The $w_{a}$ transition  also affects the background expansion rate, and thus the CMB observables (H15). In the standard cosmological model, there is CDM, which has $w_\mathrm{ CDM}=0$ at all times, and the cosmological constant (Dark Energy, $\Lambda$), which has $w_\mathrm{ DE}=-1$ at all times. The transition in $w_{a}$ means that ULAs generally exhibit behavior in an intermediate regime between DM and DE.

\begin{figure*}
\begin{center}
\includegraphics[width=0.49\textwidth]{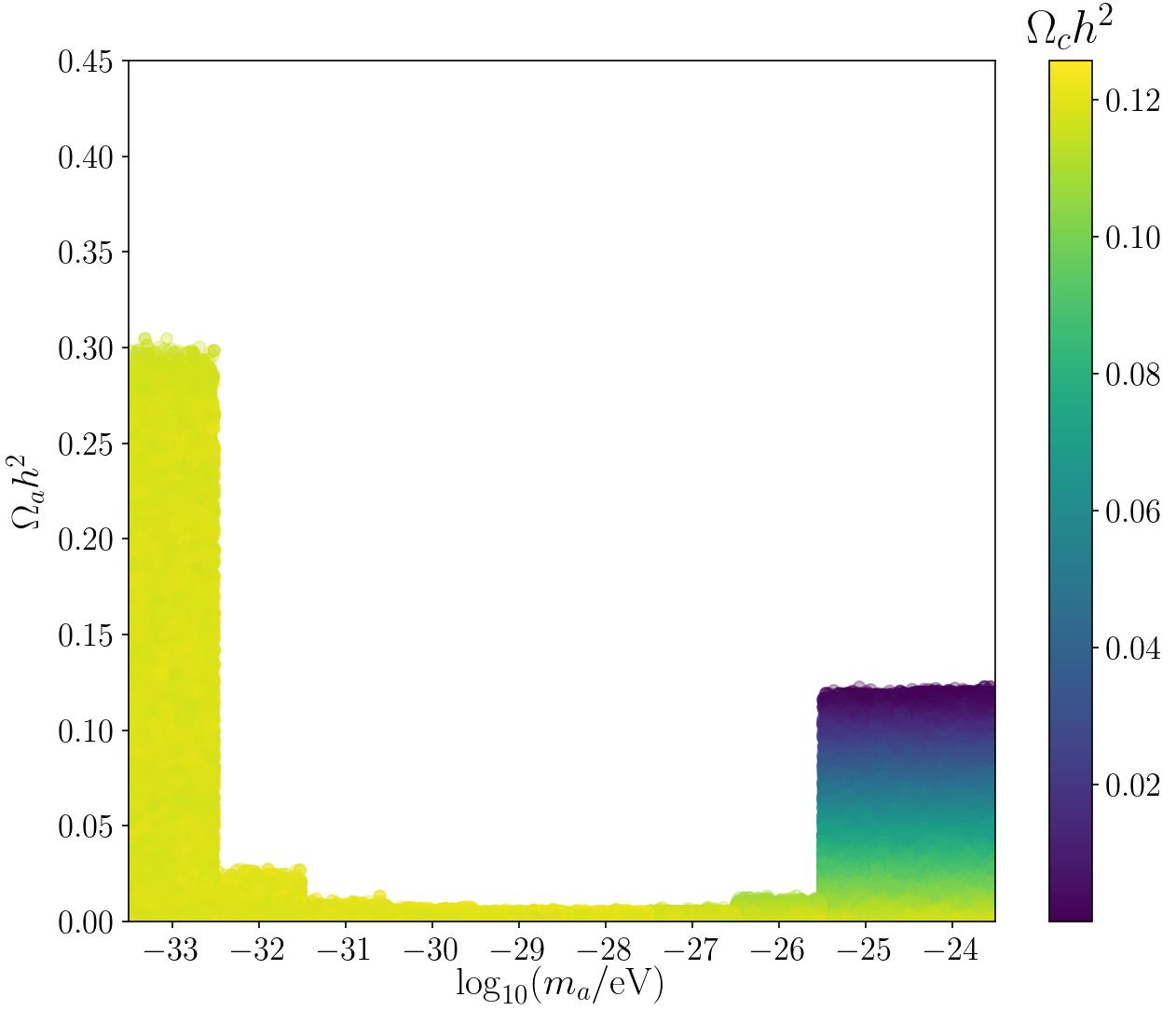}
\includegraphics[width=0.49\textwidth]{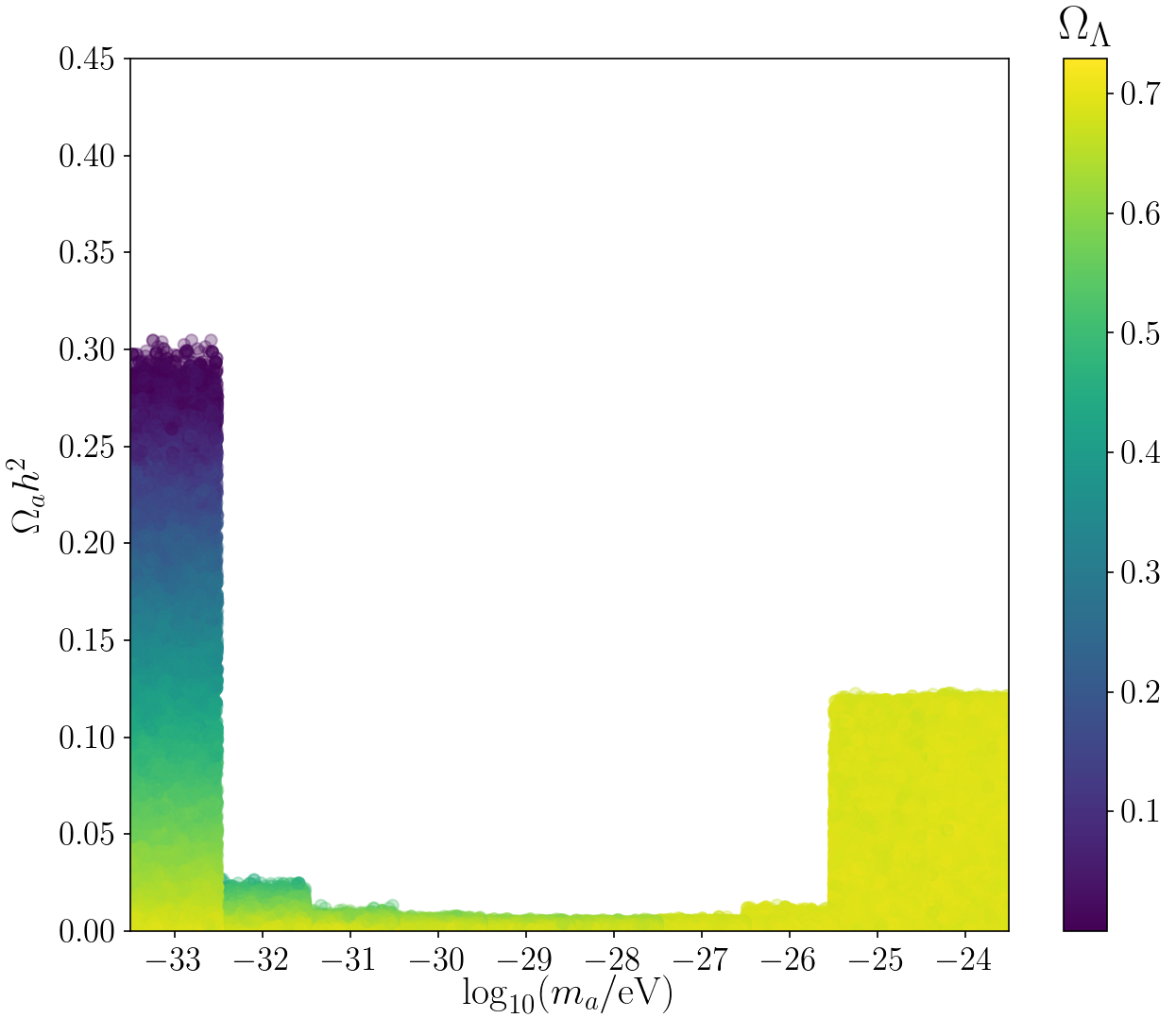} 
\caption{{\bf ULAs as DM and DE:} The ``u-plot'' is shown with MCMC points coloured by their value of $\Omega_c h^2$ (left panel) and $\Omega_\Lambda$ (right panel). This demonstrates the role of ULAs as a degenerate component of CDM for $m_a\geq 10^{-25}\text{ eV}$, and as a compoent of DE for $m_a\leq 10^{-32}\text{ eV}$. As in Fig.~\ref{fig:lens_u_plot} points are given a random spread within each mass bin.}
\label{fig:two-u}
\end{center}
\end{figure*}
\begin{figure} 
\includegraphics[width=0.49\textwidth]{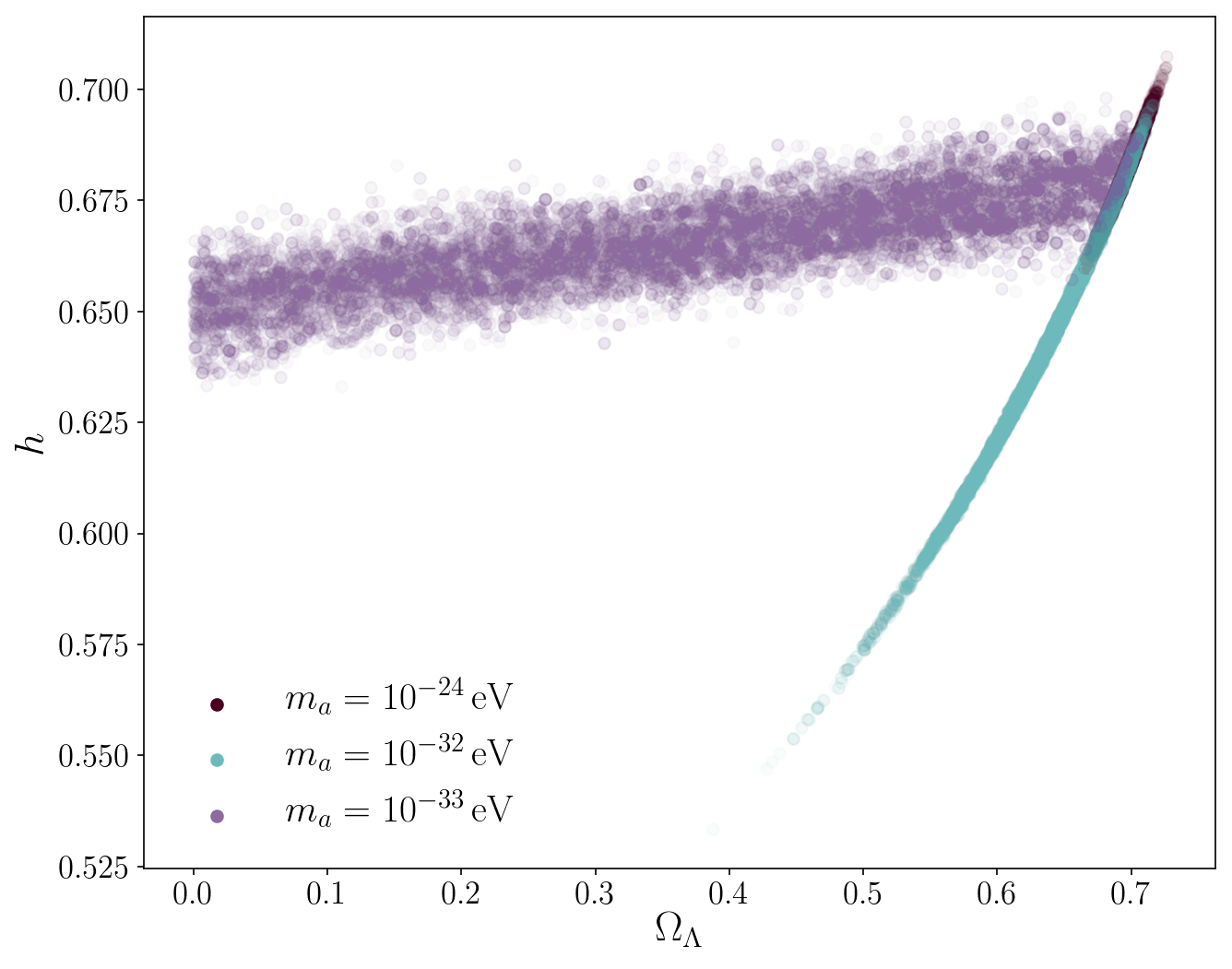} 
\caption{{\bf DE Degeneracies:} Scatter plot of $\Omega_\Lambda$ versus $h$. For $m_a=10^{-32}\text{ eV}$, $\Omega_\Lambda$ can be lowered only partially via a strong degeneracy with $h$. For $m_a=10^{-24}\text{ eV}$, there is no such degeneracy. For $m_a=10^{-33}\text{ eV}$ there is a degeneracy between $\Omega_a$ and $\Omega_\Lambda$, allowing axions to act as quintessence with only a modest reduction in $h$.}
\label{fig:omega_L-h_degeneracy}
\end{figure}

If the $w_a$-transition occurs during the radiation-dominated epoch (roughly, if $m_a\gtrsim 10^{-27}\text{ eV}$), we refer to ULAs as ``DM-like axions". If this transition happens late enough into radiation domination, the cosmic expansion is slightly altered, and perturbation growth on length scales observable with the CMB is altered by a modified diffusion (Silk) damping scale, as well as by changes to the gravitational redshift of CMB photons, known as the early integrated Sachs-Wolfe (ISW) effect. The relative heights of the CMB acoustic peaks are thus altered (H15). As the axion mass is increased beyond $m_{a}\sim 10^{-24}~\mathrm{ eV}$, this transition occurs so early that the modes observed with the CMB cannot distinguish ULAs from standard CDM. 

If the $w_a$-transition occurs during the matter or $\Lambda$-dominated eras at late times  (roughly, if $m_a\lesssim 10^{-27}\text{ eV}$), we refer to ULAs as ``DE-like axions''. If this transition happens early enough, then the expansion rate during this epoch is altered compared to a pure $\Lambda$ DE model. The change in the expansion rate affects the angular diameter distance to the CMB, changing the angular location of acoustic peaks, and also affects the time evolution of the Newtonian potential, inducing a distinctive late-time ISW effect at low multipoles (H15). After accounting for degeneracies (in particular, by adjusting $H_{0}$ to lock the distance from us to the last-scattering surface to its fiducial value), DE-like axions with lower mass deviate from the $\Lambda$ DE model at the lowest multipoles, where cosmic variance makes it harder and harder to distinguish them from $\Lambda$ as the value of $m_{a}$ is lowered. 

The spatial gradient terms in the Klein-Gordon equation lead, in the fluid approximation, to an effective sound speed for the axion density perturbations~\citep[e.g.][]{2009PhLB..680....1H,2013JCAP...01..002C}:
\be
c_s^2\equiv\frac{\delta P_a}{\delta\rho_a}=\frac{k^2}{k^2+4m_a^2 a^2} \, .
\ee
The axion sound speed leads to a suppression of structure growth for scales beyond the Jeans scale, $k>k_J$ \citep{1985MNRAS.215..575K}. For DM-like axions, this scale is imprinted in the matter power spectrum for all modes $k>k_{J,\mathrm{ eq}}$, where $k_{J,\mathrm{ eq}}$ is the Jeans scale at matter radiation equality~\citep[e.g.][]{2000PhRvL..85.1158H,2010PhRvD..82j3528M,2012PhRvD..86h3535P}. An additional effect that leads to suppressed perturbations for axions compared to CDM occurs for those modes which are sub-horizon at the $w_a$-transition, this effect being comparable to the Jeans suppression particularly for DE-like axions.

As CDM has $w_\mathrm{ CDM}=c_{s,\mathrm{ CDM}}^2=0$,  DM-like axions with sufficiently large $m_{a}$ are observationally indistinguishable from CDM. For the scales probed by CMB data, this occurs roughly when $m_a\gtrsim 10^{-24}\text{ eV}$, corresponding to $k_{J,\mathrm{ eq}}$ in the quasi-linear regime. As $w_\mathrm{ DE}=-1$, DE-like axions with very low $m_{a}$ are fully degenerate with $\Lambda$. For the scales probed by CMB data, this occurs when $m_a\lesssim 10^{-33}\text{ eV}$, with $H_0=1.5(h/0.7)\times 10^{-33}\text{ eV}$. 

In the intervening nine orders of magnitude in axion mass, the CMB provides stringent constraints on the axion energy density, leading to a ``U-shaped'' posterior distribution in the $(m_a,\Omega_a h^2)$ plane~\citep[H15,][]{2006PhLB..642..192A}. The relevant observational effects can all be computed using \textsc{axionCAMB}, and then compared with data to obtain constraints, as discussed in Sections \ref{sec:dm_constraints}-\ref{sec:iso_constraints}, Appendix \ref{appendix:sampling}, and ~\citep[H15,][]{2006PhLB..642..192A}.

The role of ULAs as DM and DE at the extreme ends of the mass range can be clearly visualised in Fig.~\ref{fig:two-u} using our MCMC constraints. We show how for $m_a\geq 10^{-25}\text{ eV}$ the value of $\Omega_c h^2$ can be lowered in correlation with $\Omega_a h^2$ being raised. On the other hand, for $m_a\leq 10^{-32}\text{ eV}$, the value of $\Omega_\Lambda$ can be lowered. 

For $m_a=10^{-32}\text{ eV}$ a lowering of $\Omega_\Lambda$ must be accompanied by a lowering in $h$ (H15). In other words, the reduction in $\Omega_\Lambda$ possible for $m_a=10^{-32}\text{ eV}$ occurs via the opening up of a partial degeneracy between $\Omega_\Lambda$ and $h$ (see Fig.~\ref{fig:omega_L-h_degeneracy}). A reduction in $h$ increases tension with low-$z$ measures of the expansion rate. ULAs can play the role of DE, allowing $\Omega_\Lambda=0$ at (almost) fixed $h$, when $m_a\leq 10^{-33}\text{ eV}$.

\section{Derivation of axion initial conditions}
\label{appendix:power_law}
To initialize fluid perturbation values, a CMB Boltzmann code like \textsc{camb} requires power-series solutions for fluid variable values in conformal time $\eta$ and $x=k\eta$ \citep{1995ApJ...455....7M,2000PhRvD..62h3508B}. These solutions are derived deep into the radiation-dominated era and super-horizon limit. The method used to obtain the expressions in \cite{1995ApJ...455....7M} and \cite{2000PhRvD..62h3508B} was not specified there.
We can obtain these solutions using a linear eigenmode analysis \citep{cambnotes,2003PhRvD..68f3505D,tristan_notes}. In the form presented here, this method is also explained in \cite{2013MNRAS.434.1619C} and \cite{2015PhRvD..91j3512H}. We now review this approach, including the evolution of the scalar field in a mixed matter-radiation background. We calculate values for all metric and fluid perturbation variables in terms of the dimensionless conformal time $\eta_\mathrm{ b}$. We begin by establishing conventions for early-time expansion history in Appendix \ref{sec:ephist}, lay out the perturbation equations of motion in Appendix \ref{sec:perteq}, obtain equation of state and sound speed approximations in Appendix \ref{appendix:early_axion}, and discuss the general solution method/power series initial conditions in Appendix \ref{appendix:close}.
\subsection{Expansion history conventions}
\label{sec:ephist}
We use an unusual scale-factor convention ($a_\mathrm{ eq}=1/4$), as well as the dimensionless conformal-time $\eta_\mathrm{ b}\equiv \mathcal{C} \eta$ with $\mathcal{C}^{2}=4\pi G \rho_\mathrm{ eq}a_\mathrm{ eq}^{4}/4$ (where $\rho_\mathrm{ eq}$ is the radiation energy-density at matter-radiation equality), used for consistency with the notation of \citep{2000PhRvD..62h3508B} and to facilitate ease of comparison with the expansions in that work. We convert to more standard conventions when needed, as for example in the \textsc{axionCAMB} code itself. Matter-radiation equality is defined by the relationship:
\begin{equation}
\rho_{a}+\rho_{b}+\rho_{c}=\rho_{\gamma}+\rho_{\nu},\end{equation} where $\rho_\gamma$ and $\rho_{\nu}$ are the energy densities of photons and neutrinos, while $\rho_{b}$ and $\rho_{c}$ are the energy densities of baryons and CDM.\\ \\
The solution to the Friedmann equation at early times ($\rho_{a}\ll \rho_\mathrm{ m},\rho_{a}\ll \rho_\mathrm{ rad}=\rho_{\gamma}+\rho_{\nu}$, $a\ll a_\mathrm{ osc}$) is
\begin{eqnarray}
a&=&\eta_\mathrm{ b}+K\eta_\mathrm{ b}^{2},\label{homosol_early_a}\\
K&=&\left\{\begin{array}{ll}
\left(1-f_\mathrm{ NR}\right)&\mbox{if $a_\mathrm{ osc}\leq a_\mathrm{ eq}$}\\
\frac{\left(1-f_\mathrm{ NR}\right)}{\left(1-f_\mathrm{ NR}\right)+f_\mathrm{ NR}a_\mathrm{ eq}^{3}/a_\mathrm{ osc}^{3}}&\mbox{if $a_\mathrm{ osc}>a_\mathrm{ eq}$}.
\end{array}\right.\label{eq:kval}\\
f_\mathrm{ NR}&=&\Omega_{a}/(\Omega_{a}+\Omega_{b}+\Omega_{c}).\label{homosol_early_b}\end{eqnarray} 

The one distinction between the early-time expansion history used here and in \cite{2000PhRvD..62h3508B} is that we have allowed for axions to make up a significant fraction $f_\mathrm{ NR}$ of the matter density today, leading to a correction term $\propto K$. In the limit that ULAs are negligible to the total early-time energy budget ($f_\mathrm{ NR}\ll 1$), we see that $K\simeq 1$ and recover the solution $a(\eta_\mathrm{ b})$ of \cite{2000PhRvD..62h3508B}. Note that in these conventions $f_\mathrm{ NR}$ is defined as the ratio of axion to non-relativistic matter energy densities today. The scale factor when axions begin to coherently oscillate, $a_\mathrm{ osc}$, is obtained by numerically solving the homogeneous Klein-Gordon equation to identify the moment when $m=3H$. 

\subsection{Fluid and Einstein equations for eigenmode analysis}
\label{sec:perteq}
We use a dimensionless wave number $\kappa=k/C$ so that $x=k\eta=\kappa \eta_\mathrm{ b}$, dimensionless velocities $\tilde{t}_{i}\equiv\theta_{i}/\left(\mathcal{C}\kappa x^{2}\right)$, and rescaled density contrasts $\tilde{\delta}_{i}\equiv\delta_{i}/x$. The axion velocity $u_{a}$ is already dimensionless, but we define $\tilde{u}_{a}=u_{a}/x^{2}$ to simplify the derivation. We also define a metric velocity $\Theta\equiv h_{m}^{\prime}$, where the derivative $^{\prime}\equiv\kappa^{-1} d/d\eta_\mathrm{ b}$. We rescale high-$l$ moments in the neutrino hierarchy using $\tilde{\sigma}_{\nu}\equiv\sigma_{\nu}/x$ and $\tilde{F}_{\nu}^{\left(3\right)}\equiv F_{\nu}^{\left(3\right)}/x^{2}$. We now reexpress the synchronous gauge fluid+Einstein equation system from \cite{1995ApJ...455....7M} and \cite{2000PhRvD..62h3508B}, using the choice of variables just described, the axion EOMs/source terms of Eqs.~(\ref{eqn:eoma})$-$(\ref{eqn:eomb}), and Eqs.~(\ref{dpdef})$-$(\ref{dqdef}) to find: 
\begin{eqnarray}
\tilde{\delta}_{\gamma}^{\prime}&=&-\tilde{\delta}_{\gamma}-\frac{4}{3}\tilde{t}_{\gamma\rm b}x^{2}-\frac{2\Theta}{3},\label{eq:fesprimo}\\
\tilde{\delta}_{\nu}^{\prime}&=&-\tilde{\delta}_{\nu}-\frac{4}{3}\tilde{t}_{\nu}x^{2}-\frac{2\Theta}{3},\\
\tilde{\delta}_{c}^{\prime}&=&-\tilde{\delta}_{c}-\tilde{t}_{c}x^{2}-\frac{\Theta}{2},\\
\tilde{\delta}_\mathrm{ b}^{\prime}&=&-\tilde{\delta}_\mathrm{ b}-\tilde{t}_{\gamma\rm b}x^{2}-\frac{\Theta}{2},\\
\tilde{t}_{\gamma\rm b}^{\prime}&=&-2\tilde{t}_{\gamma \rm b}+\frac{\tilde{\delta}_{\gamma}}{4\left[\frac{3R_{b}\left(x/\kappa \right)\left(Kx/\kappa +1\right)}{R_{\gamma}}+1\right]} \nonumber \\
&-&\frac{\frac{\left(2Kx/\kappa +1\right)}{\left(x/\kappa +1\right)}\tilde{t}_{\gamma}\frac{3R_\mathrm{ b}}{R_{\gamma}}\left(x/\kappa \right)\left(xK/\kappa +1\right)}{\left[\frac{3R_\mathrm{ b}}{R_\gamma}\left(x/\kappa \right)\left(xK/\kappa +1\right)+1\right]},\\
\tilde{t}_{\nu}^{\prime}&=&-2\tilde{t}_{\nu}+\frac{\tilde{\delta}_{\nu}}{4}-\tilde{\sigma}_{\nu},\\
\tilde{t}_{c}^{\prime}&=&-2\tilde{t}_{c}-\frac{\left(2xK/\kappa +1\right)}{\left(xK/\kappa +1\right)}\tilde{t}_{c},\\
\tilde{\sigma}_{\nu}^{\prime}&=&-\tilde{\sigma}_{\nu}+\frac{4 \tilde{t}_{\nu}x^{2}}{15}-\frac{3\tilde{F}_{\nu}^{\left(3\right)}x^{2}}{10}+\frac{2\Theta}{15}+\frac{8\left(R_{\gamma}\tilde{t}_{\gamma}+R_{\nu}\tilde{t}_{\nu}\right)}{5\left(1+xK/\kappa \right)^{2}}\nonumber \\
&+&\frac{24x\left(R_{c}\tilde{t}_{c} +R_\mathrm{ b}\tilde{t}_\mathrm{ b}\right)}{5\kappa \left(1+xK/\kappa \right)}+\frac{16\pi Gx^{4}}{5C^{2}\kappa^{4}}\left(1+Kx/\kappa\right)^{2}\rho_{a}\tilde{u}_{a},\\
\tilde{F}_{\nu}^{\left(3\right)}&=&-2\tilde{F}_{\nu}^{\left(3\right)}+\frac{6\tilde{\sigma}_{\nu}}{7},\\
\Theta^{\prime}&=&-\frac{\left(2xK/\kappa +1\right)}{\left(xK/\kappa +1\right)}\Theta-\frac{6\left(R_{\gamma}\tilde{\delta}_{\gamma}+R_{\nu}\tilde{\delta}_{\nu}\right)}{\left(1+xK/\kappa \right)^{2}}\nonumber \\
&-&\frac{12x\left(R_{c}\tilde{\delta}_{c}+R_\mathrm{ b}\tilde{\delta}_\mathrm{ b}\right)}{\kappa \left(1+xK/\kappa \right)}-\frac{32\pi G a^{2}\rho_{a} x^{2} \tilde{\delta}_{a}}{C^{2}\kappa^{2}}\nonumber \\
&-&\frac{72\pi G a^{2}\rho_{a}x^{2}}{C^{2}\kappa^{2}}\tilde{u}_{a}\left(1-c_\mathrm{ ad}^{2}\right)\left(\frac{1+2Kx/\kappa}{1+Kx/\kappa}\right),\label{fes1}\\
\eta_{m}^{\prime}&=&\frac{2x}{\left(1+xK/\kappa \right)^{2}}\left(R_\gamma\tilde{t}_\gamma+R_\nu\tilde{t}_\nu\right)+\frac{6x^{2}}{\kappa \left(1+xK/\kappa \right)}\left(R_\mathrm{ b}\tilde{t}_\mathrm{ b}+R_\mathrm{ c}\tilde{t}_{c}\right)\nonumber \\
&+&\frac{4\pi G x^{5}}{C^{2}\kappa^{4}}\left(1+Kx/\kappa\right)^{2}\rho_{a}\tilde{u}_{a},\label{fes2}\\
\tilde{\delta}_{a}^{\prime}&=&-\tilde{\delta}_{a}-\left(1+w_{a}\right)\frac{\Theta}{2}-\frac{3\left(1+2Kx/\kappa\right)}{\left(1+Kx/\kappa\right)}\left(1-w_{a}\right)\tilde{\delta}_{a}\nonumber \\
&-&9\left(1-c_{a}^{2}\right)\tilde{u}_{a}\frac{\left(1+2Kx/\kappa\right)^{2}}{\left(1+Kx/\kappa\right)^{2}},\\
\tilde{u}^{\prime}_{a}&=&\frac{2\left(1+2Kx/\kappa\right)}{\left(1+Kx/\kappa\right)}\tilde{u}_{a}+\tilde{\delta}_{a}-2\tilde{u}_{a}+\frac{w_{a}^{\prime}\tilde{u}_{a}x}{1+w_{a}},\label{eq:fes18}\\
\mathcal{A}&=&\frac{\rho_\mathrm{ a}}{a_{0}^{4}\Omega_{r}\rho_\mathrm{ crit}}.
\end{eqnarray} 
This is a linear coupled system of ODEs, which can be solved using a power-series solution, as discussed in Section \ref{appendix:close}. In the above expressions $^{\prime}=d/d\ln{x}$ and $a_{0}$ is the scale factor today under this convention:
\begin{equation}
a_{0}=\left\{
\begin{array}{ll}
\frac{\Omega_{m}}{4\Omega_{r}}\left \{\left(1-f_\mathrm{ NR}\right)+f_\mathrm{ NR}\left(\frac{a_\mathrm{ eq}}{a_\mathrm{ osc}}\right)^{3}  \right\}&\mbox{if $a_\mathrm{ osc}>a_\mathrm{ eq}$,}\\ 
\frac{\Omega_{m}}{4\Omega_{r}}&\mbox{if $a_\mathrm{ osc}\leq a_\mathrm{ eq}$}.\end{array}\right.
\end{equation}

The neutrino energy-density fraction is
\begin{equation}
R_{\nu}=\Omega_{\nu}/(\Omega_{\nu}+\Omega_{b}).
\end{equation} 
Conversely, the photon energy-density fraction is $R_{\gamma}=1-R_{\nu}$. 

Equation (\ref{fes1}) is obtained from a linear combination of the Einstein equations \citep{1995ApJ...455....7M,2000PhRvD..62h3508B}
 \begin{align}
 k^{2}\eta_{m}-\frac{1}{2}\frac{\dot{a}}{a}\dot{h}_{m}=-4\pi G a^{2}\delta \rho,\label{einstein_constraint}\\
 \ddot{h}_{m}+2\frac{\dot{a}}{a}\dot{h}_{m}-2k^{2}\eta_{m}=-24\pi G a^{2} \delta P,\label{einstein_dynamic}
 \end{align} 
where $\delta \rho$ and $\delta P$ are the total energy density and pressure perturbations, respectively.  

From the Einstein equation \citep{1995ApJ...455....7M,2000PhRvD..62h3508B}
\begin{equation}
k^{2}\dot{\eta}_{m}=4\pi G \sum_{i}\left(\overline{\rho}+\overline{P}\right)_{i}\theta_{i}+4\pi G u_{a},\end{equation}Eq.~(\ref{fes2}) is obtained,
where the sum on $i$ is over all the conventional fluid species. The axion energy-density $\rho_\mathrm{ a}$ has some time dependence, which we compute below, where we also obtain the time evolution of the axion EOS $w_{a}$ and adiabatic sound speed $c_\mathrm{ ad}$.

\subsection{Axion evolution in a matter+radiation universe}
\label{appendix:early_axion}
To obtain power-series solutions, we need the squared adiabatic sound speed $c_\mathrm{ ad}^{2}$ and scale factor $w_{a}$ as a function of conformal time. To obtain these functions, we
use Eq.~(\ref{eqn:adiabat_cs}) \& (\ref{eqn:wdef}), and evaluate $\rho_{a}$ and $P_{a}$ using the field evolution given by Eq.~(\ref{eqn:homo_eom}).  We work at early times, allowing us to make the approximation that $\rho_{a}\ll \rho_{m},\rho_{a}\ll \rho_{r}, a\ll a_\mathrm{ osc}$. We use Eqs.~(\ref{homosol_early_a})$-$(\ref{homosol_early_b}) to evaluate the conformal Hubble parameter $\mathcal{H}$. Making a power series expansion in the dimensionless conformal time $\eta_\mathrm{ b}$, we obtain the desired results from the solution for the homogeneous field $\overline{\phi}(\eta_\mathrm{ b})$:

 \begin{eqnarray}
 w_{a}&=&-1+\frac{2m^{2}\eta_\mathrm{ b}^{4}} {25\mathcal{C}^{2}}+\frac{4Km^{2}\eta_\mathrm{ b}^{5}}{75\mathcal{C}^{2}}+...,\label{weq_early}\\
 c_\mathrm{ ad}^{2}&=&-\frac{7}{3}+\frac{10K\eta_\mathrm{ b}}{9}-\frac{520K^{2}\eta_\mathrm{ b}^{2}}{189}+\frac{3445K^{3}\eta^{3}_\mathrm{ b} }{567}\nonumber \\
 &+&\left(\frac{-151465K^{4}}{11907}+\frac{2m^{2}}{27\mathcal{C}^{2}}\right)\eta^{4}_\mathrm{ b}\nonumber \\
 &+&\left(\frac{870025K^{5}}{35721}+\frac{26Km^{2}}{405C^{2}}\right)\eta_\mathrm{ b}^{5}+....,\\
 \rho_{a}&=&\rho_{a}^{\left(0\right)}\left[1-\frac{3m^{2}\eta_\mathrm{ b}^{4}}{50\mathcal{C}^{2}} -\frac{2Km^{2}\eta_\mathrm{ b}^{5}}{25\mathcal{C}^{2}}+...\right],
 \end{eqnarray}

  where $\rho_{a}$ is the asymptotic value of $\rho_{a}$ when $a\ll a_{\mathrm osc}$. Converting to dimensional conformal time via $\eta_\mathrm{ b}= \Omega_{m}H_{0}\eta/(4\sqrt{\Omega_{r}})$, we see that Eq.~(\ref{weq_early}) agrees with the the quintessence equation of state derived in \cite{cambnotes}.

 \subsection{Series solutions around linearized eigenmodes}
\label{appendix:close}
We now seek solutions to Eqs.~(\ref{eq:fesprimo})--(\ref{eq:fes18}) as power series in $x$, with source terms for the Einstein equations given by Eqs.~(\ref{dpdef})-(\ref{dqdef}). We follow the method articulated in \cite{2003PhRvD..68f3505D}, \cite{2013MNRAS.434.1619C}, and \cite{2015PhRvD..91j3512H}.

If the full system of equations can be written in the form
\begin{eqnarray}
\frac{d\vec{U}_{\vec{k}}}{d\ln x}=\left(\underline{A}_{0}+\underline{A}_{1}x+...\underline{A}_{n}x^{n}\right)\vec{U}_{k}\label{sa1}
\end{eqnarray}
where $\vec{U}_{k}$ is the Fourier transform of the vector of all fluid+metric variables of interest, then the space of linearized solutions is spanned by the eigenvectors $\vec{U}_{k}^{\alpha}$ (with eigenvalue $\alpha$) of $\underline{A}_{0}$:\begin{eqnarray}
\vec{U}_{k}(\tau)=\sum_{\alpha} c_{\alpha} x^{\alpha}\vec{U}_{k}^{\alpha}.\label{sa2}
\end{eqnarray} The coefficients $c_{\alpha}$ set the contribution of each eigenmode, and are chosen to match initial conditions. The lowest-order solutions can yield zero values for some variables, and so we expand further. We can expand each solution $\mathcal{U}_{\vec{k}}^{\alpha}(\tau)$ as an \textit{ansatz} including higher order corrections: \begin{eqnarray}
\mathcal{U}_{\vec{k}}^{\alpha}(\tau)=U_{\vec{k}}^{\alpha}x^{\alpha}+U_{\vec{k},\left(1\right)}^{\alpha}x^{\alpha+1}+...U^{\alpha}_{\vec{k},\left(i\right)}x^{\alpha+i}+....\label{sa3}.
\end{eqnarray}
We derive the corrections to the lowest-order solution by applying Eq.~(\ref{sa1}) to the \textit{ansatz}, Eq~(\ref{sa3}), obtaining a set of linear algebraic equations:
\begin{eqnarray}
\left[\left(\alpha+1\right)\mathcal{I}-\underline{A}_{0}\right]\vec{U}^{\alpha}_{\vec{k},\left(1\right)}&=&\underline{A}_{1}\vec{U}^{\alpha}_{\vec{k}}\label{u1eqa},\\
\left[\left(\alpha+2\right)\mathcal{I}-\underline{A}_{0}\right]\vec{U}^{\alpha}_{\vec{k},\left(2\right)}&=&\underline{A}_{1}\vec{U}^{\alpha}_{\kappa,\left(1\right)}+\underline{A}_{2}\vec{U}^{\alpha}_{\vec{k}},\label{u2eqa}\\
\left[\left(\alpha+3\right)\mathcal{I}-\underline{A}_{0}\right]\vec{U}^{\alpha}_{\vec{k},\left(3\right)}&=&\underline{A}_{1}\vec{U}^{\alpha}_{\vec{k},\left(2\right)}+\underline{A}_{2}\vec{U}^{\alpha}_{\vec{k},\left(1\right)}\nonumber \\&+&\underline{A}_{3}\vec{U}^{\alpha}_{\vec{k},\left(1\right)},\label{u3eqa}\\
\left[\left(\alpha+4\right)\mathcal{I}-\underline{A}_{0}\right]\vec{U}_{\vec{k},\left(4\right)}^{\alpha}&=&\underline{A}_{1}\vec{U}^{\alpha}_{\vec{k},\left(3\right)}+\underline{A}_{2}\vec{U}^{\alpha}_{\vec{k},\left(2\right)}+\underline{A}_{3}\vec{U}^{\alpha}_{\vec{k},\left(1\right)}\nonumber \\&+&\underline{A}_{4}\vec{U}^{\alpha}_{\vec{k}}.\label{u4eqa}
\end{eqnarray} 
Here $\mathcal{I}$ is the identity matrix. The solutions to this linear system can yield  corrections to the time-evolution for each eigenmode. 
  \subsection{Eigenmodes}
 To obtain the eigenmodes, we expand in $\eta_\mathrm{ b}$ and $x$, appropriate for super-horizon modes in the radiation-dominated limit. Using the assignment $\vec{U}_{k}=\left\{\tilde{\delta}_{\gamma},\tilde{\delta}_{\nu},\tilde{\delta}_{c},\tilde{\delta}_{b},\tilde{t}_{\gamma b},\tilde{t}_{\nu},\tilde{t}_{c},\tilde{\sigma}_{\nu},\tilde{F}_{\nu}^{3},\Theta,\eta,\tilde{\delta}_{a},\tilde{u}_{a}\right\}$, we obtain the matrices $\underline{A}_{0},\underline{A}_{1},\underline{A}_{2},\underline{A}_{3},\underline{A}_{4}$. To check our formalism, we restrict our equations to the no-axion case, and then recover exactly (up to $5^\mathrm{ th}$ order) the growing adiabatic, baryon isocurvature, CDM (cold DM) isocurvature, neutrino density isocurvature, and neutrino velocity isocurvature modes (as well as decaying modes) stated in \cite{2000PhRvD..62h3508B}; we apply Eqs.~(\ref{u1eqa})-(\ref{u4eqa}) to obtain this result.
The adiabatic mode has eigenvalue $\alpha=1$ and corresponds to the initial condition
\begin{eqnarray}
\delta_{\gamma}=\delta_{\nu}=\frac{4}{3}\delta_\mathrm{ c}=\frac{4}{3}\delta_\mathrm{ b},\\
\delta_{i}=(1+w_{i})\delta_{c}.
\end{eqnarray} where $\delta_{\gamma}$, $\delta_{\nu}$, $\delta_\mathrm{ c}$, and $\delta_\mathrm{ b}$ are the fractional energy over-densities in photons, neutrinos, CDM, and baryons respectively. Since at early times, $w_\mathrm{ a}=-1$, the adiabatic condition for axions implies $\delta_\mathrm{ a}=0$ initially. In synchronous gauge, the corresponding power series solution (valid at early times) is \citep{1995ApJ...455....7M,2000PhRvD..62h3508B}
\begin{eqnarray}
\delta_{\gamma}&=&\delta_{\nu}=-\frac{\left(\kappa\eta_\mathrm{ b}\right)^{2}}{3},\\
\delta_\mathrm{ c}&=&\delta_\mathrm{ b}=-\frac{\left(\kappa\eta_\mathrm{ b}\right)^{2}}{4},\\
\frac{\theta_{\gamma}}{\mathcal{C}\kappa}&=&\frac{\theta_\mathrm{ b}}{\mathcal{C}\kappa}=-\frac{\left(\kappa\eta_\mathrm{ b}\right)^{3}}{36},\\
\frac{\theta_{\nu}}{\mathcal{C}\kappa}&=&-\frac{\left(23+4R_{\nu}\right)\left(\kappa\eta_\mathrm{ b}\right)^{3}}{36\left(15+4R_{\nu}\right)},\\
\theta_\mathrm{ c}&=&\delta_{a}=u_{a}=0,\\
\sigma_{\nu}&=&\frac{2\left(\kappa\eta_\mathrm{ b}\right)^{2}}{3\left(15+4R_{\nu}\right)},\\
F_{\nu}^{\left(3\right)}&=&\frac{4\left(\kappa\eta_\mathrm{ b}\right)^{3}}{21\left(15+4R_{\nu}\right)},\\
h_{m}&=&\frac{\left(\kappa \eta_\mathrm{ b}\right)^{2}}{2},\\
\eta_{m}&=&1-\frac{\left(5+4R_{\nu}\right)\left(\kappa \eta_\mathrm{ b}\right)^{2}}{12\left(15+4R_{\nu}\right)},
\end{eqnarray}where the metric perturbations $h_{m}$ and $\eta_{m}$ are defined as described in \cite{1995ApJ...455....7M} and \cite{2000PhRvD..62h3508B}, as are the fluid perturbations.  These initial conditions were used in \cite{2015PhRvD..91j3512H} to obtain modified adiabatic initial conditions in the presence of ULAs, and then applied to generate power spectra and parameter-space constraints to ULAs. This solution is valid up to corrections of order $(k\eta)^{4}$ for metric and standard fluid perturbations, and $\eta/\eta_\mathrm{ eq}$ for the axion variables themselves. The normalization of perturbations is arbitrary here, but is eventually set by the primordial power spectrum inside \textsc{axionCAMB}. This method recovers the standard CDM, baryon, neutrino density, and neutrino velocity isocurvature mode initial conditions, as described in \cite{2000PhRvD..62h3508B}.

We use this method to obtain the initial conditions of Section \ref{isinit:theory} for the ULA isocurvature mode, Eqs.~(\ref{aia})-(\ref{aig}). These initial conditions are then used in \textsc{axionCAMB} to generate theoretical power spectra for comparison with data, necessary for the constraints obtained in this work. For the ULA isocurvature mode,
$\alpha=-1$ \& $\delta_{a}=1$; as for the adiabatic mode, this normalization is eventually rescaled for consistency with the desired primordial entropy power spectrum inside \textsc{axionCAMB}. We ignore a ULA velocity isocurvature mode because it quickly damps away (as $\eta_\mathrm{ b}^{-4}$).

\section{Sampling}
\label{appendix:sampling}

We use the \emcee~\citep{emcee} sampler within \cosmosis. \emcee~uses the \cite{goodman-weare} affine-invariant MCMC ensemble sampling algorithm. 

Sampling with \emcee~consists of setting up an ensemble of ``walkers'', which are correlated MCMC chains. The positions of all the walkers in the parameter space at any time play the role of the covariance matrix. Thus, starting the walkers in sensible places with respect to the degeneracies of the parameter space is crucial to achieving timely convergence. Careful considerations of starting position and priors as a function of axion mass are both necessary in the difficult ULA parameter space. In particular, we treat axions in the ``belly of the U'' (i.e. $10^{-29}\text{ eV}\leq m_a \leq10^{-26}\text{ eV}$) differently from those outside it. The priors on the relevant parameters are given in Table~\ref{table:priors}.

\subsection{Starting the Walkers}

We use 50 walkers in ten different mass bins. The position of each walker is labelled by a 8(9)-dimensional parameter vector for adiabatic (isocurvature) initial conditions. 

When starting the chain initially, we begin the walkers with random positions drawn in the following manner. In the belly, we take the walker co-ordinates for all the standard cosmological parameters in Eq.~\eqref{eqn:param_vec_standard} drawn from uncorrelated Gaussian distributions using the \cite{2016A&A...594A..13P} best-fit values. The axion density co-ordinate, $\Omega_a h^2$, is drawn from a uniform distribution in logarithmic space:
\be
\log_{10}\left[\Omega_a h^2 \right]\in \mathcal{U}[-4,-2] \, .
\ee
This strategy sufficiently restricts the parameter space to near maximum likelihood to facilitate rapid convergence.
\begin{figure*}
\begin{center}
\includegraphics[width=0.49\textwidth]{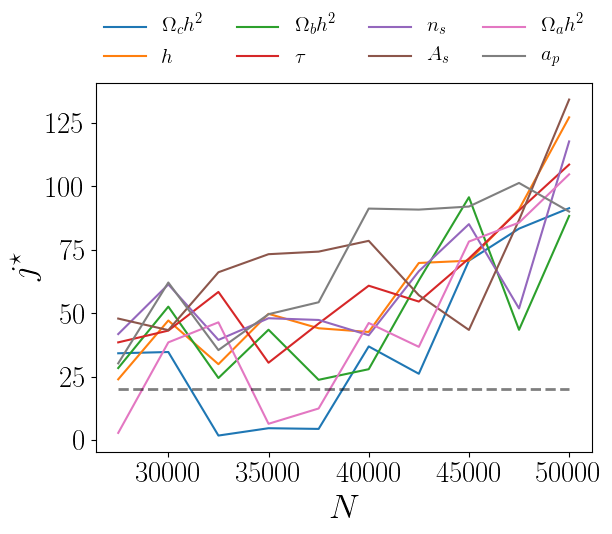}
\includegraphics[width=0.49\textwidth]{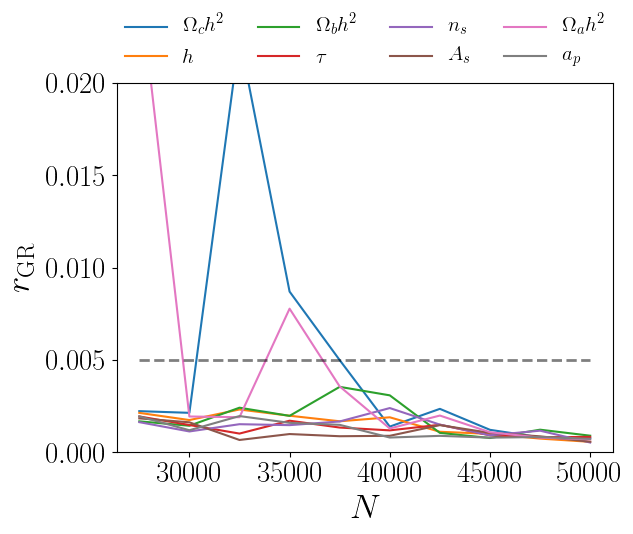} 
\includegraphics[width=0.49\textwidth]{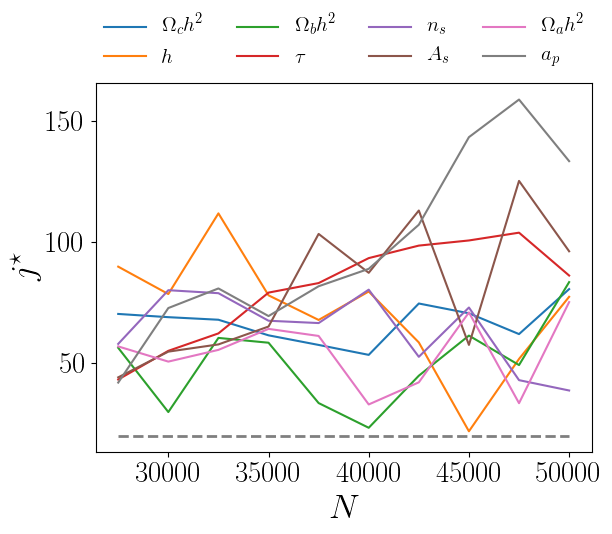}
\includegraphics[width=0.49\textwidth]{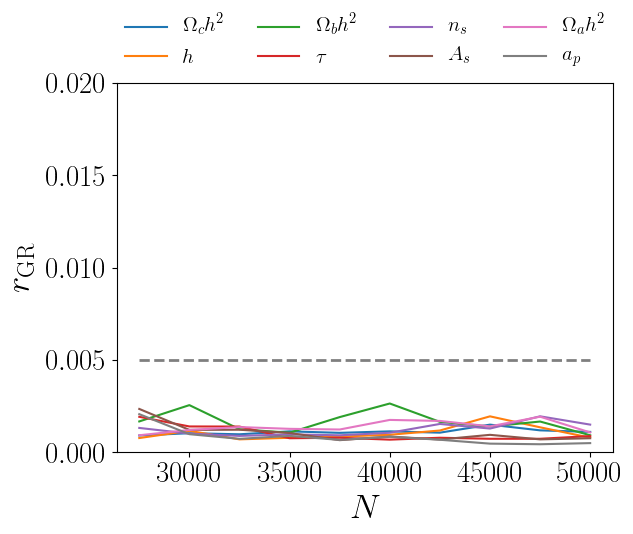} 
\caption{{\bf Convergence Tests:} Tests are shown as a function of chain length, $N$, with fixed burn-in of $N/3$. The top row shows the test for an axion mass of $m_a=10^{-24}\text{ eV}$, while the bottom row shows $m_a=10^{-26}\text{ eV}$. Convergence is more rapid in the lighter case, where there is no strong degeneracy between $\Omega_a h^2$ and $\Omega_c h^2$.}
\label{fig:convergence_2432}
\end{center}
\end{figure*}
For the axions that are most degenerate with CDM, $m_a\geq 10^{-25}\text{ eV}$, we take the first five parameters of $\Xi_\mathrm{ standard}$ in Eq.~\eqref{eqn:param_vec_standard} to be drawn as above, but we take:
\begin{eqnarray}
\Omega_c h^2&\in &\mathcal{U}[0,0.15]\nonumber \\
 \Omega_a h^2&\in & \mathcal{U}[0,0.15]\, .
\end{eqnarray}
This strategy allows more massive axions the opportunity to explore the full degeneracy between the axion and CDM densities.

For the case of the `belly-like' axions, we place a more restrictive (while still conservative) upper prior on the axion energy density. In H15, the 95\% C.L upper bound was given as $\Omega_ah2 < 0.006.$ We relax this to $\Omega_ah^2 < 0.05$ to allow for any shift in the upper bound given new data. However, the resultant bounds on the axion density reduce by a factor of two ($\Omega_ah^2< 0.003$) to the \textit{Planck} 2013 bounds presented in H15. For these intermediate-mass axions, one need not apply a different bound on the CDM or dark energy densities, as the expected limits on these parameters are comparable to the vanilla constraints.

The dark energy like axions require relaxing the bounds on the density to exceed that of the CDM constraints, since the in this case the axions are degenerate with the dark energy density. As such, we set $\Omega_ah^2 
< 0.5,$ while still allowing $\Omega_ch^2 < 0.15$.

\subsection{Convergence}
\begin{figure*}
\begin{center}
\includegraphics[width=0.49\textwidth]{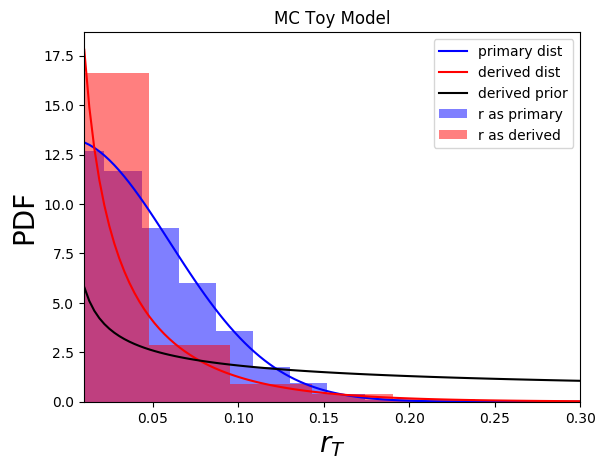}
\includegraphics[width=0.49\textwidth]{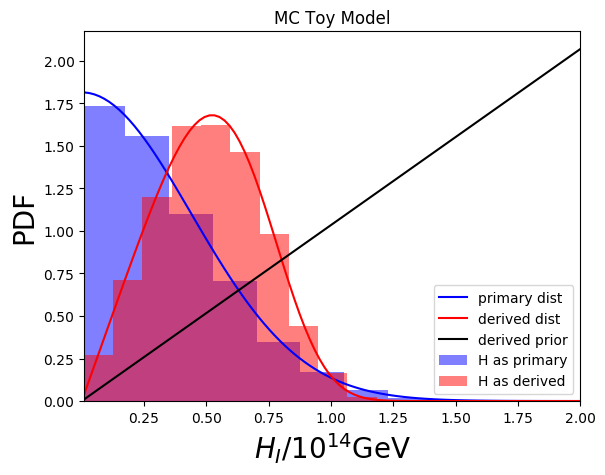} 
\caption{{\bf Change of Variables and Prior Dependence:} Histograms show the Monte Carlo samples, and solid lines show the analytic models used. \textit{Left Panel:} Tensor-to-scalar ratio. \textit{Right Panel:} Hubble scale during inflation. \label{fig:mc_toy_model}}
\end{center}
\end{figure*}
We assess convergence of our chains using the spectral method of \cite{2005MNRAS.356..925D}. The spectral method judges a parameter to be converged based on a fit to the power spectrum of a chain. The power spectrum is fit as:
\be
P(k)=P_0\frac{(k^\star/k)^\alpha}{1+(k^\star/k)^\alpha}\, .
\ee
The parameter $P_0$ gives an estimate of the Gellman-Rubin statistic, $r_\mathrm{ GR}=P_0/N$ for a chain of $N$ samples. This estimate is reliable as long as the ``knee'' in the power spectrum occurs at a sufficiently large value of $j^\star=k^\star(N/2\pi)$. We estimate that a chain is converged if both $j^\star>20$ and $r_\mathrm{GR}<0.005$ (the standard value is $r_\mathrm{GR}<0.01$, but we use a more conservative value here for illustration). The \cite{2005MNRAS.356..925D} test is implemented in \cosmosis~for Metropolis-Hastings chains. We implement an independent version which works on ``flattened'', i.e. effectively independently sampled, \emcee~chains (see the \emcee~documentation).

We illustrate the convergence tests in Fig.~\ref{fig:convergence_2432}. We plot $r_{\mathrm GR}$ and $j^\star$ for each parameter in an adiabatic run as a function of chain length $N$ with fixed amout of burn-in $N/3$. Convergence is typically very rapid, with all parameters converged within a few tens of thousands of steps. Convergence is more rapid for lower mass axions than high mass axions, due to the wide $(\Omega_a h^2,\Omega_c h^2)$ degeneracy that must be explored at higher mass. This can be seen by the pink and blue coloured lines in the top right panel of Figure~\ref{fig:convergence_2432} showing the convergence for $\Omega_ch^2$ and $\Omega_ah^2,$ which have $r_\mathrm{GR}> 0.01$ for $N<40 000$ steps. For the lighter axion convergence is already achieved for $N<30 000$ steps.
\subsection{Prior Dependence in $H_I$ Constraints}
\label{appendix:hi_prior}

We have investigated the dependence of our constraints on $H_I$ to changes in prior and choice of primary parameters. We considered the following possibilities:
\begin{eqnarray}
\log_{10}(H_I/\mathrm{ GeV}) &\in& \mathcal{U}[11,14.5]\nonumber\, , \\
H_I/(10^{14}\mathrm{GeV}) &\in& \mathcal{U}[10^{-4} , 4]\nonumber\, , \\
r &\in& \mathcal{U}[0,0.5].
\end{eqnarray}
While the log-flat prior on $H_I$ appears sensible from the point of view that we should maintain ignorance of the inflationary energy scale, it has the undesirable effect of introducing dependence of the constraints on the range of the prior.\footnote{This is discussed in detail for the case of axion quintessence in \cite{2017JCAP...01..023S}.} 

The log-flat versus uniform priors on $H_I$ place different prior weight on different types of model, with the log-flat prior placing more prior weight on low-scale inflation. For the log-flat prior there is no well defined lower limit and infinite prior volume extending to $H_I=0$. Thus the 95\% upper limit on $H_I$ for the log-flat prior depends strongly on the lower limit of the prior. Since $r$ is a derived parameter in the case with $H_I$ a primary parameter, this also leads to prior dependence in the upper limit on $r$, which can be in excess of an order of magnitude. 

We thus reject the log-flat prior on $H_I$ and turn to the question of whether $H_I$ or $r$ should be used as our primary parameter. In this case the posterior in $H_I$ or $r$ considered as a derived parameter is affected by the usual change of variables:
\be
f(x)\mathrm{ d}x=g(y)\mathrm{ d}y\, .
\ee
We illustrate the effect of the resulting Jacobian for a Monte Carlo toy model with one parameter in Fig.~\ref{fig:mc_toy_model} (we also investigated this effect in full MCMC constraints with CMB data and drew the same conclusions). In the left panel we show samples assuming $r$ is drawn from a one-sided Gaussian with zero mean and $\sigma_r=0.06$. We also show the derived constraint on $r$ assuming $H_I$ is drawn from a one-sided Gaussian with zero mean and $\sigma_H=0.44\times 10^{14}\text{ GeV}$, using the relation Eq.~\eqref{eqn:r_H_relation} with $A_s=2.3\times 10^{-9}$. The right panel shows the opposite case for $H_I$. The error on $H_I$ in the toy model is chosen to give the same 2$\sigma$ limit substituting $r=0.12$ into Eq.~\eqref{eqn:r_H_relation}. 

The change of variables affects each parameter in a very different way. The Jacobian for the $H_I\rightarrow r$ transformation is divergent near the origin, and falls off at large values. This has the effect of causing the derived $r$ distribution to be more peaked near zero. The 95\% C.L. limit is $r^\mathrm{( d)}<0.115$ compared to $r^\mathrm{ primary}<0.118$. On the other hand the Jacobian for the $r\rightarrow H_I$ transformation goes to zero at the origin, and rises at large values. This has the effect that the distribution on $H_I$ as a derived parameter demonstrates a peak away from the origin and a looser constraint. The 95\% C.L. limit is $H_I^\mathrm{ derived}<0.874\times 10^{14}\text{ GeV}$ compared to $H_I^\mathrm{ primary}<0.860\times 10^{14}\text{ GeV}$. 

The above discussed effect can interpreted in a Bayesian manner as a prior dependence.\footnote{Of course the effect from the Jacobian in the change of variables is also present in the frequentist case for a PDF on a parameter. On the other hand, the Jacobian does not affect the profile likelihood, though does affect the numerical sampling of it.} The uniform prior on $r$ places very little prior weight near $H_I^\mathrm{ derived}=0$, being proportional to the Jacobian. On the other hand, the uniform prior on $H_I$ places more weight near $r^\mathrm{(d)}=0$. This is caused by the specific form of the non-linear mapping between the parameters. A flat prior on the logarithm of $r$ would not show a preference concerning low $H_I$, since the relation between $\log H_I$ and $\log r$ is linear, but in this case the upper posterior limit depends on the lower prior limit for a one-sided bound.  

Using $H_I$ as a primary parameter preserves the one-sided shape of the distribution for $r^\mathrm{(d)}$, while using $r$ as a primary parameter introduces a shape change and bias in the derived distribution on $H_I$, see Fig.~\ref{fig:mc_toy_model}. We thus choose to use $H_I$ as our primary parameter with a uniform prior. It is the physical parameter we seek to constrain, which fixes the physical amplitudes of both the tensor and isocurvature spectra. This choice leads to upper limits to both $H_I$ and $r$ that are unaffected by the range of the prior. In addition this leads to unambiguous one-sided distributions for the primary and derived parameter. A one-sided limit on $H_I$ translates into a one sided limit on $r^{\mathrm(d)}$, which is slightly tighter than the case when $r$ is considered as the primary parameter.

\bibliographystyle{mnras}
\bibliography{axion_review}
\end{document}